\pdfoutput=1
\documentclass[useAMS,usenatbib,usegraphicx]{mn2e}
\usepackage{pslatex,color}
\usepackage{amsmath,amsfonts}
\newcommand{\Hfull}{\mathcal{H}}
\newcommand{\Hsec}{\mathcal{H}_{\idm{sec}}}
\newcommand{\Hrel}{\mathcal{H}_{\idm{GR}}}

\newcommand{\Hpert}{\mathcal{H}_{\idm{pert}}}
\newcommand{\Hmut}{\mathcal{H}_{\idm{NG}}}

\newcommand{\HKepl}{\mathcal{H}_{\idm{kepl}}}

\newcommand{\R}{\mathcal{R}\,}

\newcommand{\RP}{${\cal S}$}
\newcommand{\RPs}{${\cal P}_{S}$}
\newcommand{\RPc}{${\cal P}_{C}$}

\newcommand{\nAMD}{\mathcal{A}}
\def\vec#1{{\mathbf{#1}}}
\def\idm#1{{\mbox{\scriptsize #1}}}

\newcommand{\icrit}{i_{\idm{crit}}}
\newcommand{\imut}{i_{\idm{mut}}}
\newcommand{\mJ}{\mbox{m}_{\idm{J}}}
\newcommand{\au}{\mbox{au}}
\newcommand{\msun}{\mbox{m}_{\odot}}
\definecolor{darkred}{rgb}{0.8,0.0,0.2}
\def\edi#1{{\color{black} #1}}
\def\ed#1{{\color{black} #1}}
\def\corr#1{{\color{black} #1}}
\title[The relativistic dynamics of circumbinary planets]
{The non-resonant, relativistic dynamics of circumbinary planets}
\author[C. Migaszewski and K. Go\'zdziewski]{Cezary
Migaszewski$^{1}$\thanks{E-mail: c.migaszewski@astri.uni.torun.pl} and Krzysztof
Go\'zdziewski$^{1}$\footnotemark[1]\thanks{E-mail:
k.gozdziewski@astri.uni.torun.pl}\\
$^{1}$Toru\'n Centre for Astronomy, Nicolaus Copernicus University, 
Gagarin Str. 11, 87-100 Toru\'n, Poland}
\begin{document}
%
\date{Accepted 2010 September 13. Received 2010 September 10; in original form 2010 June 30}
\pagerange{\pageref{firstpage}--\pageref{lastpage}} \pubyear{2008}
\maketitle
\label{firstpage}
%
%
\begin{abstract}
%
We investigate the non-resonant, 3-D (spatial) model of the
\edi{hierarchical system composed of point-mass stellar (or sub-stellar)
binary and a low-mass companion (a circumbinary planet or a brown dwarf)}. 
We take into account the leading relativistic corrections to the Newtonian
gravity.  \edi{The secular model of the system relies on the expansion of
the perturbing Hamiltonian in terms of the ratio of semi-major axes
$\alpha$, averaged over the mean anomalies.} We found that the low-mass
object in a distant orbit may excite large eccentricity of the inner binary
when the mutual inclination of the orbits is larger than about of 60~deg. 
\edi{This is related to strong instability caused by a phenomenon which acts
similarly to the Lidov-Kozai resonance (LKR).} \edi{The secular system may
be strongly chaotic and its dynamics unpredictable over the long-term time
scale}.  Our study shows that in the Jupiter-- or brown dwarf-- mass regime
of the low-massive companion, the restricted model does not properly
describe the long-term dynamics in the vicinity of the LKR.  The
relativistic correction is significant for the parametric structure of a few
families of stationary solutions in this problem, in particular, for the
direct orbits configurations \edi{(with the mutual inclination 
less than 90 degrees)}.  We found that the dynamics of hierarchical systems
with small $\alpha \sim 0.01$ may be qualitatively different in the realm of
the Newtonian (classic) and relativistic models.  This holds true even for
relatively large masses of the secondaries.
\end{abstract}
%
%
%
%
\section{Introduction}
%
The extrasolar planets are discovered routinely\footnote {The recent 
detections and literature are frequently updated by Jean Schneider in 
www.exoplanet.eu}. Recently, about of 500 low-mass companions to stars of 
different spectral types are known. Most of them are bounded to  single 
stars. Moreover, there is also a growing number of planetary candidates in 
binaries as well as multi-stellar systems  \citep[see][for the statistical 
properties of planets in binaries]{Eggenberger2010}. Generally, following 
nomenclature in \citep{Rabl1988}, we can consider two classes of such 
multiple systems. In {\em the satellite case} or the S-type configuration, a 
planet revolves around one of the primaries in the binary,  and the second 
primary is much more distant. In {\em the cometary} or {\em circumbinary} 
configuration (C-type from hereafter), the planet has a wide orbit around 
the inner, massive binary. 

Current theories of planet formation in multiple stellar systems \citep
[e.g.,][and references therein]{Tekada2009} show that the inclination of a 
planetary orbit to the orbital plane of the binary may be non-zero in the 
C-type and the S-type systems. It is well known that in the S-type 
configurations, when the mutual inclination of circular orbits is larger 
than the critical value $i_{\idm{crit}} \sim 40^{\circ}$, the inner orbit 
undergoes large--amplitude oscillations of the eccentricity, which is in 
anti-phase with the mutual inclination. This  dynamical phenomenon is well 
known as the Kozai (or Lidov--Kozai) resonance \citep
[][]{Lidov1962,Kozai1962}. We will call it the LKR from hereafter. Keeping in 
mind the two types of possible orbital configurations, this instance of the 
LKR  may be also  understood  as  {\em the inner LKR} \citep[see, 
e.g.,][]{Krasinsky1972, Krasinsky1974}. Actually, many authors explain large 
eccentricities of some planetary candidates due to forcing by a 
distant star or massive companion \citep[see, 
e.g.,][]{Tamuz2008,Fabrycky2007}. In fact, the amplification of  
eccentricity and inclination may also appear in the C-type systems. The 
critical inclination is then  $\sim 60^{\circ}$, and it may be attributed to 
{\em the outer LKR} \citep[see, e.g.,][]{Krasinsky1972, 
Krasinsky1974,Migaszewski2010}, and this will be addressed further in 
this work. 

In the literature, the problem is most often considered as the restricted 
one, which means that the planet is a mass-less particle not perturbing the 
motion of the binary. It has been studied in many  papers, with different 
analytical and numerical techniques. The restricted model help us to 
simplify the analysis, nevertheless, the assumption of negligible influence 
of the planet on the motion of primaries may be not valid if the planet is 
large. In fact, low-mass objects in a few Jupiter-mass range are quite 
common. If the mutual interactions are significant, as we will show 
further in this work, the binary orbit may be  strongly perturbed by a 
distant, \edi{relatively massive planet or a brown dwarf moving in inclined, 
wide orbit, even if the mass of inner companion is {\em ten times} larger than the 
perturber.}

In this work, we focus on the unrestricted C-type problem by means of the 
secular model in terms of the semi-major axes ratio, $\alpha$. We assume 
that $\alpha$ is small ($<0.1$), hence we focus on hierarchical systems. In 
such a case, we  face very different times scales of the orbital evolution. 
Typically, the inner binary has the orbital period  counted in months or 
years but the period of the outer planet is at least $\sim 100$ times 
longer. Hence the short-term mean motion resonances (MMRs) are not present
in the system. 
The  secular evolution of the mean orbits, depending on the mutual 
interactions, is even much longer, and spans Myr time-scale. To follow the 
orbital evolution in all these time-scales in details, one could integrate 
the equations of motion numerically. Unfortunately, this direct, brute-force 
approach requires  too large CPU overhead. 

Because we are primarily interested in the long-term evolution of the 
hierarchical systems, the problem may be simplified with the help of the 
averaging principle. The Hamiltonian of the hierarchical system may be 
splitted onto integrable Keplerian part of the inner binary, and  for the 
perturbation part of the mutual interaction with the low-mass companion. 
Because the system is non-resonant, the perturbing Hamiltonian may be 
averaged out over the mean anomalies or the mean longitudes, which play the 
role of the fast angles.  After the averaging, we obtain the secular 
Hamiltonian describing the long-term evolution of the mean, slowly varying 
orbits. To obtain the secular model, we extend a simple averaging scheme 
(the so called {\em mixed anomalies method}) in our earlier paper \citep 
{Migaszewski2008} to the non-coplanar case.  The perturbing Hamiltonian is 
expanded in  power series with respect to $\alpha$, and then these 
series  may be averaged out term by term. We derived such averaged expansion 
of the secular Hamiltonian to the $10$-th order in $\alpha$. It is a more 
general version of the third-order (octupole-level) theory, studied in the 
planetary context in \citep{Ford2000,Lee2003} and of the integrable, second 
order (quadrupole-level) approximation in many papers \cite[e.g.,][and 
references 
therein]{Harrington1968,Krasinsky1972,Lidov1976,Ferrer1994,Farago2010}.

Our work is closely related to remarkable study by \cite{Michtchenko2006}, 
to which we will refer many times, as well as to our earlier paper \citep 
{Migaszewski2009a} devoted to the analysis of equilibria in the 
three-dimensional problem of two planets. Moreover, this work extends these 
papers in two important aspects. The averaging of the secular problem is 
done analytically, which simplifies and optimizes the computations, helping 
us to avoid numerical artifacts. Here, we also consider a generalization of 
the Newtonian model, by accounting for the leading non-Newtonian point-mass 
corrections to the perturbing Hamiltonian, i.e., the relativistic, 
post-Newtonian (PN) correction. It can be also easily averaged out over the 
mean anomalies. The PN corrections are particularly important if the 
frequencies of the slow angles, which they induce, are comparable with the 
frequencies causes by the {\em Newtonian} interactions \citep
[e.g.,][]{Adams2006,Fabrycky2007}. We shown already \citep{Migaszewski2009b} 
that accounting for the relativistic corrections  in the co-planar case of 
two-planet system may lead to qualitatively different global view of the 
phase space in both models. Hence, the PN correction might be also important 
in the 3D problem. Indeed, as we will show below, this apparently subtle 
effect induces global, qualitative changes of the structure of the phase 
space.

\edi{It should be emphasised here, that a large number of physical and 
orbital parameters fully characterising planetary configurations contradicts 
our desire to study the problem in possibly qualitative way, with the help 
of particular geometric tools. Hence, we restrict the work to specific 
ranges of these parameters, focusing on a ``typical'' binary with relatively 
large mass ratio of the primaries, as well as the circumbinary object in 
Jupiter/brown-dwarf mass range. Moreover, considering corrections to the 
Newtonian point-mass gravity, we only consider the relativistic effects, 
which, in turn, limit the orbital parameters of the binary. The conservative 
and dissipative tidal distortions are neglected here, though they might 
dominate in compact binaries, or in configurations with very hot-Jupiter or 
super-Earth planets \citep 
{Fabrycky2007,Mardling2007,Batygin2009,Ragozzine2009,Mardling2010}. In the 
range of semi-major axes $\sim 0.025$~au, the planetary tidal bulge raises 
apsidal rotation of the inner orbit which may may reach a few degrees per 
year, exceeding the effects of general relativity and the rotational stellar 
quadrupole by more than an order of magnitude \citep{Ragozzine2009}. 
However, in general, as we explain below, the rotational distortions 
introduce extra-degrees of freedom to the model (assuming that the stellar 
and planetary spins may be arbitrary), that cannot be treated in terms of 
geometric tools natural to investigate two-degrees of freedom Hamiltonian 
dynamics. Still, although the tidal effects could be basically treated in 
this formalism too, it would introduce new parameters (bodies' radii, Love 
numbers), hence we postpone investigations of this more general and complex 
model in future papers. Overall, as we show below, in the parameter ranges 
investigated here (semi-major axis of the binary  $\sim 0.1$--$0.2$~au and 
larger), the general relativity is dominant over the rotational and tidal 
corrections to the mutual Newtonian interactions. Yet we shall also 
demonstrate that our results may be quite easily scaled down to the regime 
of masses and semi-major axes typical for multi-planet configurations, and 
investigated in the coplanar case mostly. 
}

A plan of this paper is as follows. In Sections~2 and~3,  we derive the 3D 
secular model of the planetary system, in that the mean motion resonances of 
low order are absent, following the co-planar case considered in \citep
{Migaszewski2008}. We try to keep the presentation self-contained, therefore 
we recall basic facts regarding the dynamics of the secular model, which 
might be found in other papers already published. Section~3 describes a test 
of the secular approximation, and recalls the notion of the so called  
representative planes of initial conditions, as well as a scheme of 
investigating families of stationary solutions in the secular model. 
Section~4 is for the analysis of the eccentricity evolution and chaotic 
dynamics. Section~5 is devoted to a parametric survey of the equilibria in 
the classic (point-mass Newtonian) model. In Sect.~6, we study influence 
of the PN corrections on these solutions. In Conclusions, we summarize the 
results and sketch perspectives of the further research. 
%
%
\section{The secular 3D model of $N$-bodies}
%
We consider the general, spatial  model of the $N$-planet system
around a central star. It may be described
in terms of the Hamiltonian written with respect to
canonical Poincar\'e variables \citep{Michtchenko2006} in the form of
$
\Hfull = \HKepl + \Hpert,
$
where
\begin{equation}
\HKepl = \sum_{i=1}^{N} {\bigg( \frac{\mathbf{p}_i^2}{2 \beta^*_i} 
- \frac{\mu^*_i \beta^*_i}{r_i} \bigg)},
\label{HKepl}
\end{equation}
stands for the integrable part comprising of the direct sum of the relative,
Keplerian motions of point-mass secondaries $m_i$, $i=1,\ldots,N$, with respect
to the primary mass $m_0$.
We also define the mass
parameters ${\mu^*_i=k^2~(m_0+m_i)}$, where $k$ is the Gauss gravitational
constant, and ${\beta^*_i=(1/m_i+1/m_0)^{-1}}$ are the so called reduced masses.
The term $\Hpert$ stands for the  perturbing function  of the Keplerian
motions.  We assume that the perturbation is a sum of two terms:
\begin{equation}
\Hpert = \Hmut + \Hrel,
\label{Hpert}
\end{equation}
where $\Hmut$ is related to a {\em small} Newtonian mutual interactions between
$m_i$ and $m_j$, and we assume that $\Hpert/\HKepl \ll 1$. That may be
accomplished either by keeping $m_i$ small (then we have the {\em planetary
regime}) or permitting that secondary masses  are relatively large (even
comparable with the central object) and simultaneously requiring large
separations between particular orbits (the {\em stellar regime}). The term of
$\Hrel$ is for the leading general relativity (PN) corrections to the potential
of the central star  and the innermost companion. Here we focus on 
the non-resonant
systems, with well separated orbits, hence we may neglect the relativistic
post-Newtonian perturbations of the outer bodies. If the
semi-major axes ratios $\alpha_{i,j}=a_i/a_j<0.1$ are small, the
relativistic corrections for the distant objects are by orders of magnitude
smaller   than the PN perturbation of the inner binary. 

Following the notion of the Poincar\'e coordinates, the $\Hmut$ perturbation
may be written as follows:
\begin{equation}
\Hmut = \sum_{i=1}^{N-1} \sum_{j>i}^{N} {\bigg(
 - \underbrace{\frac{k^2 m_i m_j}{\Delta_{i,j}}}_{\textrm{\small direct part}} +
\underbrace{\frac{\mathbf{p}_i \cdot \mathbf{p}_j}{m_0}}_{\textrm{\small indirect
part}}\bigg)},
\label{Hmut}
\end{equation}
where  ${\mathbf{r}_i}$, $i=1,...,N$ are for the position vectors of $m_i$
relative to the central body, ${\mathbf{p}_i}$ are for their conjugate momenta
relative to the {\em barycenter} of the whole $(N+1)$-body system, 
${\Delta_{i,j}=\|\mathbf{r}_i-\mathbf{r}_j\|}$ denote the relative distance
between bodies $i$ and $j$.

After \citep{Richardson1988}, or developing the PN Hamiltonian  from the
general Lagrangian in \citep{Brumberg2007},  the post-Newtonian potential in the
PN formulation,  $\Hrel \equiv \beta^* \Hrel'$, where $\Hrel'$ has the following
form:
\begin{equation}
\Hrel' = \gamma_1 \vec{P}^4 + \gamma_2 \frac{\vec{P}^2}{r} + \gamma_3
\frac{\left(\vec{r} \cdot \vec{P}\right)^2}{r^3} + \gamma_4 \frac{1}{r^2},
\label{Hrel}
\end{equation}
with coefficients $\gamma_1,\gamma_2,\gamma_3,\gamma_4$:
\[
   \gamma_1 = - \frac{\left(1 - 3 \nu\right)}{8 c^2}, \quad
   \gamma_2 = - \frac{\mu^* \left(3 + \nu\right)}{2 c^2}, \quad
   \gamma_3 =   \frac{(\mu^*)^2}{2 c^2}, \quad
   \gamma_4 = - \frac{\mu^* \nu}{2 c^2},
\]
where $c$ stands 
for the speed of light in a vacuum, $\mu^* = k^2 (m_0 + m_1)$,  $\nu
\equiv m_0 m_1 / (m_0 + m_1)^2$, $\vec{P}$ is  the astrocentric momentum of the
innermost secondary (normalized through the reduced mass): 
\begin{equation} 
\vec{P} = \vec{v} + \frac{1}{c^2}
\left[ 4\gamma_1 (\vec{v} \cdot \vec{v}) \vec{v} + 
\frac{2\gamma_2}{r} \vec{v} + \frac{2\gamma_4}{r^3} (\vec{r} \cdot \vec{v})
\vec{r} \right],
\end{equation} 
and $ \vec{v} \equiv \dot{\vec{r}}$ stands for the astrocentric velocity of
the innermost object (still, assuming that the relativistic corrections from the
other bodies in the system are neglected). Hence,  $\vec{P} = \vec{v}$ with the
accuracy of $\vec{O}({c^{-2}})$ and then the relativistic  Hamiltonian is
conserved up to the order of $O({c^{-4}})$.

It is well known that the equations of motion of the $N$-body system with $N 
\geq 3$ are not integrable. However, making use of the assumptions above, we 
may apply the averaging proposition \citep{Arnold1993} to remove the short 
order perturbations, and to derive the equations of the secular dynamics, 
governing the long-term evolution of the {\em mean} orbital elements. 

To perform the averaging, the perturbation must be expressed in terms of the
canonical action--angle variables $(\vec{I},\vec{\phi})$:
\begin{equation}
 \Hfull(\vec{I},\vec{\phi}) = \HKepl(\vec{I}) + \Hpert(\vec{I},\vec{\phi}),
\label{poincare} 
\end{equation}
and we assume that $\Hpert(\vec{I},\vec{\phi}) \sim \epsilon \HKepl(\vec{I})$, 
where $\epsilon \ll 1$ is a small parameter. 
Here, we use the mass-weighted Delaunay elements \citep[e.g.,][]{Murray2000}:
\begin{eqnarray}
\label{dvars}
{l_i \equiv \mathcal{M}_i}, & \quad {L_i=\beta^*_i~\sqrt{\mu^*_i~a_i}},\nonumber\\
{g_i \equiv \omega_i}, & \quad  {G_i=L_i~\sqrt{1-e_i^2}},\\
{h_i \equiv \Omega_i}, & \quad {H_i=G_i~\cos~I_i},\nonumber
\end{eqnarray}
where $\mathcal{M}_i$ are the mean anomalies,  $a_i$ stand for canonical
semi-major axes,  $e_i$ are the eccentricities,   $I_i$ denote inclinations,
$\omega_i$ are the arguments of pericenter, and $\Omega_i$ denote the longitudes
of ascending node. The Hamiltonian of the $N$-planet system written in terms of the
these Delaunay variables (Eq.~\ref{dvars}) has the form of:
\begin{equation}
\Hfull = -\sum_{i=1}^N \frac{(\mu^*_i)^2 (\beta^*_i)^3}{2 L_i^2} +
\Hpert\underbrace{(L_i,l_i,G_i,g_i,H_i,h_i)}_{i=1,\ldots,N}.
\label{Hdvars}
\end{equation}
In this formulation, $l_i$ are the fast angles. They may be eliminated through
the averaging that is accomplished with:
\begin{equation}
\Hsec=\frac{1}{(2 \pi)^N}\underbrace{\int_0^{2 \pi} \ldots 
\int_0^{2\pi}}_{i=1,\ldots,N}{\Hpert \, d\mathcal{M}_1 \ldots d\mathcal{M}_N}.
\label{eq:secular}
\end{equation}
We should remember here that $\Hsec$ is valid only if (1) $\Hpert \sim 
\epsilon \HKepl$ (where $\epsilon \ll 1$ means a small parameter), and the 
averaging of the {\em unperturbed} Keplerian orbits is equivalent to 
performing the first step of the perturbation calculus \citep
{FerrazMello2007}, (2) there is no mixed resonances between the inner binary 
and the outer companion (e.g., between slow frequencies of the inner 
orbit and the mean motion of the outer orbit). \edi{We checked
that planetary systems studied in this paper obey these assumptions
within respective parameter ranges. These calculations rely on the averaged model, and
will be given below (see the end of Sect.~3).
}
Because the secular Hamiltonian $\Hsec$ does not depend on mean
anomalies $\mathcal{M}_i$, the conjugate momenta $L_i$ are
integrals of the secular problem. Obviously, the mean semi-major axes are also
constant, hence \corr{we get rid of $N$-degrees of freedom (DOF)}.
%
%
\section{Averaging the 3D model of $N$-bodies}
%
In \citep{Migaszewski2008}, we describe a simple scheme of the averaging the 
perturbing function $\Hpert$  (Eq.\ref{Hpert}) in coplanar case, which makes 
use of the very basic properties of the Keplerian motion, the mixed 
anomalies algorithm. This method may be 
easily applied to the 3D problem.  At first, we consider the {\em direct} 
part of the mutual interaction between the planets, $\Hmut$ (Eq. \ref
{Hmut}). The {\em indirect} part averages out to a constant and does not 
contribute to the secular dynamics \citep{Brouwer1961}.

The secular Hamiltonian may be written as a sum of terms 
representing mutual interactions between all pairs of bodies $i<j$, where
$a_i<a_j$:
\begin{equation}
\left<\Hmut\right> =
\sum_{i=1}^{N-1}  \sum_{j>i}^{N}{\left<\Hmut^{(i,j)}\right>}.
\label{eq:secular2}
\end{equation}
For a particular pair of planets, we calculate the following integral:
\begin{equation}
\left<\Hmut^{(i,j)}\right> = 
 \frac{1}{(2\pi)^2} \int_0^{2\pi} \int_0^{2\pi} 
-{\frac{k^2 m_i m_j}{\Delta_{i,j}} d\mathcal{M}_i d\mathcal{M}_j}.
\label{eq:rsec}
\end{equation}
Hence, the problem may be reduced to  averaging the inverse of the distance
$\Delta_{i,j}$ between two particular planets over their mean anomalies:
\begin{equation}
\Delta_{i,j} = \sqrt{r_i^2 + r_j^2 - 2 \vec{r}_i \cdot \vec{r}_j},
\label{squareroot1}
\end{equation}
where $\vec{r}_i$ and $\vec{r}_j$ must be expressed in a common reference 
frame $\mathcal{F}$.  The same vectors, written in the orbital reference 
frames $\mathcal{F}_i$ of each planet, are $\vec{r}_i\big|_{\mathcal{F}_i} = 
[x_i,y_i,0]^T$, and expressed in the common reference frame, they have the 
form of: $\vec{r}_i = \mathbb{A}_i \vec{r}_i\big|_{\mathcal{F}_i}$. Here,  
the rotation matrix $\mathbb{A}_i \equiv \mathbb{A}_i (\omega_i, \Omega_i, 
I_i)$ is the matrix product of  elementary Eulerian rotations  \citep
{Murray2000}:
\[
\mathbb{A}_i (\omega_i, \Omega_i,I_i) 
= \mathbb{P}_3(-\Omega_i) \, \mathbb{P}_1(-I_i) \, \mathbb{P}_3(-\omega_i).
\] 
Formulae~\ref{squareroot1} may be rewritten as follows:
\begin{equation}
\Delta_{i,j} = r_j \sqrt{1 - 2 \frac{1}{r_j} \vec{r}_i \cdot
\frac{\vec{r}_j}{r_j} + \left(\frac{r_i}{r_j}\right)^2}.
\label{squareroot2}
\end{equation}
Following \citep{Migaszewski2008}, we express the radius vector $\vec{r}_i$ 
of the inner body in each planetary pair with respect to the eccentric 
anomaly. The radius vector of the outer body in the pair is parametrised by 
the true anomaly.   This choice implies that $\Delta_{i,j}^{-1}$ expanded in 
Taylor series with respect to $\alpha$ is \ed{a trigonometric polynomial}. To 
compute the integral in Eq.~\ref{eq:rsec}, we also change the integration 
variables:
\[
d\mathcal{M}_i \equiv \mathcal{I}_i dE_i, \ \mbox{and} \
d\mathcal{M}_j \equiv \mathcal{J}_j df_j,
\]
where auxiliary functions appear:
\begin{equation}
\mathcal{I}_i = 1 - e_i \cos{E_i},\quad \mathcal{J}_j = \frac{(1 -
e_j)^{3/2}}{(1 + e_j \cos{f_j})^2}.
\end{equation}
Finally, the averaged mutual perturbation has the same general form as in the 
coplanar case \citep{Migaszewski2008}:
\begin{eqnarray}
\label{expansion}
& & \left<\Hmut^{(i,j)}\right> = -\frac{k^2 m_i m_j}{a_j} \times  \\
&& \quad \times \left[1 + \sqrt{1-e_j^2} \sum_{l=2}^{\infty}
{\mathcal{X}_{i,j}^l
\mathcal{R}^{(i,j)}_l(e_i,e_j,\omega_i,\omega_j,\Omega_i,\Omega_j, I_i,I_j)}\right],
\nonumber
\end{eqnarray}
although explicit expressions for $\R^{(i,j)}_l$ are obviously  different in 
the 3D model. The zeroth-order term in Eq.~\ref{expansion} is reduced to a 
constant and does not influence the secular equations of motion. Also the 
first order term vanishes identically. The remaining terms $\R^{(i,j)}_l$ 
have rather complex form.  In the simplest three-body system ($i \equiv 1, j 
\equiv 2$), we may express them in the Laplace reference frame. In this 
case, $\Delta{\Omega} = \pm \pi$ and $G_1 \sin{I_1} = G_2 \sin{I_2}$ \citep
[see e.g.,][]{Michtchenko2006}. It is also natural to introduce the mutual 
inclination, $i_{\idm{mut}} \equiv I_1 + I_2$.  Then the $\R^{(i,j)}_l$
-terms of the order of $2$ and $3$ may be identified with the {\em 
quadrupole} and {\em octupole} terms,  respectively \citep[see, 
e.g.,][]{Ford2000,Lee2003,Farago2010}. The quadrupole-level term is the 
following:
\begin{equation}
\label{quadropole} 
\R^{(1,2)}_2 = \frac{1}{8} 
  D_1 \left(2 + 3 e_1^2\right)  
-\frac{15}{16} e_1^2 D_2 \cos{2\omega_1}, 
\end{equation}
where $C_I \equiv \cos{i_{\idm{mut}}}$, and 
$
D_1 = (3 C_I^2 - 1)/2,\quad 
D_2 = C_I^2 - 1. 
$
The third order (octupoloe-level) term reads as follows:
\begin{eqnarray}
\label{octupole}
\R^{(1,2)}_3 &=& -\frac{15}{64} D_6 e_1 e_2 \cos{\Delta{\varpi}} 
\left(3 e_1^2 + 4\right) \nonumber \\
& &  -\frac{525}{256} D_3 D_2 e_1^3 e_2 \cos (3 \omega_1-\omega_2) \nonumber \\
& &  -\frac{525}{512} D_4 D_2 e_1^3 e_2 \cos (3 \omega_1+\omega_2) \\
& & +\frac{15}{128} D_5 e_1 e_2 \cos (\omega_1+\omega_2) \nonumber \\
& &  +\frac{45}{512} D_5 e_1^3 e_2 \cos (\omega_1+\omega_2)\nonumber,
\end{eqnarray}
where coefficients $D_j$ are:
\begin{eqnarray*}
&&D_3 = (1 + C_I)/2, \quad 
D_4 = 1 - C_I,\nonumber\\ 
&&D_5 = -15 C_I^3 + 5 C_I^2 + 11 C_I - 1, \\
&&D_6 = (15 C_I^3 + 5 C_I^2 - 11 C_I - 1)/8.
\end{eqnarray*}
Equations~\ref{quadropole} and \ref{octupole} are written similarly to terms
appearing in the coplanar model [see equations (22) and (23) in
\citep{Migaszewski2008}]. Clearly,
if $i_{\idm{mut}}=0$ then $D_1=D_3=D_6=1$, $D_2=D_4=D_5=0$, and formulae
$\R^{(i,j)}_2$, $\R^{(i,j)}_3$ coincide with those ones of
in the coplanar problem.

An explicit expansion of $\Hsec$ shows that the quadrupole-order term in 
$\alpha$ introduces the evolution of eccentricity $e_1$,  and in this 
approximation, the outer eccentric  $e_2$ is constant.  The variation of  
the outer eccentricity may be only introduced through the third order 
(octupole) and higher terms.  Indeed, up to the quadrupole approximation, 
the secular Hamiltonian does not depend on $\omega_2$ (the cyclic angle),  
and the eccentricity of the outer body becomes an additional integral of 
motion. In this case, the problem can be reduced to one DOF 
and is integrable \citep{Lidov1976}. This feature has been accounted for in 
many recent papers, moreover, the apparently subtle third-order perturbation 
to the Keplerian model, or the first order perturbation to the integrable 
quadrupole-order  approximation may introduce qualitative changes of the 
dynamics.

We calculated the secular expansion (Eq.~\ref{expansion}) up to the 10-th 
order\footnote{This expansion is available on request in the form of a raw 
MATHEMATICA input file; it will be also available on-line, after publishing 
this manuscript.}. One should be aware that by increasing the order of this 
expansion, we do not necessarily improve the approximation of the secular 
model of the real system, because this model is still limited by the first 
order perturbation theory. In Section~3.2 we will examine the accuracy of 
the secular expansion in more details. 

Finally, the averaged relativistic correction possesses the same form as in 
coplanar case \citep{Migaszewski2008}. Moreover, we include this 
perturbation only to the  mutual interaction  of masses $m_0$ and $m_1$: %
\begin{equation} 
\left<\Hrel\right> = -\frac{3 (\mu^*_1)^4 (\beta_1^*)^5}{c^2 L_1^3 G_1} +
\mbox{const},
\label{avbHrel}
\end{equation}
as it was explained above.

\edi{
Having the averaged model in hand, we may calculate the secular
frequencies of the inner companion, and compare them with the mean motion of the outer
object ($n_2$). For the relativistic advance of the inner periastron we have:
\[
\frac{f_{1,\idm{rel}}}{n_2} = \frac{3\mu^*_1}{c^2 a_1} \sqrt{\frac{\mu^*_1}{\mu^*_2}}
\alpha^{-3/2} \frac{1}{1-e_1^2},
\]
and for the apsidal motion forced by the mutual interaction of the
inner and outer companion (in the quadrupole approximation):
\[
\frac{f_{1,\idm{mut}}}{n_2} =    \sqrt{\frac{\mu^*_1}{\mu^*_2}}
\frac{m_2}{m_0} \alpha^{3/2} \frac{1}{\sqrt{1-e_2^2}} \Lambda_1 + 
   \frac{m_1}{m_0} \alpha^{2} \frac{1}{(1-e_2^2)^2} \Lambda_2, 
\]
where $\Lambda_{1,2}$ are the following functions of the geometric elements:
\begin{eqnarray*}
\Lambda_1 &=& \frac{3}{8} (1-e_1^2) \left[ (3 C_I^2-1) - 5 (C_I^2-1) \cos 2\omega_1\right] \\
&+& \frac{3}{8} C_I^2 \left[ (2 + 3 e_1^2) - 5 e_1^2 \cos 2\omega_1\right], \\
\Lambda_2 &=& \frac{3}{8} C_I \left[ (2 + 3 e_1^2) - 5 e_1^2 \cos 2\omega_1\right].
\end{eqnarray*}
Assuming now that $\mu^*_{1,2} \sim k^2 m_0$ (the central mass dominates),
we may obtain the following estimates of the secular frequencies in terms
of the characteristic units in our model: the relativistic frequency relative
to $n_2$ is
\[
\frac{f_{1,\idm{rel}}}{n_2} = 2\times 10^{-5} \left( \frac{m_2}{1 m_{\sun}}\right)
\left( \frac{0.2 \mbox{au}}{a_1}\right)
\left( \frac{0.04 \mbox{au}}{\alpha}\right)^{3/2} \,  \frac{1}{1-e_1^2}, 
\]
while the mutually forced frequency relative to $n_2$:
\begin{eqnarray*}
\frac{f_{1,\idm{mut}}}{n_2} &=& 8\times 10^{-5} \left( \frac{m_2}{10 m_{\idm{J}}}\right)
\left( \frac{1 m_{\sun}}{m_0}\right)
\left( \frac{\alpha}{0.04}\right)^{3/2} \\ &\times& \frac{1}{(1-e_2^2)^{3/2} \sqrt{1-e_1^2}} \Lambda_1 \\
&+& 2\times 10^{-4} \left( \frac{m_1}{100 m_{\idm{J}}}\right)
\left( \frac{1 m_{\sun}}{m_0}\right)
\left( \frac{\alpha}{0.04}\right)^{2}  \frac{1}{(1-e_2^2)^{2}} \Lambda_2.
\end{eqnarray*}
These frequencies may be compared  with the tidal apsidal frequency induced
by the primary and the inner body bulge (Migaszewski \& Go\'zdziewski 2010, in preparation):
\begin{eqnarray*}
\frac{f_{1,\idm{tid}}}{n_2} 
&=& \bigg\lbrace
4 \times 10^{-8}
\left( \frac{m_1}{100 m_{\idm{J}}} \right)
\left( \frac{1 m_{\sun}}{m_0} \right)
\left( \frac{R_0}{1 R_{\sun}} \right)^5
\left( \frac{k_{L,0}}{0.03} \right)  \\
&+& 7 \times 10^{-9}
\left( \frac{m_0}{1 m_{\sun}} \right)
\left( \frac{100 m_{\idm{J}}}{m_1} \right)
\left( \frac{R_1}{2 R_{\idm{J}}} \right)^5
\left( \frac{k_{L,1}}{0.15} \right)  \bigg\rbrace \\
& \times & 
\left( \frac{0.2 \mbox{au}}{a_1} \right)^5 
\left( \frac{0.04}{\alpha} \right)^{3/2}
\frac{1 + 3 e^2/2 + e^4/8}{(1-e^2)^5},
\end{eqnarray*}
where $R_0$ is the stellar radius, $R_1$ is the radius of the inner 
secondary and $k_{L,0}$, $k_{L,1}$ are tidal Love numbers of these bodies. 
Let us choose a reference model through setting characteristic parameters of 
$a_1 \sim 0.2$~au, $\alpha \sim 0.04$, $m_0 \sim 1 m_{\sun}$,  $m_1 \sim 100 
m_{\idm{J}}$, $R_0 \sim 1 R_{\sun}$, $R_1 \sim 2 R_{\idm{J}}$. Assuming that 
the bodies are modeled by polytropes with indices of $3$ and $2$, respectively, 
we compute their Love numbers, $\sim 0.03$ for the primary, and $\sim 0.15$ for 
the inner secondary. Then setting $e_1 \sim 0$, we obtain that the relativistic 
frequency is comparable with the mutual, Newtonian frequency, while the mean 
motion of the outer secondary is orders of magnitude larger than both of them 
(hence no mixed resonances are possible). Simultaneously, the assumptions of 
the averaging principle are well fulfilled. This guarantees that the 
evolution of the mean (secular) system closely follows the real configuration 
over the time scale of order $\sim 1/\epsilon$, where $\epsilon$ is the small 
parameter of the perturbation.
}

\edi{
Under the same 
assumptions, the tidal frequency is orders of magnitude smaller than the 
leading frequencies of the mutual (Newtonian) and relativistic
corrections. This means that the tidal effect is negligible, indeed,
as far as the model parameters do not strongly deviate from 
the characteristic values, as defined above. 
}

%
%
\subsection{A global, 2-dim representation of the phase space}
%
Because the general planetary $N$-body problem is very complex,  we restrict the
further analysis to its simplest, non-trivial case of three bodies
(\edi{``non-trivial'' in the sense of its non-integrability}). We shall
consider configurations of the host star and two planets or C-type systems 
comprising of a binary and a more distant body (a planet).  

We recall that the secular Hamiltonian $\Hsec$ of the three body problem 
does not depend on $\mathcal{M}_1, \mathcal{M}_2$, therefore the conjugate 
actions $(L_1, L_2)$ are constant. The Hamiltonian $\Hfull$ written in the 
Laplace reference frame  depends on $\Delta{\Omega}=\Omega_2-\Omega_1 \equiv 
\pm \pi$ only, not on $\Omega_1$ and $\Omega_2$ separately.  Hence, the 
following canonical transformation \cite[e.g.,][]{Michtchenko2006}:
\begin{equation}
\begin{array}{l}
(\omega_1,G_1)\\
(\omega_2,G_2)\\
(\Omega_1,H_1)\\
(\Omega_2,H_2)
\end{array}
\quad \Rightarrow \quad
\begin{array}{l}
(\omega_1,G_1)\\
(\omega_2,G_2)\\
\left(\theta_1=\frac{1}{2}(\Omega_1+\Omega_2), J_1=H_1+H_2 \right)\\
\left(\theta_2=\frac{1}{2}(\Omega_1-\Omega_2), J_2=H_1-H_2 \right)
\end{array}
\label{trans}
\end{equation}
removes $\Omega_1, \Omega_2$ from the secular Hamiltonian. After this 
transformation it does not depend on $\theta_1$, therefore $J_1 \equiv 
|\mathbf{C}|=C=\mbox{const}$, where $\mathbf{C}$ is the total angular momentum 
of the system.  Moreover, $\theta_2=\pm \pi/2=\mbox{const}$ (after the Jacobi 
reduction of nodes) and $J_2$ may be expressed as a function of $G_1, G_2$ 
and $J_1$ in the following form: 
\[
J_2 = (G_1^2 - G_2^2)/J_1.
\] 
Therefore, for constant values of the angular momentum $J_1\equiv C$ and the 
secular energy $\Hsec$, the secular dynamics are reduced to two DOF  
Hamiltonian system.  Instead of the total angular momentum $C$, we shall use 
the so called Angular Momentum Deficit ($AMD$) \citep{Laskar2000}:  
\[ 
AMD = L_1 + L_2 - C, 
\] 
or its {\em normalized} value of $\nAMD \equiv AMD / (L_1 + L_2)$, $\nAMD \in
[0,1]$ \citep{Migaszewski2009a}.
Because $L_1$, $L_2$ and $C$ are integrals of the secular system, the 
relativistic correction, Eq.~\ref{avbHrel} does not change $\nAMD$, and the 
DOF number does not change. Because $\left<\Hrel\right>$ depends on $G_1$ 
only, thus it affects only the temporal evolution of $\omega_1$.  We note 
here that the perturbation induced by the quadrupole moment of the star, 
which was discussed \citep{Migaszewski2009b} in the coplanar case, also 
depends on $H_1$ in the 3D problem, i.e. on the orbital inclination to the 
equatorial plane of the star. This perturbation introduces an additional 
frequency to $\Omega_1$ and  then $\Delta{\Omega}$ is not constant anymore. 
It means that we could not perform the reduction of nodes and the secular 
problem would have three DOF.  \corr{This also means that the Laplace 
reference frame, defined in terms of the total orbital angular momentum, 
does not possess any constant orientation in space}. Being aware of this 
problem, we do not consider the dynamical flattening of the star and/or of 
the innermost planet. The two DOF model is then less general but the 
dynamics are better tractable, thanks to the geometrical tools, which can be 
applied to study this basic, low-dimensional problem.

If we fix the secular Hamiltonian in the form of $\Hsec \equiv 
\Hsec(G_1,G_2,\omega_1,\omega_2)$, then  $\nAMD$ may be considered as a free 
parameter of the secular model.  Moreover, the phase space is {\em 
four-dimensional}, and to represent the phase space of the system globally in 
terms of two-dimensional sections, which are easy to visualize, we 
follow a concept of {\em the representative plane of initial conditions}~
\citep {Michtchenko2004}, the $\Sigma$-plane from hereafter. The $\Sigma$
-plane may be chosen in different ways, although all representations may be 
fixed and defined through the following conditions:
\begin{equation}
\frac{\partial{\Hsec}}{\partial{\omega_1}} = 0, \quad 
\frac{\partial{\Hsec}}{\partial{\omega_2}} = 0. 
\label{eq:zero}
\end{equation}
These conditions imply that all phase trajectories of the secular system 
cross the $\Sigma$-plane \citep{Michtchenko2006,Libert2007b}, see also our 
explanation in \citep{Migaszewski2009a}.  In accord with the symmetries in 
the secular 3D model, the solutions to these equations are the following 
four pairs of angles:
\begin{equation}
(\omega_1, \omega_2) = \lbrace(0,0); (0,\pm\pi); (\pm\pi/2,\pm\pi/2);
(\pm\pi/2,\mp\pi/2)\rbrace,
\end{equation}
that also define four distinct quarters of the $\Sigma$-plane, numbered with 
\ed{Roman numbers II, I, IV, and III, respectively}, see \citep
{Michtchenko2006} for details.  \edi{Let us note that
no other combinations of the angles are permitted by Eqs.~\ref{eq:zero}. This feature of the
secular system flows from the symmetry of the secular
Hamiltonian with respect to the apsidal lines of the mean orbits, and may be also justified by the explicit form of the equations of motion 
derived from the expansion of the perturbing
Hamiltonian, see \citep{Michtchenko2006,Migaszewski2009a} for details.}

In this sense, the $\Sigma$-plane may be thought as an analogue of the 
Poincar\'e cross section. The conditions fixing the characteristic plane may 
be also rewritten as follows:
\[
\cos{\omega_1}=\cos{\omega_2}=0 \ \cup  \ \sin{\omega_1}=\sin{\omega_2}=0.
\] 
Further, we shall use three, basically equivalent geometric representations of
the $\Sigma$-plane, which cover certain combinations of the quarters
(the solution pairs of the pericenter arguments):
\begin{itemize}
\item the \RPs-plane is defined with  $\cos{\omega_1}=\cos{\omega_2}=0$, and
\begin{equation}
\textrm{\RPs} = \lbrace x = e_1 \sin{\omega_1}, y = e_2 \sin{\omega_2}, e_1
\in [0,1), e_2 \in [0,1)\rbrace,
\end{equation}
\item the \RPc-plane is defined with  $\sin{\omega_1}=\sin{\omega_2}=0$, and
\begin{equation}
\textrm{\RPc} = \lbrace x = e_1 \cos{\omega_1}, y = e_2 \cos{\omega_2}, e_1
\in [0,1), e_2 \in [0,1)\rbrace,
\end{equation}
\item and, finally, two incarnations of the \RP-plane:
\begin{equation}
\textrm{\RP} = \lbrace x = e_1 \cos{\Delta\varpi}, y = e_2 \cos{2\omega_1}, e_1
\in [0,1), e_2 \in [0,1)\rbrace, 
\end{equation}
\begin{equation}
\textrm{\RP'} = \lbrace x = e_1 \cos{2\omega_1} , y = e_2 \cos{\Delta\varpi}, e_1
\in [0,1), e_2 \in [0,1)\rbrace,
\end{equation}
\end{itemize}
that was defined originally in \citep{Michtchenko2006}.  
It can be shown, that the  \RPs{}- and
\RPc{}-planes carry out the same information as the \RP{}-plane. However, the
\RP{}-plane has a discontinuity along the $x$-axis, and the former representations
are sometimes more convenient to the analysis of solutions of the secular
system.
%
%
\subsection{A test of the analytic model}
\begin{figure*}
 \centerline{
 \vbox{
    \hbox{\includegraphics [width=42mm]{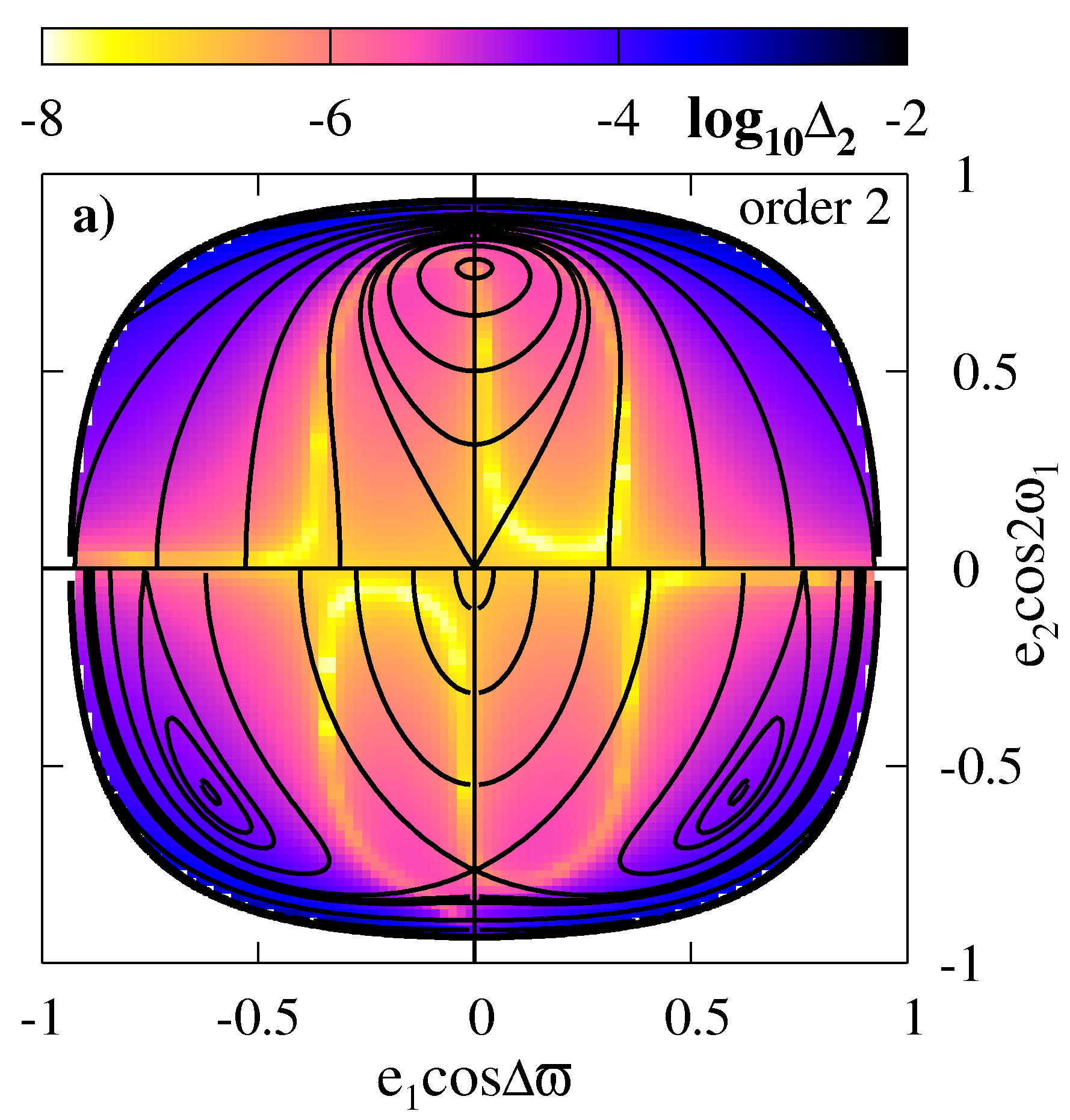}
          \includegraphics [width=42mm]{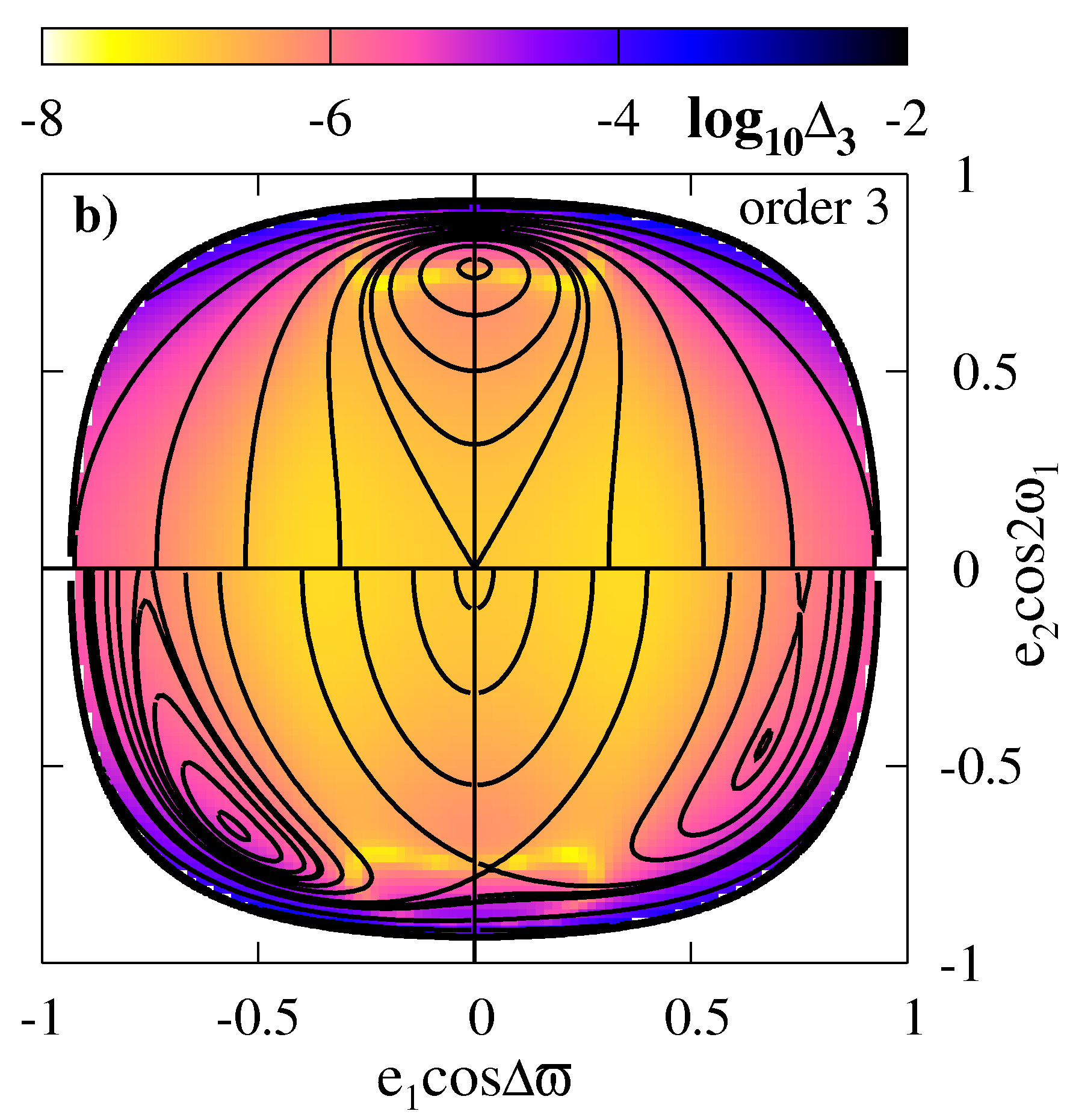}
          \includegraphics [width=42mm]{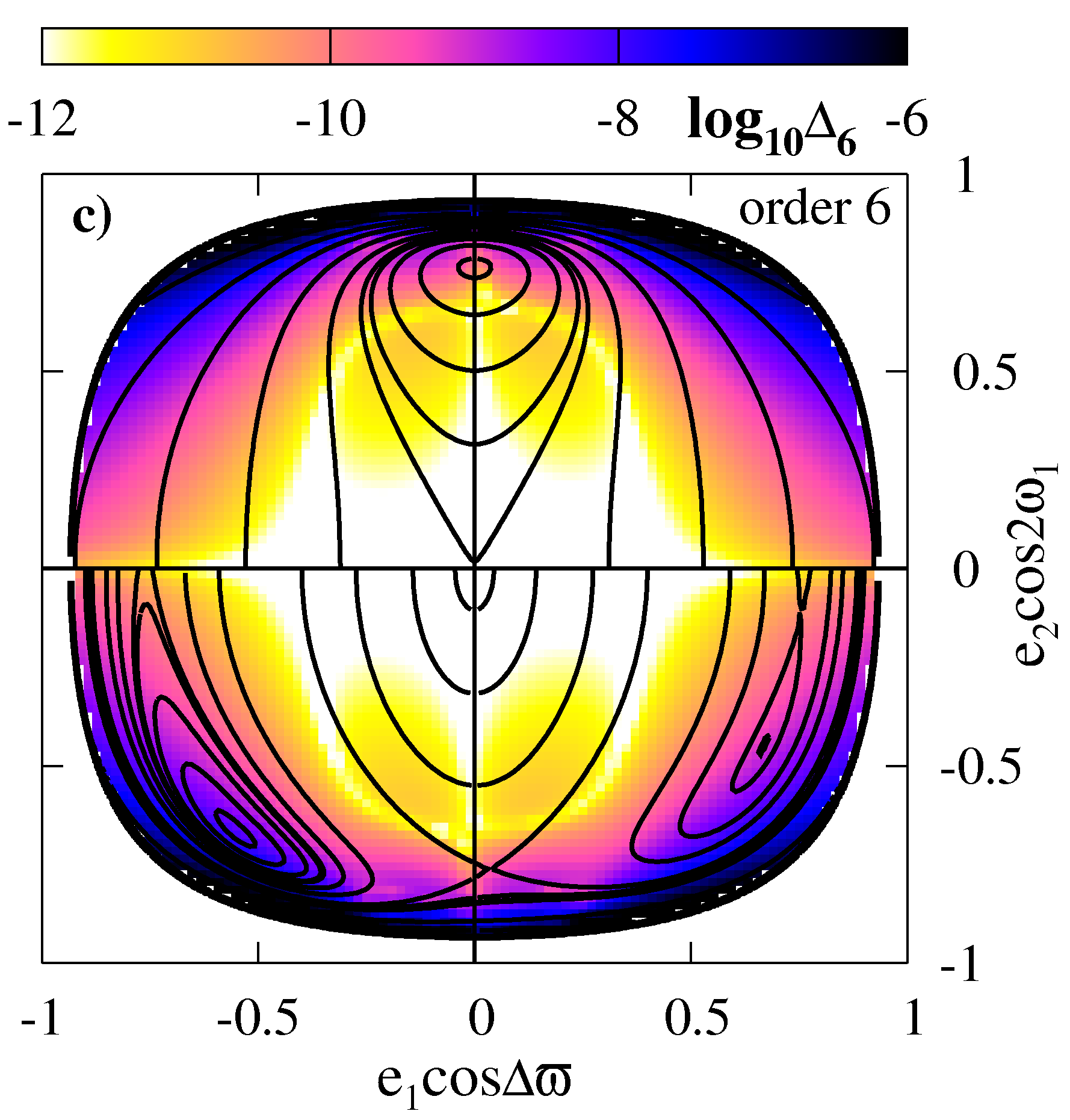}
          \includegraphics [width=42mm]{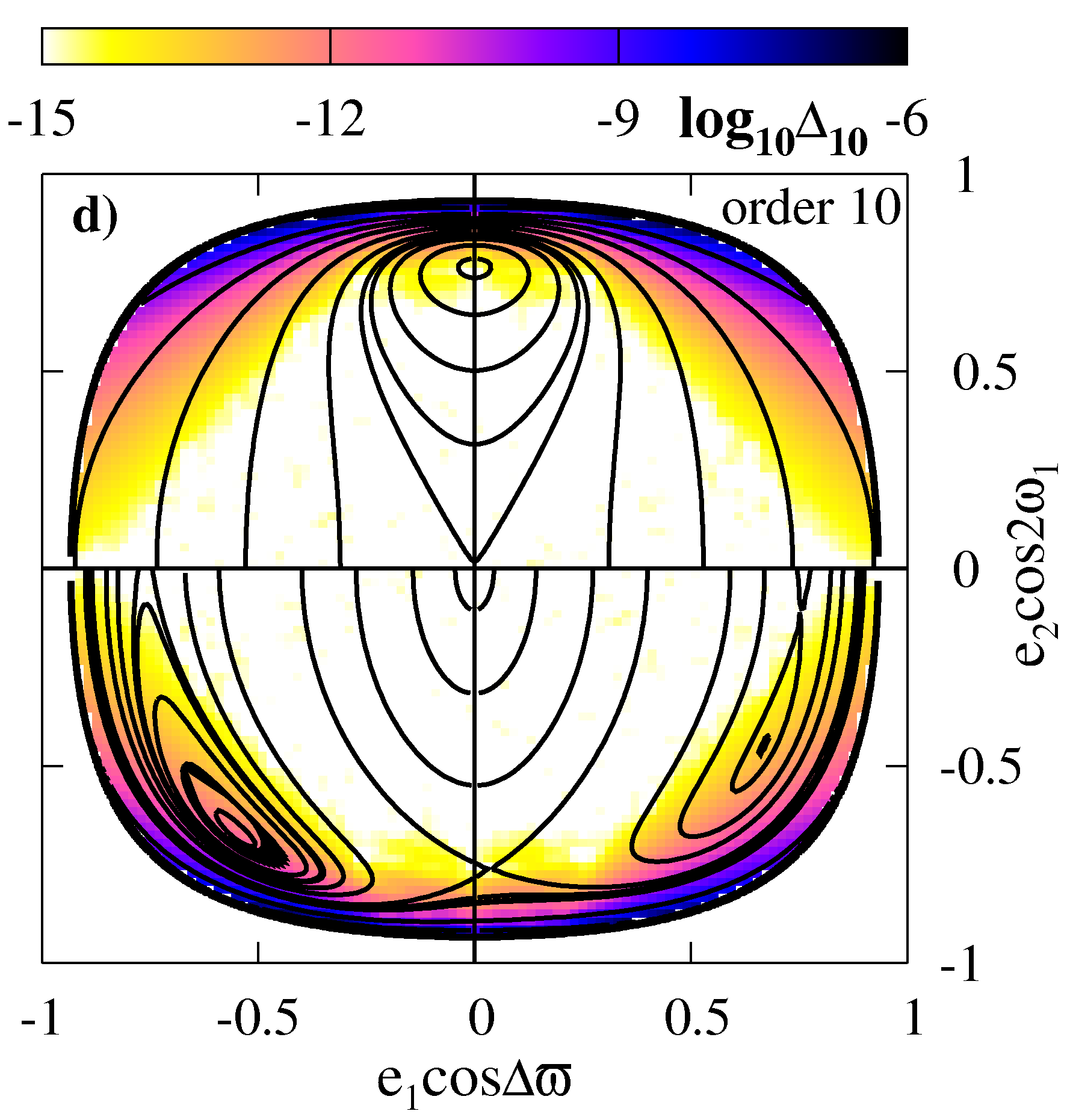}
	 }
      }
 }
\caption{
The precision of the analytic theory in terms of $\Delta_l$, Eq.~\ref{deltal},
which is color-coded in the \RP-plane. Parameters of the planetary system are as
follows:  $\alpha = 0.04, \mu = 5, \nAMD = 0.32$. Panels from (a) to (d) are
for the expansions in $\alpha$  of the second, third, sixth, and tenth order,
respectively.
}
\label{fig1}
\end{figure*}

%
We left a test of the accuracy of the secular expansion to this end, because 
the introduced $\Sigma$-planes are useful to illustrate the results of this 
test in a global manner. We select initial conditions in the \RP{}-plane, 
and the secular energy computed with the help of the analytic expansion is 
compared with the results of numerical averaging developed in \citep
{Gronchi1998,Michtchenko2004}, which are exact up to the numerical 
quadrature error. We consider the non-relativistic case only, because the 
secular relativistic correction is exact (with the first non-zero 
post-Newtonian term included), and it does not influence the precision of 
analytic formulae. Figure~\ref{fig1} shows the levels of $\Hsec$, marked 
with solid curves in the \RP{}-plane. Each panel is for a different order of 
the analytic approximation. The relative difference between values of the 
mean  Hamiltonians, derived through the analytic (``A'') and numerical 
(``N'') algorithms may be defined as follows:
\begin{equation}
\Delta_l \equiv \left\|\frac{\Hsec^{\textrm{A}}(l) -
\Hsec^{\textrm{N}}}{\Hsec^{\textrm{N}}}\right\|,
\label{deltal}
\end{equation}
where $l$ is the order of the analytic expansion in $\alpha$. The  results 
of this comparison are illustrated in Fig.~\ref{fig1} that shows the levels 
of $\Delta_l$ computed  for a hierarchical system with $\alpha=0.04$ and $\mu
\equiv m_1/m_2 = 5$.  The quadrupole-level 
model reproduces the secular Hamiltonian \ed{as the even function with 
respect to both $x \equiv e_1 \{\sin,\cos\} \omega_1$ and $y \equiv e_2 
\{\sin,\cos\}\omega_2$}. The higher order approximations of $\Hsec$ broke 
this symmetry. We have shown in \citep{Migaszewski2009a} that the shape of 
$\Hsec$ significantly depends on $\alpha$. This is more important in the 
spatial problem, because even for relatively small $\alpha$, the quadrupole 
model distorts the structure of the phase-space (see Sect.~4.1).
An inspection of Fig.~\ref{fig1} reveals that the octupole model 
reconstructs the secular Hamiltonian and its shape in the \RP{}-plane very 
well, because the largest deviations $\Delta_3 \sim 10^{-3}$  appear only 
for $e_1 \sim 1$. In other parts of the representative plane, $\Delta_3 \sim 
10^{-5}$. The  high-order expansions are obviously even more precise.
\edi{Following estimates of the secular frequencies in the
relevant parameter ranges (see Sect.~3), the averaging principle assures us that
the orbital evolution of the secular system 
follows closely the ``real'' orbits. Hence,
we may quite safely skip a comparison of the results from both approaches by direct
numerical integrations.}

The  octupole model introduces the
first order perturbation to the integrable quadrupole model, hence, because
it is very precise in the range of small $\alpha$,  we further focus
on this most simple, non-trivial case.
%
%
\subsection{Equilibria in the secular 3D problem}
%
In this work, we focus on the simplest class of solutions that are the 
equilibria (or stationary solutions). These solutions imply the basic 
structure of the phase space. By determining their stability, we may derive 
the local structure of neighboring phase trajectories through relatively 
simple analysis. The equilibria of the non-resonant system in terms of the 
quadrupole approximations were studied in the past, \cite 
[e.g.,][]{Krasinsky1974,Lidov1976}. In our earlier work \citep
{Migaszewski2009a},  we classified families of equilibria emerging in the 
two-planet, non-planar problem, in terms of the quasi-analytic averaging, 
and basically exact secular model.  We studied a few families of equilibria 
known in the literature, e.g., the zero-eccentricity solutions and the 
Lidov-Kozai resonance. We also found new families of these equilibria, in 
particular,  the so called \textit{chained orbits} solution that could be 
hardly derived with the perturbative approach, although the results in \cite
{Gronchi1998} might be applied here. This work focus on the planetary regime 
of parameters $\mu, \alpha$, the mass ratio $\mu$ was 
restricted to the range of $[0.1,2]$ and $\alpha \in [0.1,0.667]$. Moreover, 
we explored the whole  permitted range of $\nAMD \in [0,1]$. Yet we learned 
that the semi-analytic approach has serious disadvantages. All calculations, 
including the integrations of the equations of motions, and the stability 
analysis must be performed with the help of numerical algorithms. This may 
introduce large errors and hinders the analytic, qualitative analysis of the 
problem.

In this section, we extend the study of stationary solutions in \citep
{Migaszewski2009a} to a wider range of the  mass ratio, $\mu > 2$. 
Simultaneously, we consider smaller $\alpha < 0.1$. That makes it possible 
to apply the analytic model described and tested in Sect.~3. Because 
planets  emerge most likely from remnants of a thin protoplanetary disk, we 
also restrict our attention to mutual inclinations up to $\imut \sim 
90^{\circ}$ (direct orbits).  In that range, we may find families of 
equilibria related to the zero-eccentricity orbits, and the LK resonance 
classified in \citep{Migaszewski2009a} as  solutions of family IVa, 
accompanied by families~IIIa, IIIb, and IVb+. Our analysis is also 
restricted to the initial conditions in  quarters III and IV of the \RP{}
-plane, in which these solutions may only ``reside''.  This region of the 
parameter space has been studied \citep [e.g.,][]{Krasinsky1974} with the 
quadrupole-level model \edi{which is somehow the next to trivial,
non-interacting Keplerian 
approximation of the three-body orbits}. However, as we show below, this 
approximation may introduce artifacts due to \edi{generally non-realistic} symmetry of 
the secular Hamiltonian. Obviously,  to avoid this problem, higher order 
expansions are required. We also attempt to  extend the results of \cite
{Ford2000}, who applied the octupole-level theory to the analysis of 
hierarchical planetary systems with very small $\alpha \sim 0.01$.
\begin{figure*}
 \centerline{
 \vbox{
    \hbox{\includegraphics [width=56mm]{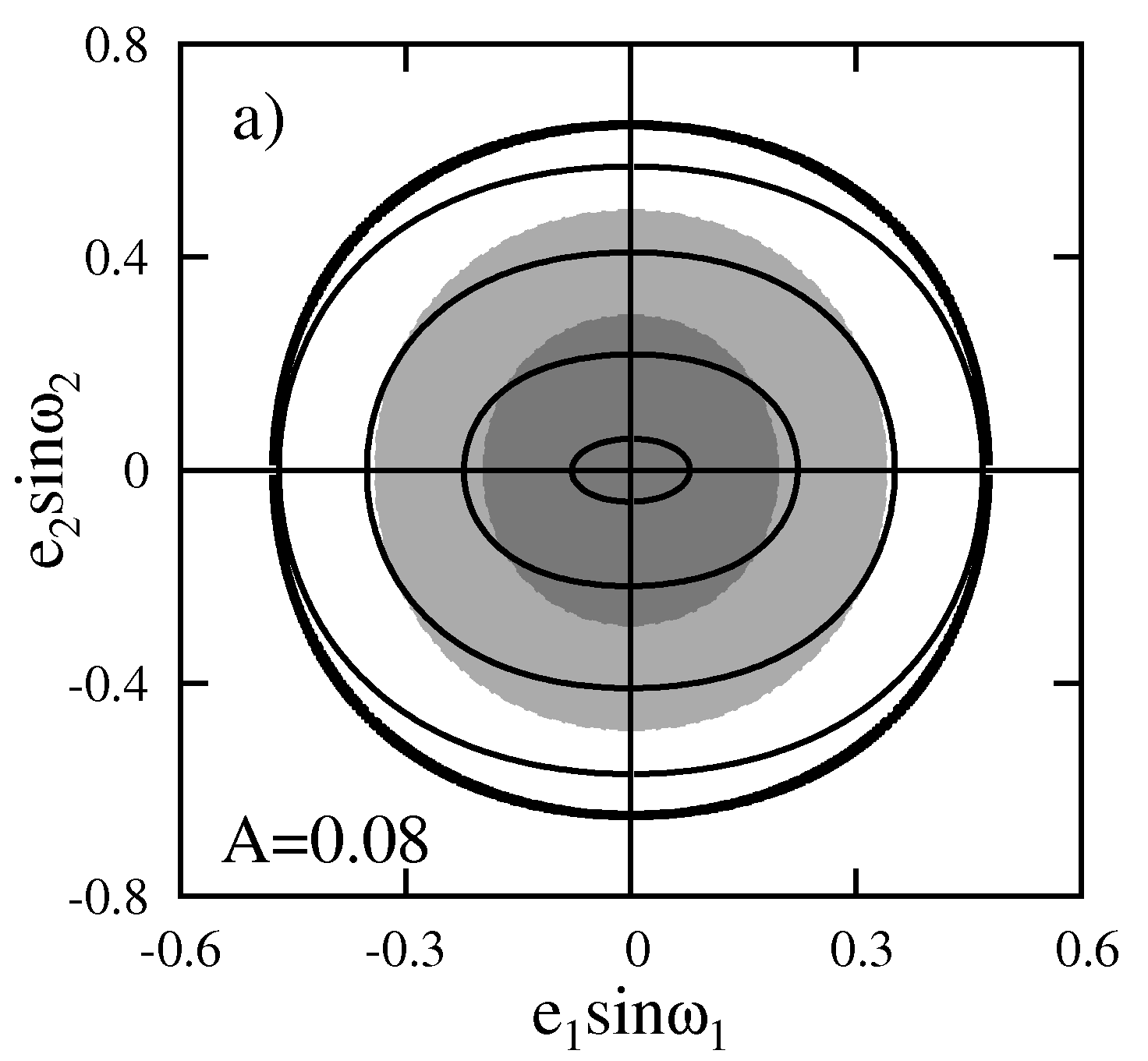}\hspace*{2mm}
          \includegraphics [width=56mm]{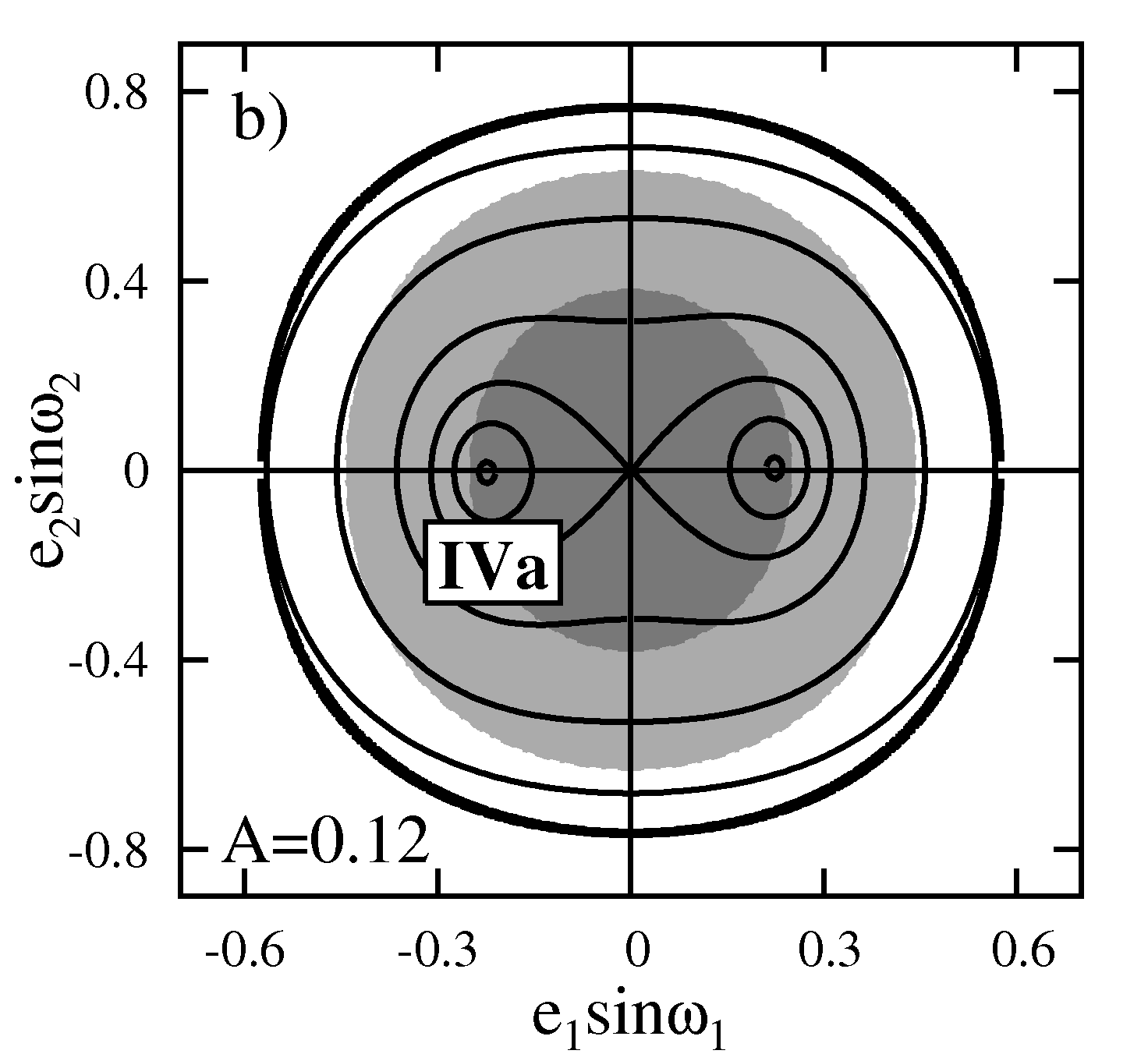}\hspace*{2mm}
          \includegraphics [width=56mm]{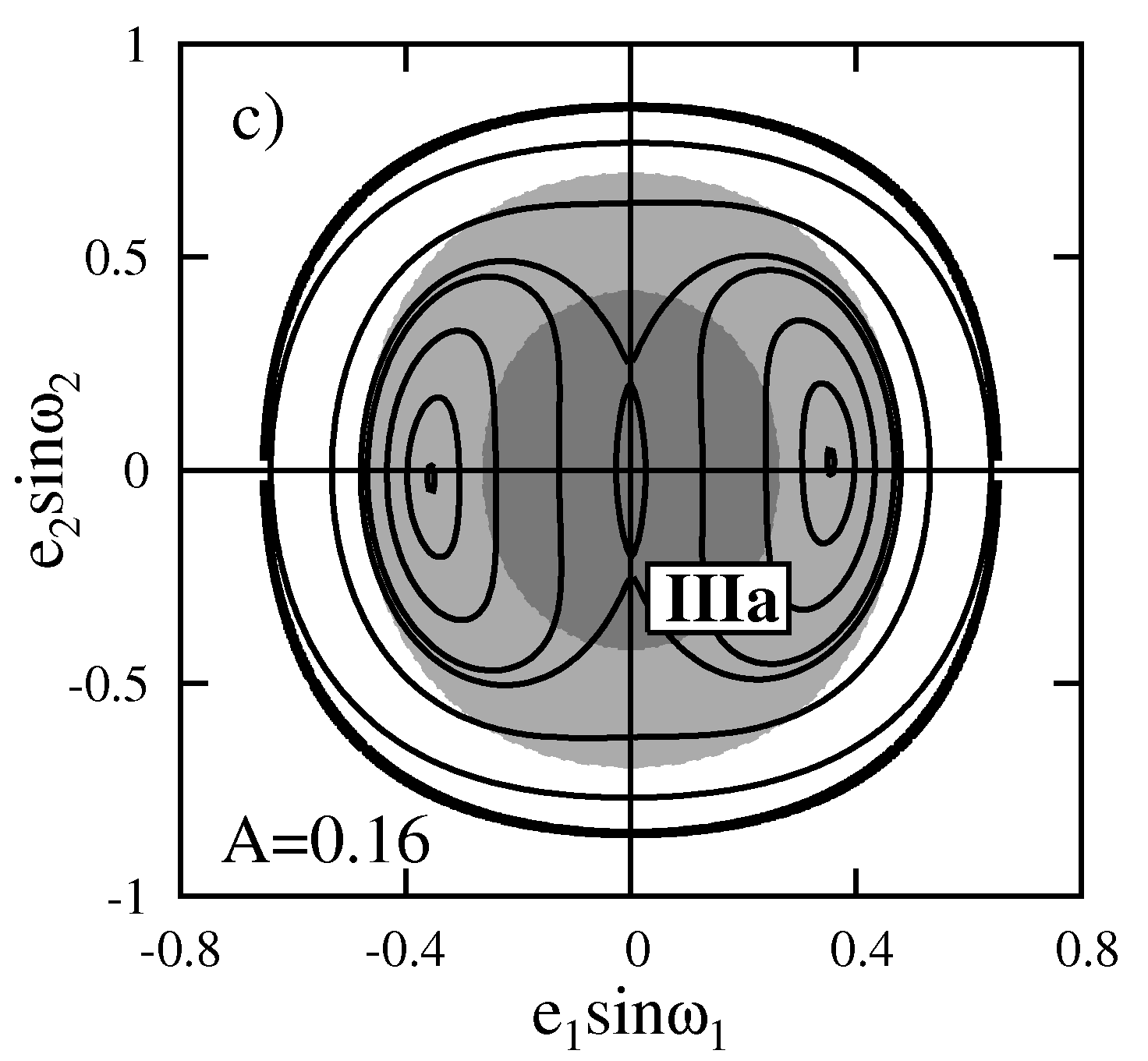}}
    \hbox{\includegraphics [width=56mm]{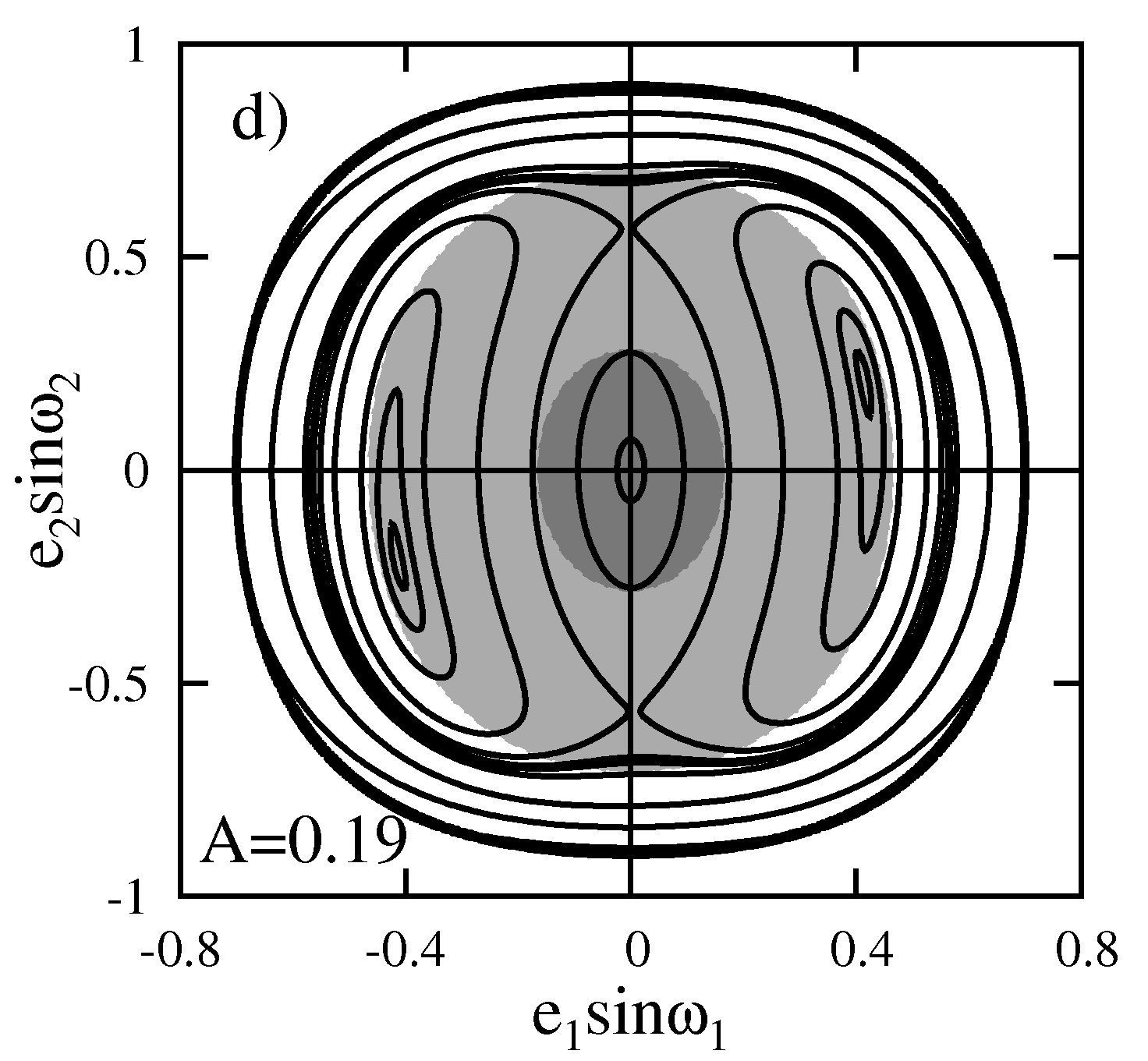}\hspace*{2mm}
          \includegraphics [width=56mm]{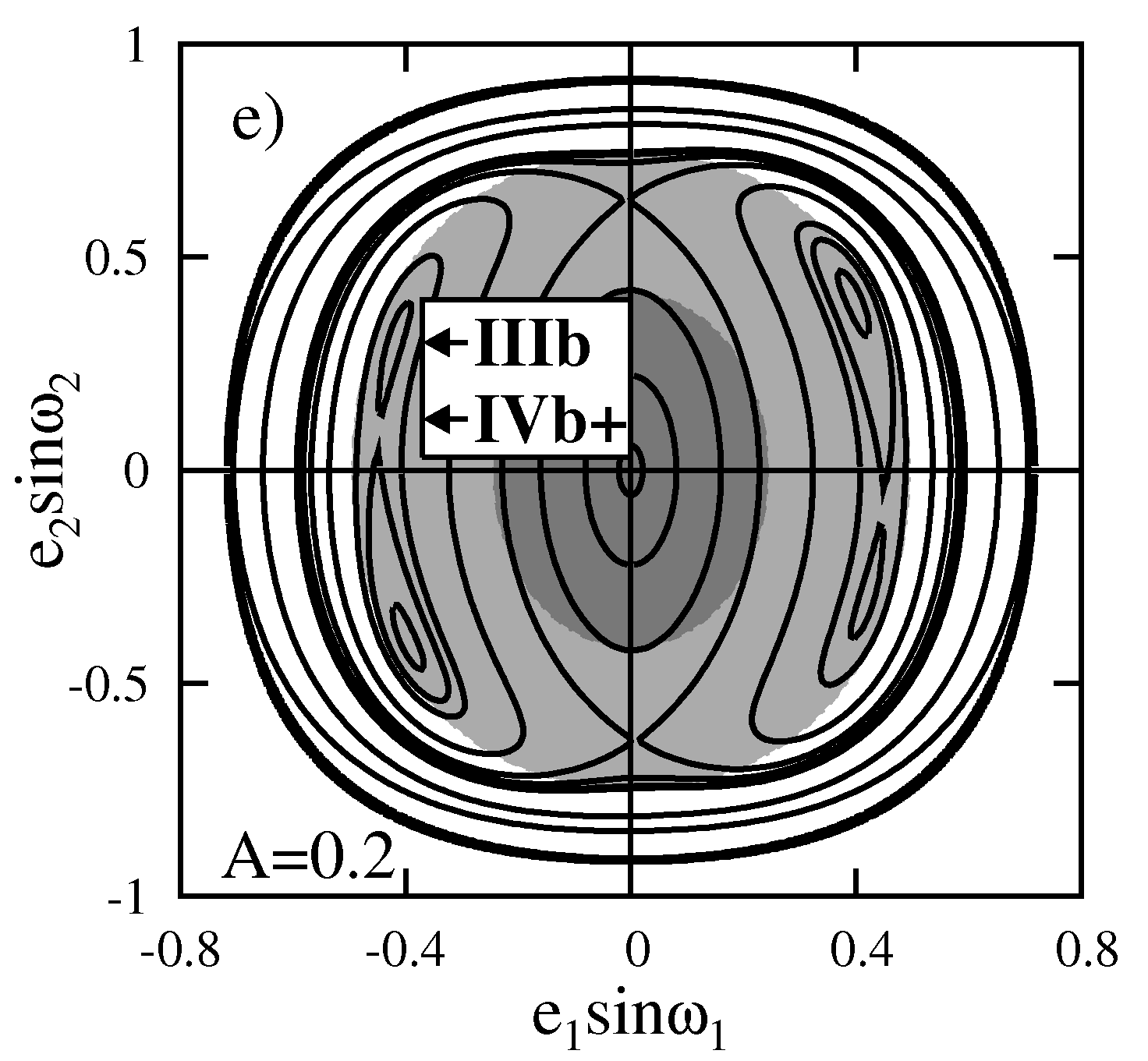}\hspace*{2mm}
          \includegraphics [width=56mm]{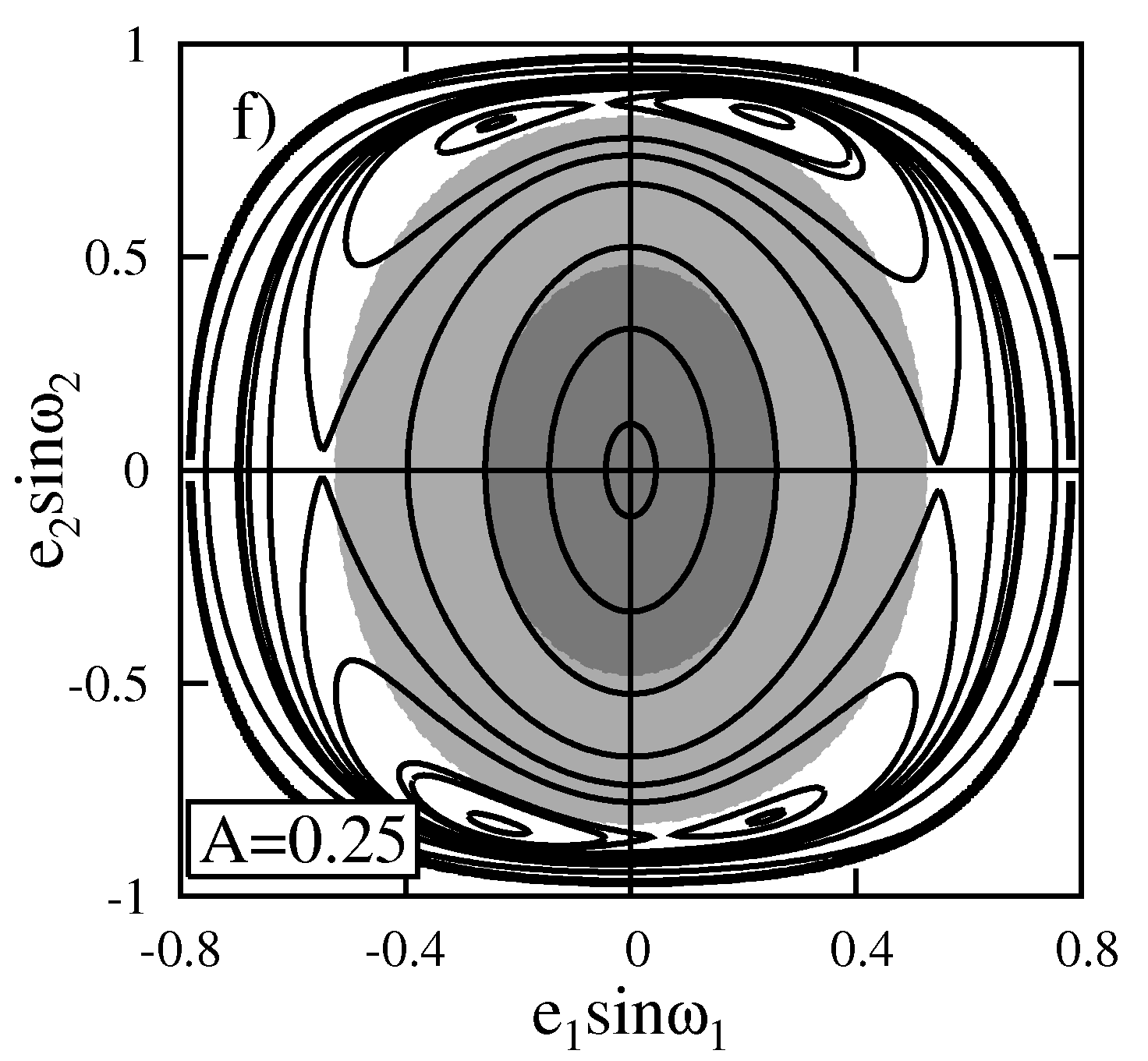}}
     }
}
\caption{
Levels of $\Hsec$ at the \RPs-plane calculated  for $\alpha=0.01$ and 
$\mu=20$. Panels from a) to f) are for different values of  
$\nAMD=\{0.08,0.12,0.16,0.19,0.2,0.25\}$, respectively. These values of 
$\nAMD$ imply the mutual inclination at the origin 
$i_0=\{49^{\circ},61^{\circ},70^{\circ},\ed{77^{\circ}}, 79^{\circ}, 
\ed{89^{\circ}}\}$, respectively.   The intensity of shaded ares encodes  
the mutual inclination in prescribed ranges, darker shade is for larger 
$\imut$. The inclinations ranges are $\imut=(35^{\circ},45^{\circ}), 
(40^{\circ},55^{\circ}), (50^{\circ},65^{\circ}), 
\ed{(60^{\circ},75^{\circ})}, (60^{\circ},75^{\circ}), 
\ed{(70^{\circ},85^{\circ})}$ in subsequent panels.
}
\label{fig2}
\end{figure*}

\begin{figure}
\centerline{
\vbox{
    \hbox{\includegraphics [width=85mm]{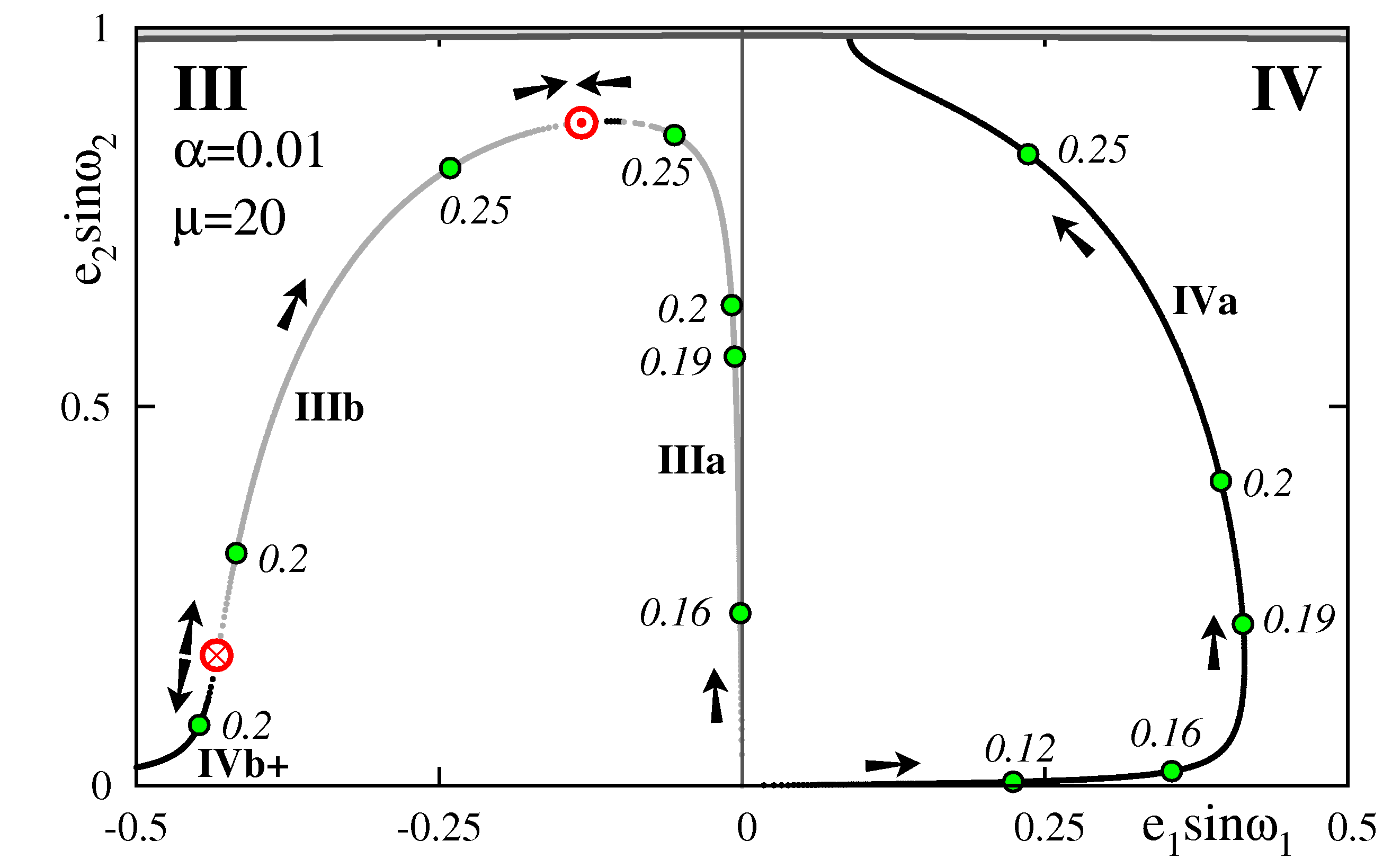}}
         }
}
\caption{
Families of stationary solutions (IVa, IIIa, IIIb, IVb+). Semi-major axes 
ratio $\alpha=0.01$, mass ratio $\mu=20$. Dark dots are for stable 
equilibria, grey dots are for unstable equilibria. Crossed and dotted 
circles mark bifurcations. Small arrows shows the direction of  particular 
stationary solution with increasing $\nAMD$. Green dots are for the 
positions of equilibria for particular values of $\nAMD = 0.12, 0.16, 0.19, 0.2, 
0.25$ (the energy levels are
shown in Fig.~\ref{fig2}). Solutions are for classic model, and were obtained with the 
help of the octupole theory.
}
\label{fig3}
\end{figure}

To show the generic properties of the relevant families of equilibria, we 
begin with an example that is illustrated in Figure~\ref{fig2}. Panels in 
this figure show levels of $\Hsec$ at the \RPs-plane for $\alpha=0.01$ and 
$\mu=20$ and a few different values of $\nAMD$.  Panel~\ref{fig2}a was 
derived for $\nAMD=0.08$, and it reveals the zero-eccentricity equilibrium. 
For larger $\nAMD=0.12$ (Fig.~\ref{fig2}b) the figure-eight structure  of 
Lidov--Kozai equilibrium appears, and is labeled with IVa. \corr{We recall, 
that the mutual inclination of circular orbits that corresponds to the LK 
bifurcation will be called the {\em critical inclination}, $i_{\idm{crit}}$ 
although, in general, any such value of the mutual inclination that leads to 
a bifurcation of equilibria has the sense of being ``critical'' \citep
{Krasinsky1972,Krasinsky1974}.} For larger value of  $\nAMD=0.16$ (Fig.~\ref
{fig2}c) the LKR ``moves'' towards larger $e_1$ (simultaneously, $e_2 \sim 0$
)  and a new saddle point appears. We call this solution a member of family 
IIIa. \ed{For larger $\nAMD=0.19$ (Fig.~\ref{fig2}d), these structures
still expand, and for $\nAMD=0.2$ (Fig.~\ref{fig2}e)}  two new equilibria 
emerge: one is associated with quasi-elliptic point (family IIIb from 
hereafter), and a saddle of family IVb+. \corr{Apparently, it emerges from a 
point near the IVa solution in the \RPs-plane, nevertheless it does {\em 
not} correspond to  a bifurcation of this equilibrium, see \citep
{Migaszewski2009a}}. Solution IVa moves towards larger $e_2$, see 
Fig.~\ref{fig2}f).

The {\em parametric paths} of these equilibria  in terms of $\nAMD{}$, may 
be depicted in the \RPs{}-plane (Fig.~\ref{fig3}).  Black and grey curves are 
for the stable and unstable equilibria, respectively. The relevant families 
of equilibria are labeled with Roman numbers and Latin letters. 
The 
direction of ``motion'' of particular solutions along the $\nAMD$-axis is 
marked with arrows.  For a reference, positions of the equilibria for a few 
discrete values of $\nAMD=0.12, 0.16, 0.19, 0.2, 0.25$ (corresponding
to subsequent panels in Fig.~\ref{fig2}), are marked with green dots 
and labeled. Following a particular evolution path of equilibrium IVa, we 
see that it appears for $\nAMD \sim 0.1$ through a bifurcation of the 
origin. When $\nAMD$ increases, this solution moves along $e_2 \sim 0$ 
towards larger $e_1$. For $\nAMD \sim 0.17$, it reaches the maximal $e_1 
\sim 0.4$, and it turns back, towards smaller $e_1$ with simultaneous 
increase of $e_2$. When $\nAMD$ increases even more, this solution reaches 
the border of convergence of the analytic expansion. We call that border 
{\em the anti-collision line} [see \citep{Migaszewski2008}].

The parametric evolution of equilibria in quadrant III is more complex. The 
first non-trivial stationary solution appears for $\nAMD \sim 0.15$, which 
is unstable, saddle point IIIa. It evolves along $e_1 \sim 0$ and then $e_2$ 
increases to large values. For $\nAMD \sim 0.195$, two new solutions IIIb and 
IVb+  emerge from the elliptic structure  related to the LKR (see Fig.~\ref
{fig2}e).  One of them is stable (solution IVb+) and  the another one is 
unstable (solution IIIb). They appear around $(e_1 \sim 0.45, e_2 \sim 0.15)$
.  When $\nAMD$ increases, the solution IVb+ moves towards $e_1 \rightarrow 
1$ and $e_2 \rightarrow 0$.  Simultaneously,  equilibrium~IIIb is shifted
towards increasing $e_2$ and decreasing  $e_1$. Then also IIIa increases 
$e_2$ and leaves off the $e_1=0$ axis. Finally, for $\nAMD \sim 0.26$, 
equilibria IIIa and IIIb merge at one point $(e_1 \sim 0.13, e_2 \sim 0.8)$.
%
%
\section{The secular chaos}
%
Following \citep{Michtchenko2006}, we shown in \citep{Migaszewski2009a} that 
the averaged 3D model may involve strongly chaotic motions, in the secular 
time-scale. To study in details, how the model parameters influence the 
structure of the \RP-plane, and how it relates to the long-term chaotic 
phenomena, we apply the fast indicator approach.  Among many numerical tools 
of this kind, we choose the so called  {\em coefficient of the diffusion of 
fundamental frequencies} $\sigma$ introduced by \citep{Laskar1990}. To check 
whether a phase space trajectory of a quasi-integrable Hamiltonian system is 
regular (quasi-periodic) or irregular (chaotic), one integrates the 
equations of motion over two subsequent  intervals of time, e.g., $[0,T]$ 
and $[T,2T]$. Next, we resolve the frequencies in the discrete orbital 
signal with the help of refined FFT analysis \citep{Laskar1990}, obtaining 
two estimates of a given frequency, say $f_T$ and $f_{2T}$, over these two 
intervals of time. The coefficient of the diffusion of the fundamental 
frequency is then defined through:
\[
 \sigma =  \left\| \frac{f_T}{f_{2T}}-1 \right\|.
\]
Clearly, if the signal does not change over time, $\sigma \sim 0$ and this 
means that the phase trajectory is quasi-periodic (stable). If $\sigma$ is 
significantly different from 0, the trajectory is chaotic and regarded 
unstable. In our calculations, we used a variant of the frequency analysis 
developed by \citep{Sidlichovsky1996}, which is called the Frequency 
Modified Fourier Transform (FMFT). We also used publicly available code of 
the FMFT algorithm, kindly provided by David {Nesvorn{\'y}} on his personal 
web-page\footnote{http://www.boulder.swri.edu/$\sim$davidn/}. 

Because the secular evolution is associated with 
$\omega_i$ angles, we compute 
the $\sigma$ coefficient on the basis of complex time-series 
$
 \{ G_i(t) \exp \mbox{i}\omega_i(t) \}, \ldots i=1,2,
$
where $\mbox{i}$ is the imaginary unit. In this signal, the osculating 
eccentricity and pericenter argument for each planet \edi{represent 
rescaled  canonical action-angle variables}. Hence, resolving its Fourier components, we may 
determine the leading amplitudes (proper eccentricities)  and the fundamental 
frequencies of pericenter angles. 

Next, we did massive integrations of the secular equations of motion. The 
initial conditions were selected at the grid of $200 \times 100$ data points 
of the \RPs-plane. At each point of the dynamical map, we integrated the 
secular trajectory over $\sim 10^4$ secular periods with respect to the {\em 
smaller} frequency (typically, one of the fundamental frequencies is much 
larger than the other one). Having the computed phase trajectory, we then 
find an estimate $\sigma$, as well as the maximal eccentricities attained by 
both orbits during the integration time (the so called $\max e$ indicator) 
as well as \corr{the amplitude of variation of the mutual inclination}  
$\Delta i \equiv (\max i_{\idm{mut}} - \min i_{\idm{mut}})$ attained during the 
integration time. These geometrical characteristics are very useful to 
understand the source of instabilities indicated and detected by the 
diffusion coefficient $\sigma$. 
\begin{figure*}
 \centerline{
 \vbox{
    \hbox{\includegraphics [width=84mm]{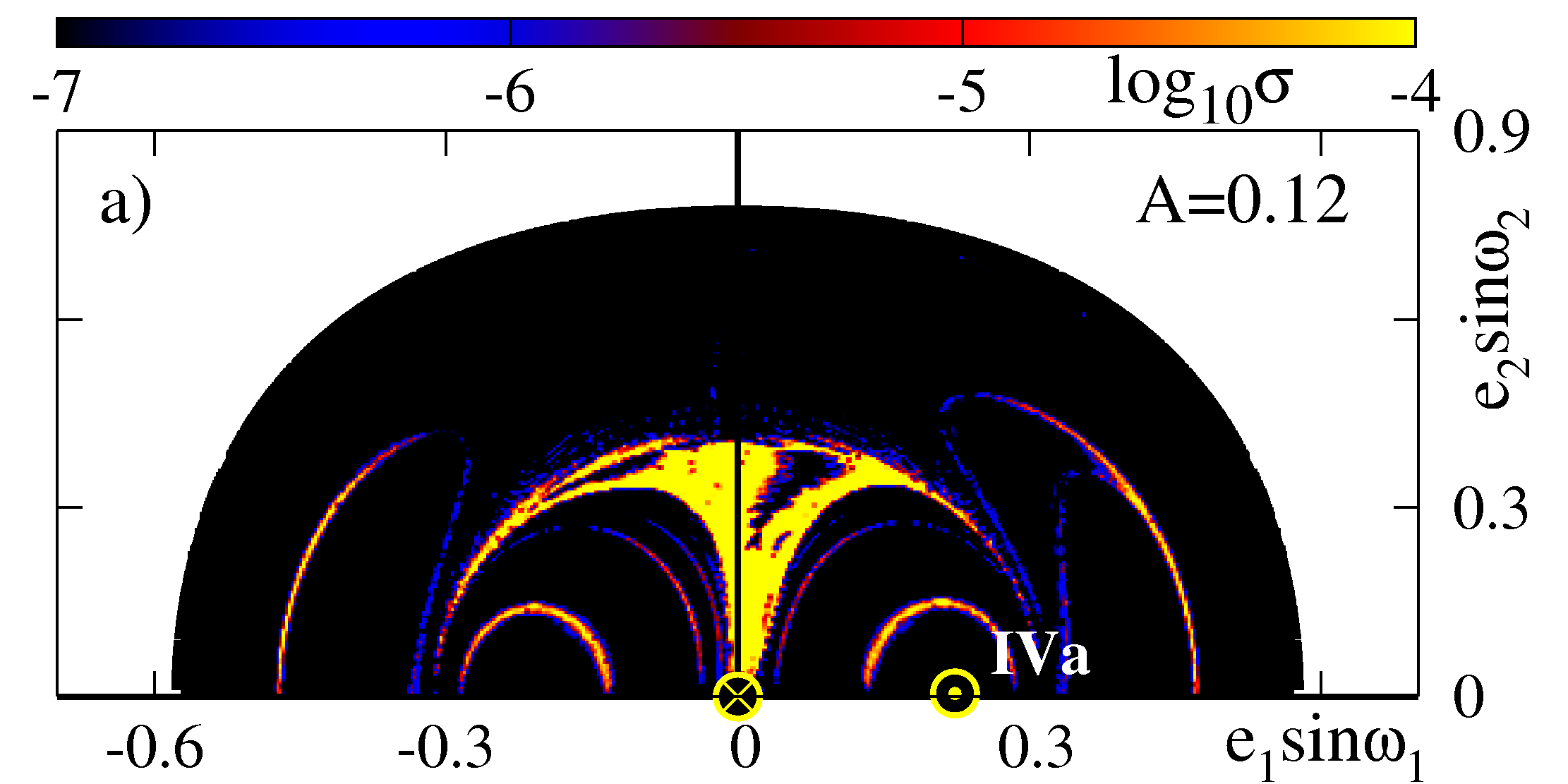}\hspace*{2mm}
          \includegraphics [width=84mm]{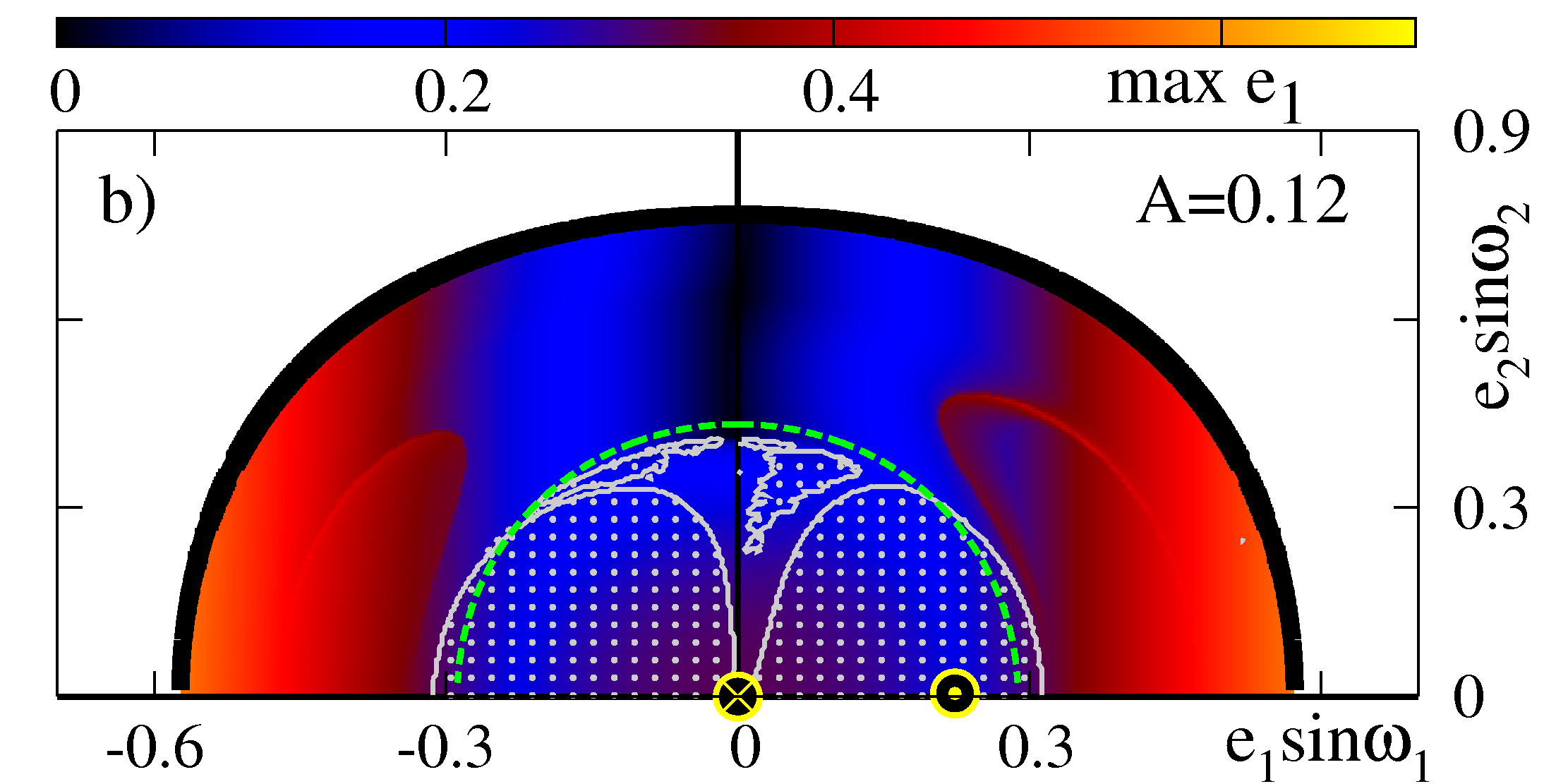}
	  }
    \hbox{\includegraphics [width=84mm]{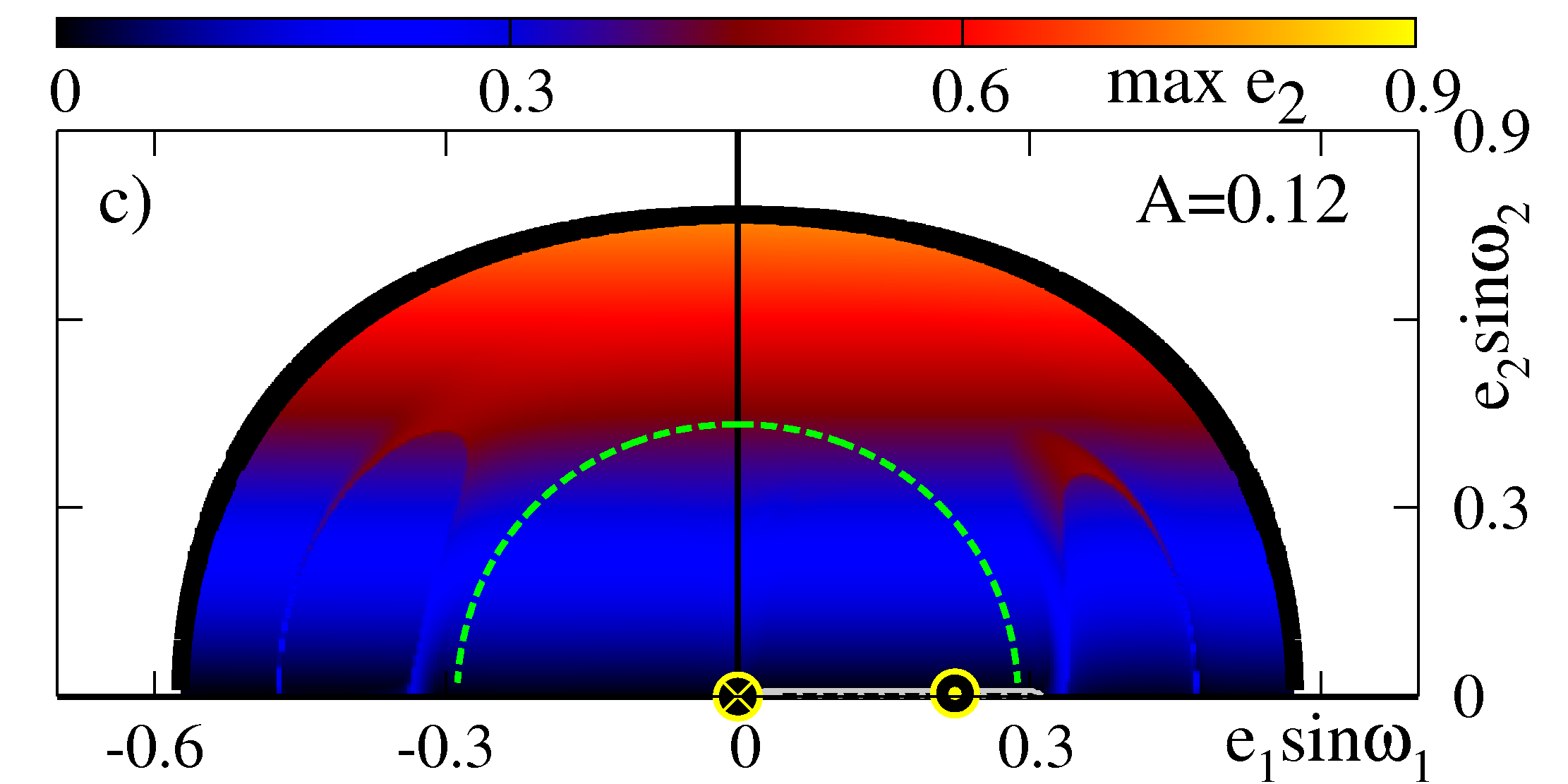}\hspace*{2mm}
          \includegraphics [width=84mm]{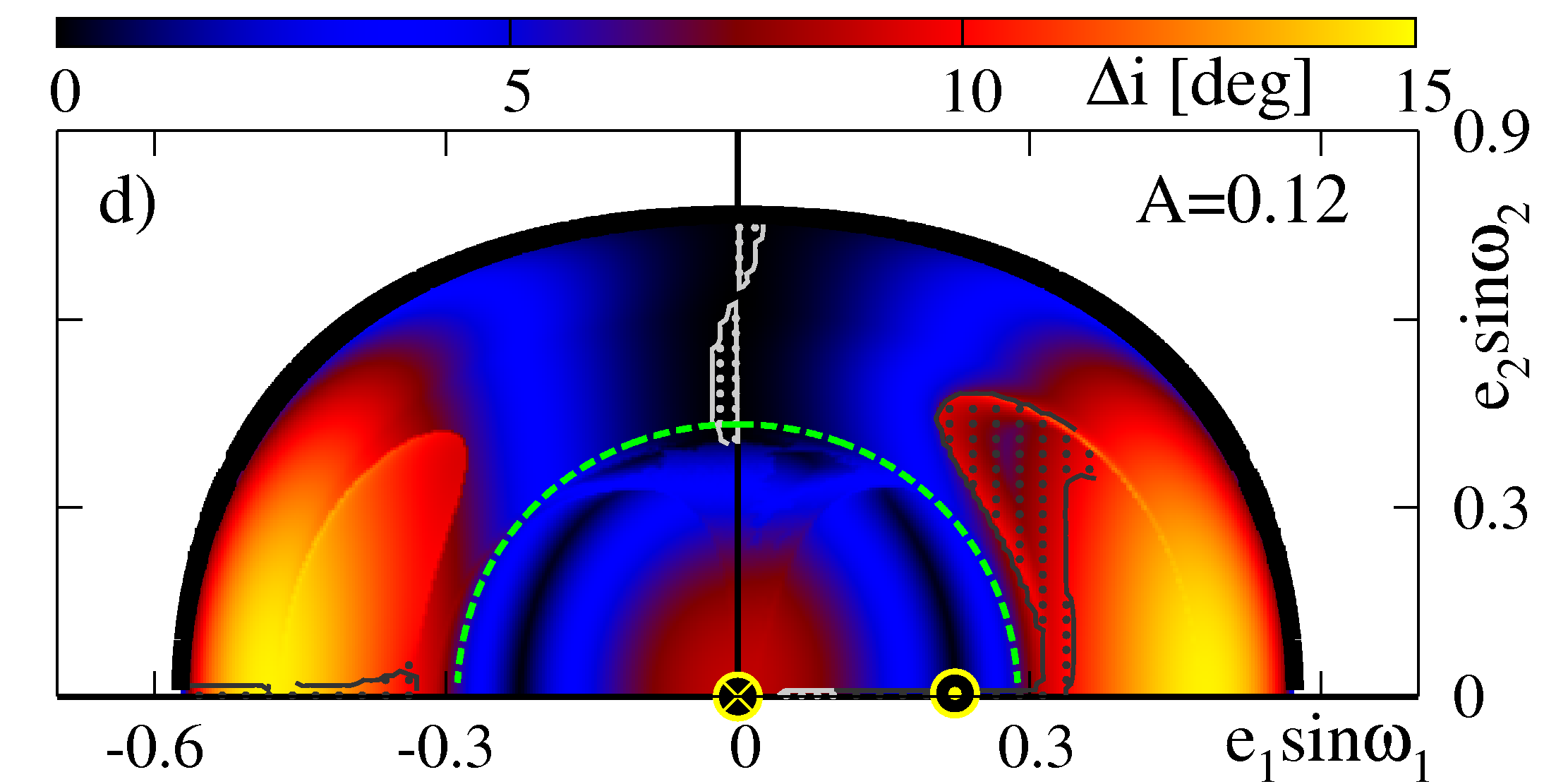}
	  }
         }
 }
 \caption{
Dynamical maps for the classic (point-mass Newtonian) model, shown in the 
\RPs-plane. Panels from a) to d) are for the coefficient of the diffusion of 
fundamental frequencies ($\sigma$), maximal eccentricity of the inner and 
outer planet ($\max e_1, \max e_2$), respectively, and the amplitude of 
variation of the mutual inclination ($\Delta i$). Dots mark areas of 
librations of $\omega_1$ (panel b), $\omega_2$ (panel c), and 
$\Delta\varpi$ (panel d). Stationary solutions are marked with 
circles: dotted circles are for stable equilibria (corresponding to the 
maximum of the secular Hamiltonian), crossed circles are for unstable 
equilibria, empty circles are for the linearly stable equilibria. These 
solutions are labeled, in accord with \citep{Migaszewski2009a}. Parameters 
of the system: $\alpha=0.01, \mu=20, \nAMD=0.12$, see also Fig.~\ref{fig2}b.
}
\label{fig4}
\end{figure*}
\begin{figure*}
 \centerline{
 \vbox{
    \hbox{\includegraphics [width=84mm]{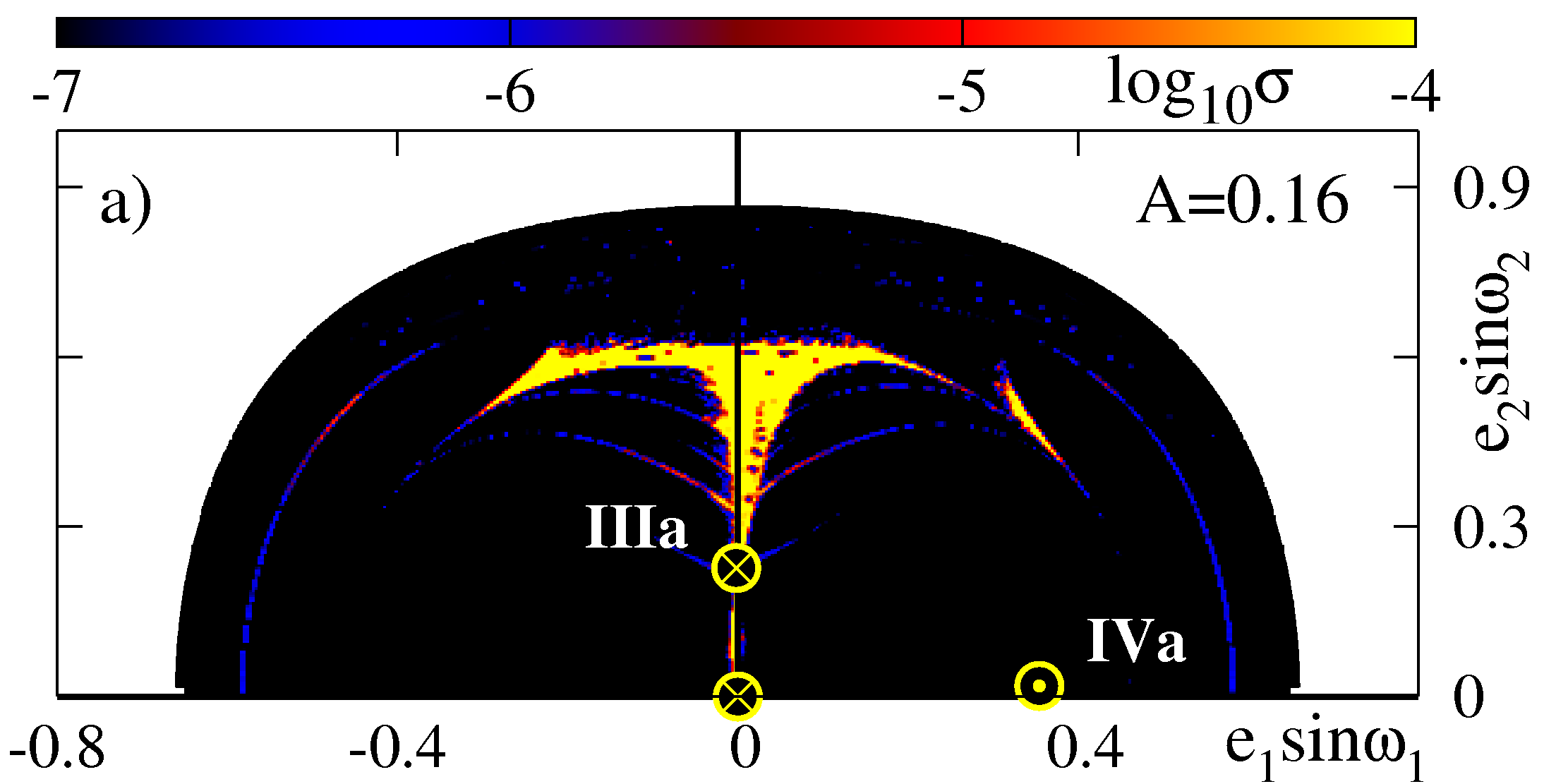}\hspace*{2mm}
          \includegraphics [width=84mm]{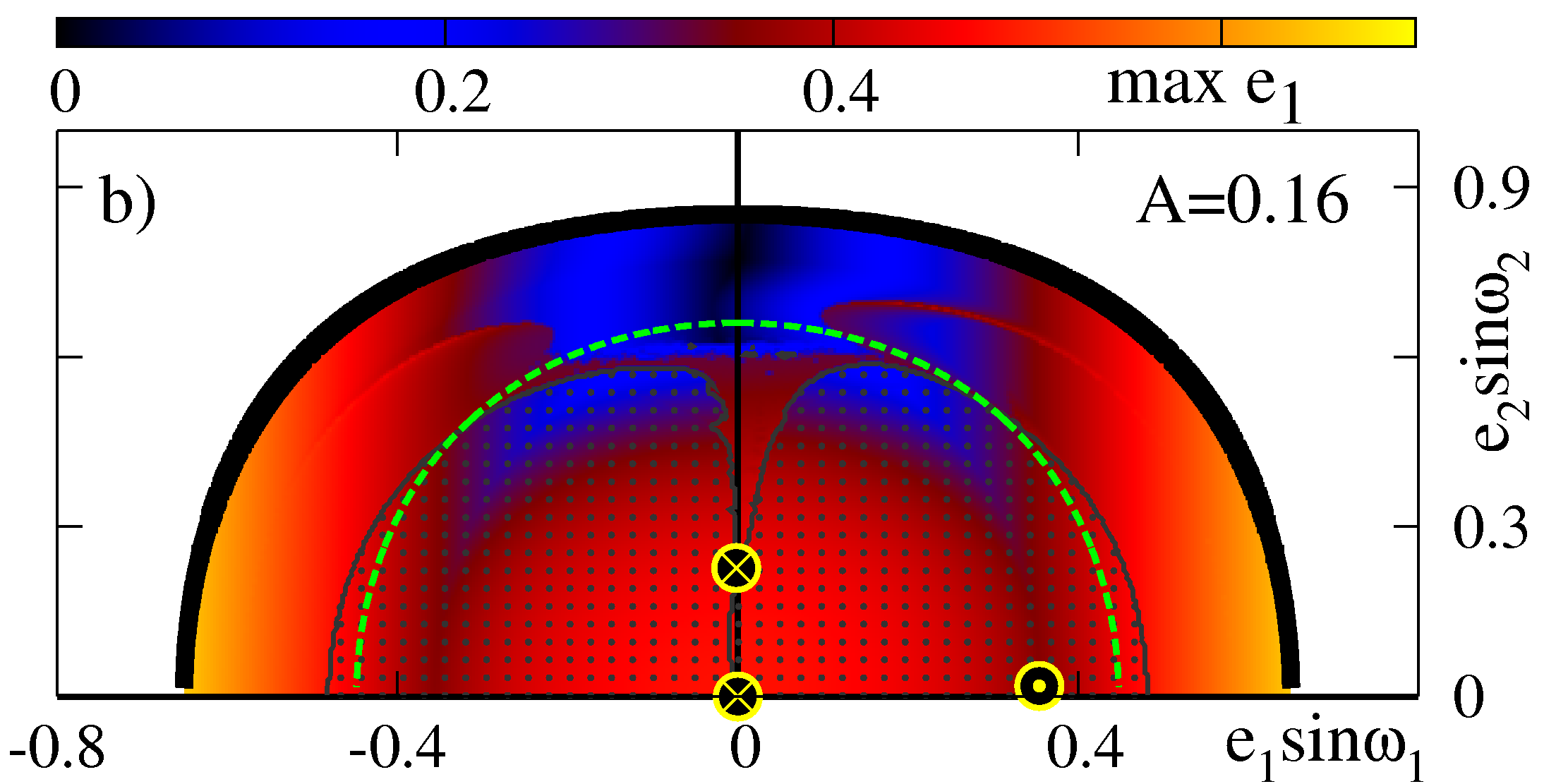}
	  }
    \hbox{\includegraphics [width=84mm]{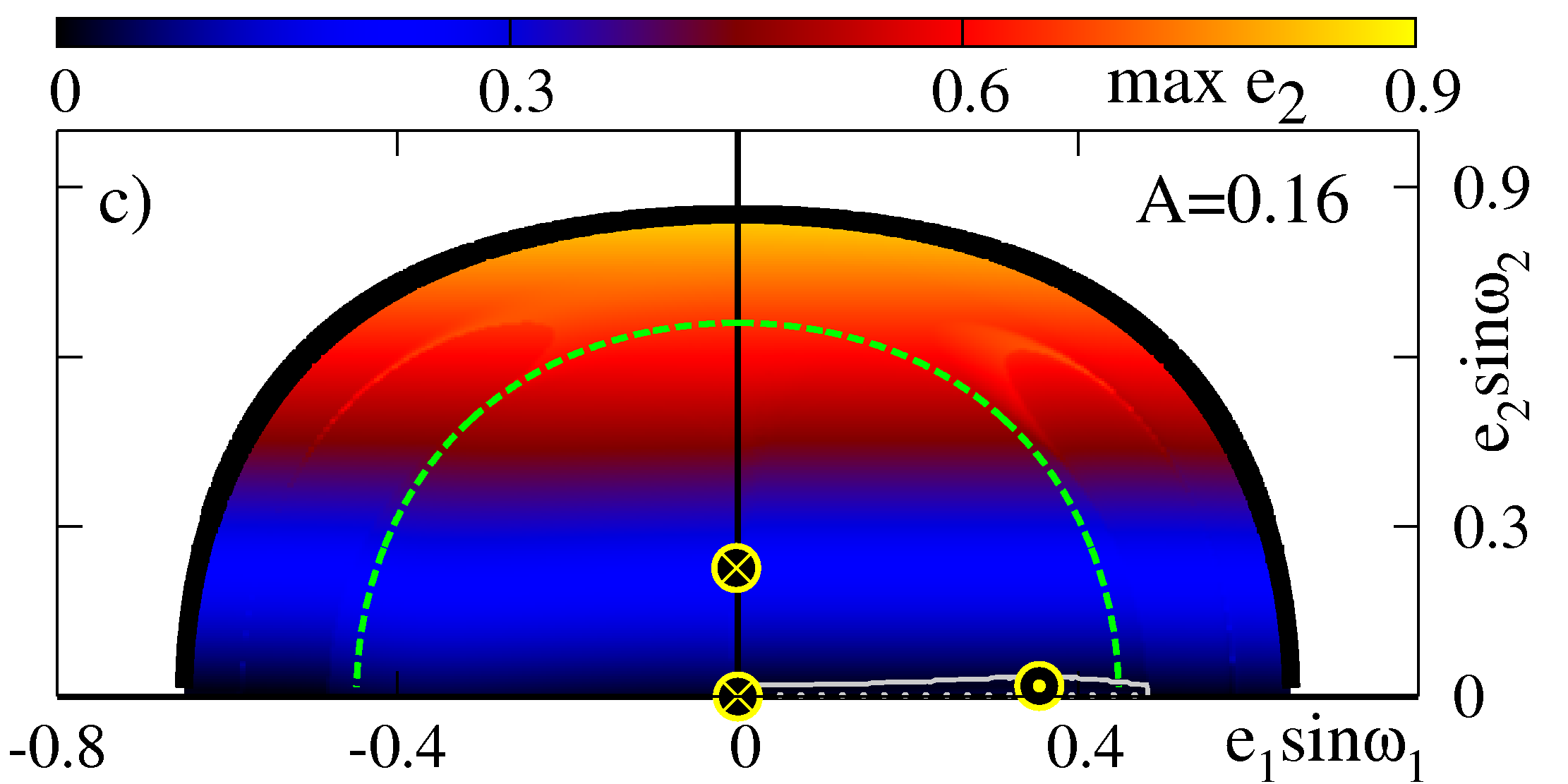}\hspace*{2mm}
          \includegraphics [width=84mm]{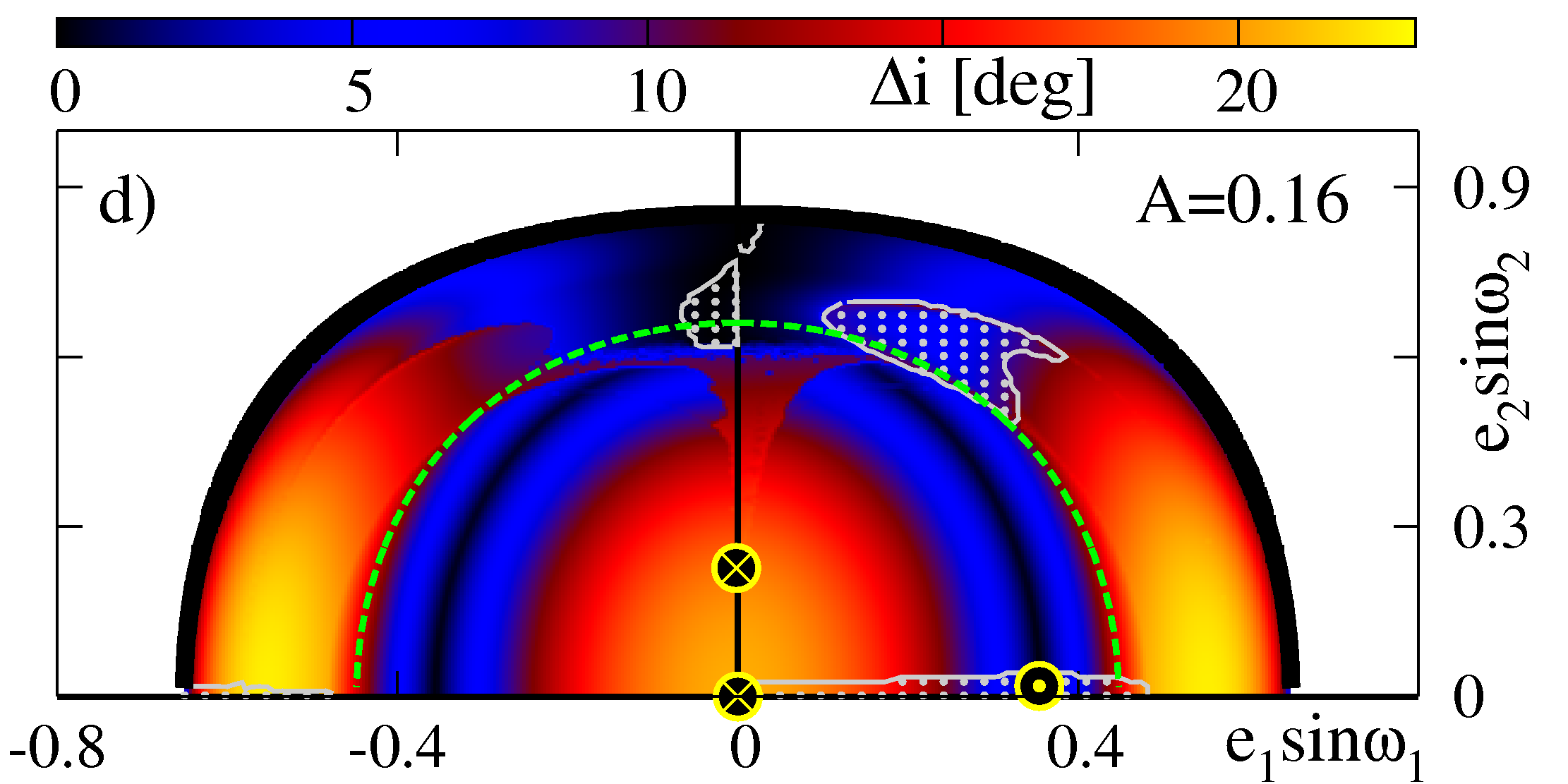}
	  }
 }
 }
 \caption{
 Dynamical maps for the Newtonian, point mass model in the \RPs-plane,
 $\nAMD=0.16$. See the caption to Fig.~\ref{fig4} for 
 more details, and also Fig.~\ref{fig2}c.
}
\label{fig5}
\end{figure*}

\begin{figure*}
 \centerline{
 \vbox{
    \hbox{\includegraphics [width=84mm]{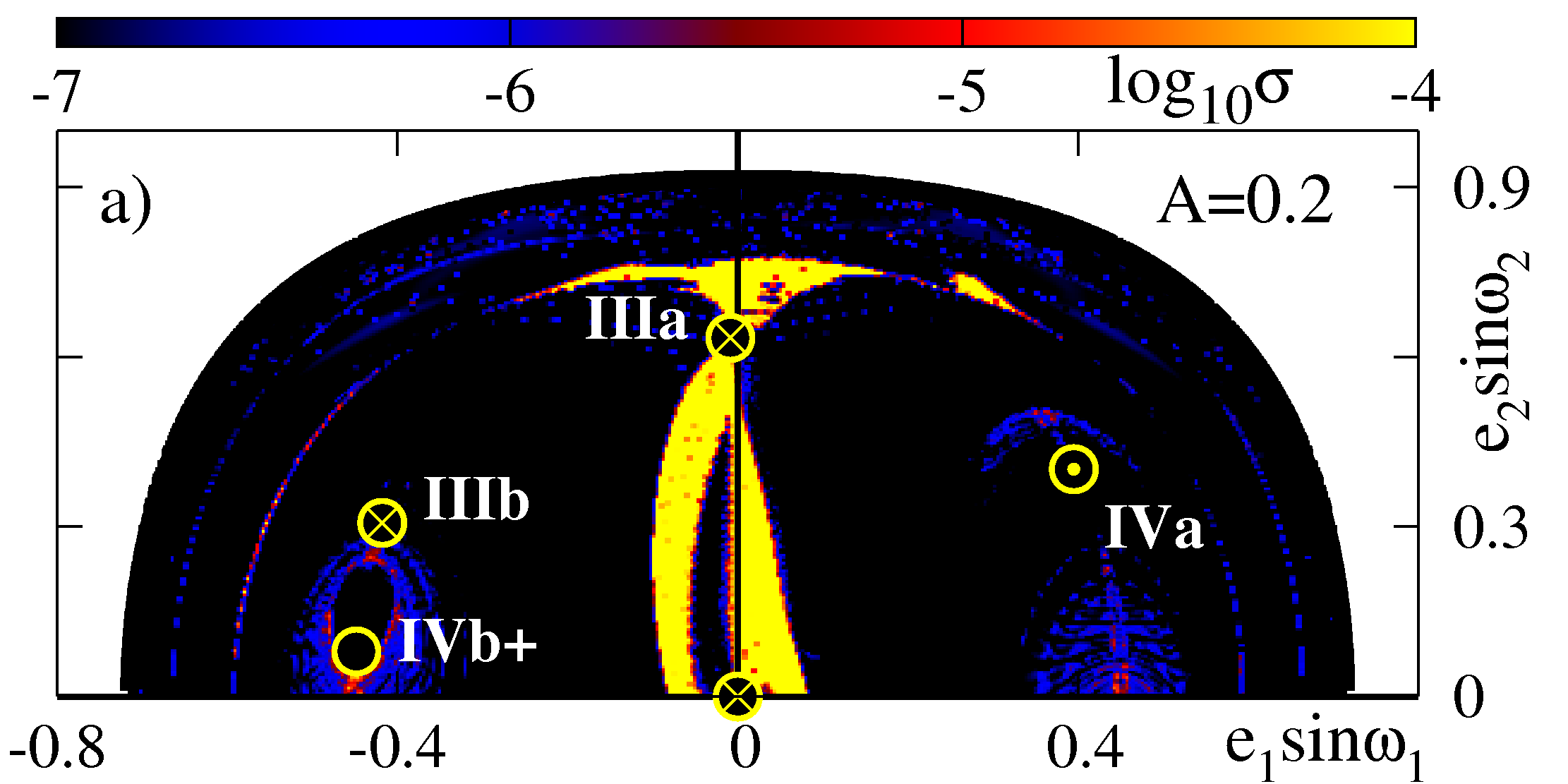}\hspace*{2mm}
          \includegraphics [width=84mm]{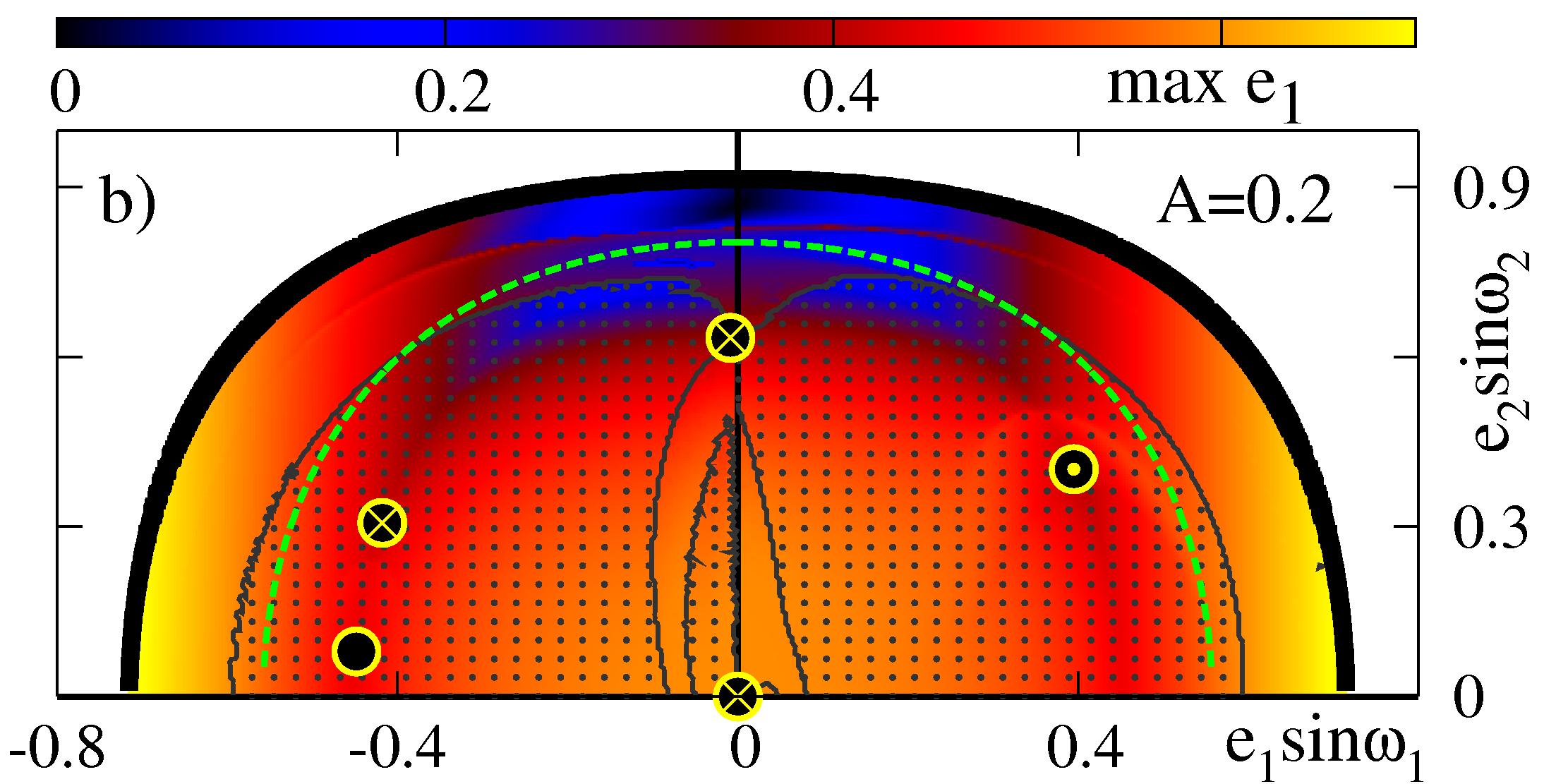}
	  }
    \hbox{\includegraphics [width=84mm]{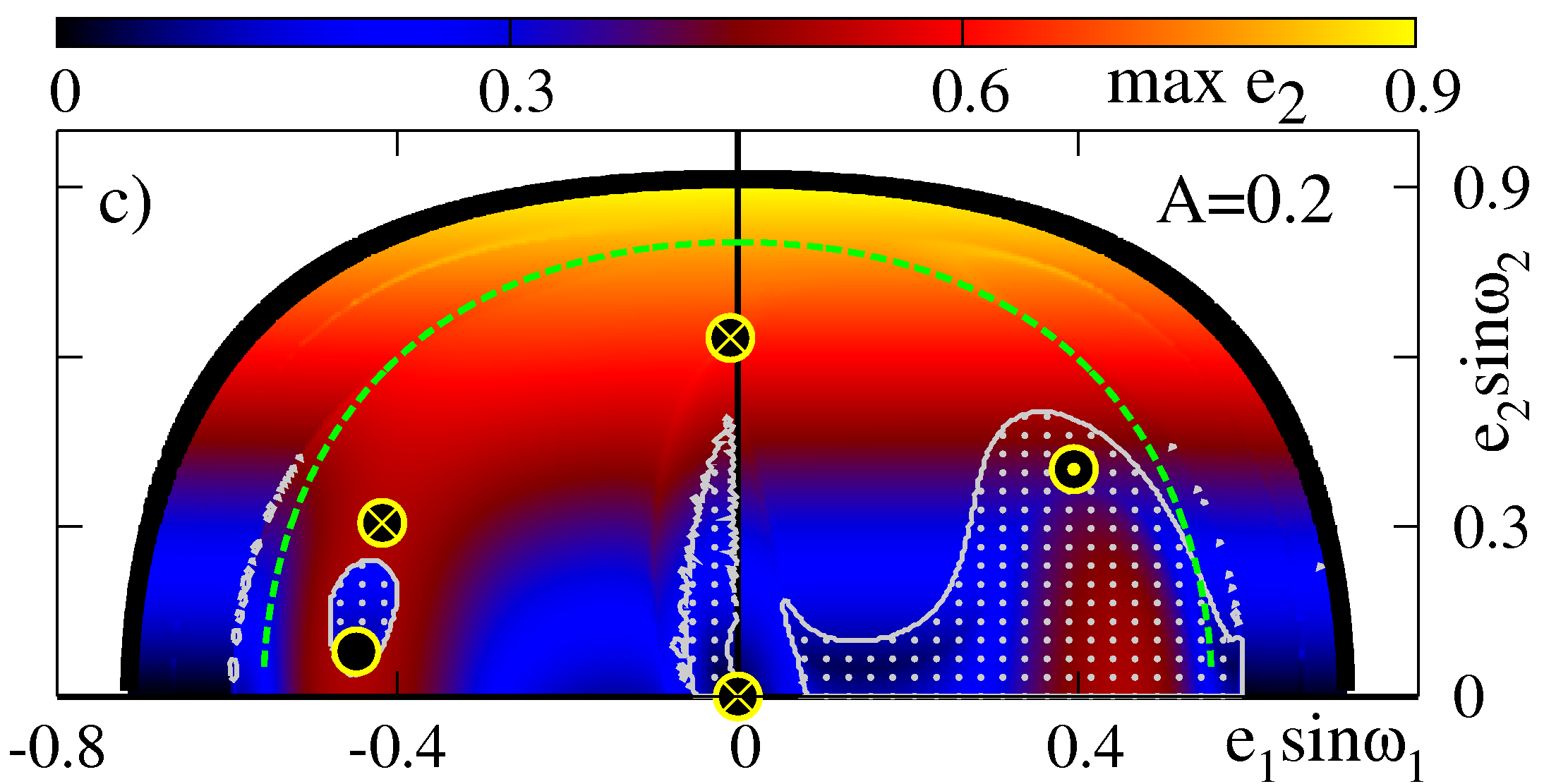}\hspace*{2mm}
          \includegraphics [width=84mm]{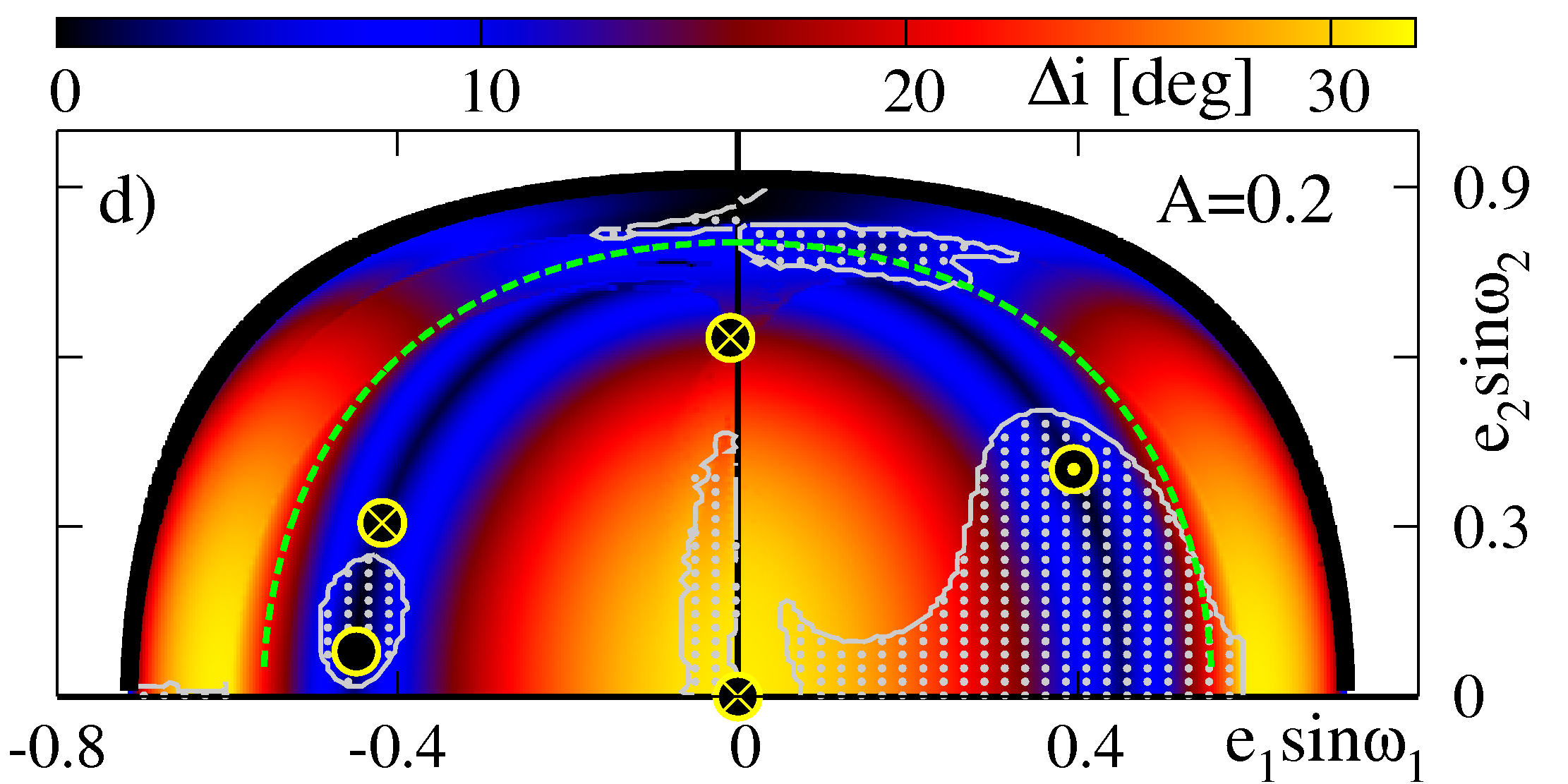}
	  }
 }
 }
 \caption{
 Dynamical maps for the Newtonian, point-mass model, in the \RPs-plane,
  $\nAMD=0.20$. See the caption to Fig.~\ref{fig4} for 
 more details, also Fig.~\ref{fig2}e.
}
\label{fig6}
\end{figure*}
Figures~\ref{fig4}--\ref{fig6} illustrate the results derived for the same 
system that energy levels are shown in Fig.~\ref{fig2}. Subsequent 
figures are for  $\nAMD=0.12, 0.16, 0.2$, respectively 
(see Figs.~\ref{fig2}b,c,e for the respective levels of  $\Hsec$). (We note, 
that for $\nAMD=0.08$ the view of the phase space is basically 
very simple and the 
motions are regular everywhere). The right-hand panel of each figure is 
for $\sigma$. The dynamical  character of the phase-space trajectories is 
color-coded: black color means quasi-regular evolution of the secular 
system ($\sigma \sim 10^{-6}$--$10^{-8}$), and yellow colour is for strongly 
chaotic  motions ($\sigma \geq 10^{-4}$). The $\sigma$-maps reveals that 
almost the whole phase space is filled up with regular motions, and chaos 
appears only in some small regions in the \RPs-plane. For a reference, the 
coordinates of equilibria IVa, IIIa, IIIb, IVb+ and 0 (the equilibrium at 
the origin)  are marked with circles. Dotted circle is for Lyapunov stable 
solution IVa, crossed circles are for unstable solutions 0, IIIa and IIIb, 
respectively. Equilibrium IVb+ is linearly stable and is marked with open 
circle.  Clearly, trajectories close to unstable equilibria IIIa and 0 are 
chaotic. Let us note that unstable equilibrium IIIa lies in a very narrow, chaotic
zone too.

Besides dynamical maps for $\sigma$,  Figs.~\ref{fig4}--\ref{fig6} illustrate 
geometric characteristics or orbits in the \RPs-plane, in terms of $\max 
e_{1,2}$ indicator  for the inner and the outer orbit, respectively, and for 
\ed{the amplitude of variation of the mutual inclination, $\Delta i$}. In all 
these dynamical maps, the green dashed contours mark the mutual inclination 
equal to the critical inclination of the LKR bifurcation \citep
{Krasinsky1972,Krasinsky1974}. Also regions, in which the secular angles $\omega_1$, 
$\omega_2$, and $\Delta\varpi$ oscillate, are shown: the gray/white/black 
dots surrounded by appropriate curves bound initial conditions that lead to 
librations of $\omega_1$ around $\pm \pi/2$ in the $\max e_1$-maps, 
librations of $\omega_2$ around $\pm \pi/2$ in the $\max e_2$-maps, and 
librations $\Delta\varpi$ around $0, \pi$ in the $\Delta i$-maps. We note that $\omega_1$ 
librates in almost all regular trajectories with the initial $\imut > \icrit$
.  If this condition of the Lidov-Kozai mechanism is fulfilled  then 
$\omega_1$ librates around $\pm \pi/2$.  In chaotic zones of the \RP-plane, 
these angles  do not oscillate, even if condition $\imut > \icrit$ holds 
true. Angle $\omega_2$ librates around $\pi/2$ only when $e_2 \sim 0$, for  
$\nAMD=0.12, 0.16$. This region  extends significantly for larger $\nAMD=0.2$
.   In such a case, there are three zones in which {\em both} angles 
$\omega_1$ and $\omega_2$ librate: a region connected to equilibrium IVa,  the 
neighborhood of linearly stable solution IVb+, and a region of small $e_1$ 
\ed{(though there is no stationary solution associated with these
librations)}. 
The later area is surrounded by strongly chaotic  motions shown in the 
$\sigma$-plane.

Clearly, the equilibria permitted by relatively large $\nAMD$ affect the 
secular dynamics, which become very complex. This may be  better seen in the 
$\max e$-maps. Subsequent panels in Figs.~\ref{fig4}--\ref{fig6} show that 
the inner eccentricity may be strongly excited if $\imut > \icrit$. For 
relatively small $\nAMD=0.12$, the inner orbit, which is initially 
quasi-circular, becomes moderately eccentric, with $\max e_1 \sim 0.3$. Note 
that in this specific case, {\em the inner body} is $20$ times more massive 
then the outer companion. Simultaneously, as Fig.~\ref{fig4} shows, the 
outer eccentricity $e_2$ is not amplified. We may conclude that in the 
regime of the  outer LKR, when the inner body is much more massive than the 
outer body, the secular  evolution may lead to strong perturbations of the 
inner orbit. Hence, even the mass hierarchy is reversed, a strong excitation 
of the inner eccentricity is still possible, similarly to the LKR in the 
restricted problem. \corr{This dynamical phenomenon may explain a large 
eccentricity of the binary, when a distant, low-mass and dark companion 
cannot be detected due to very long orbital period.}

\corr{The dynamical maps in Figs.~\ref{fig4}--\ref{fig6} reveal some zones, 
in which the outer eccentricity is strongly amplified. This eccentricity is 
obviously constant in terms of the quadrupole model. This feature of the 
octupole model shows that the small perturbation may cause extended, 
geometric changes of the mean orbits. The amplification of $e_2$, with
simultaneously almost constant relative inclination, $\Delta i \sim\mbox{const}$ 
(see Figs.~\ref{fig6}c,d) seems
appear due to the bifurcation of equilibria IVb+ and IIIb for $\nAMD \sim 0.195$.}

%
\subsection{A model explaining the secular chaos}
%
The origin of chaotic secular dynamics may be explained by the presence of 
separatrices, which encompass \corr{different types of motions, librations and 
rotations of angles $\omega_1$, $\omega_2$  and $\Delta\varpi$}. A classification of
these libration modes in the secular problem of
two planets modes has been introduced in \citep{Michtchenko2006}. 
The separatrices appear due to equilibria and periodic solutions in the
integrable, or close to integrable secular models, which might be 
understood as the quadrupole approximation and/or the co-planar
configuration.

\begin{figure*}
 \centerline{
 \vbox{
    \hbox{\includegraphics [height=56mm]{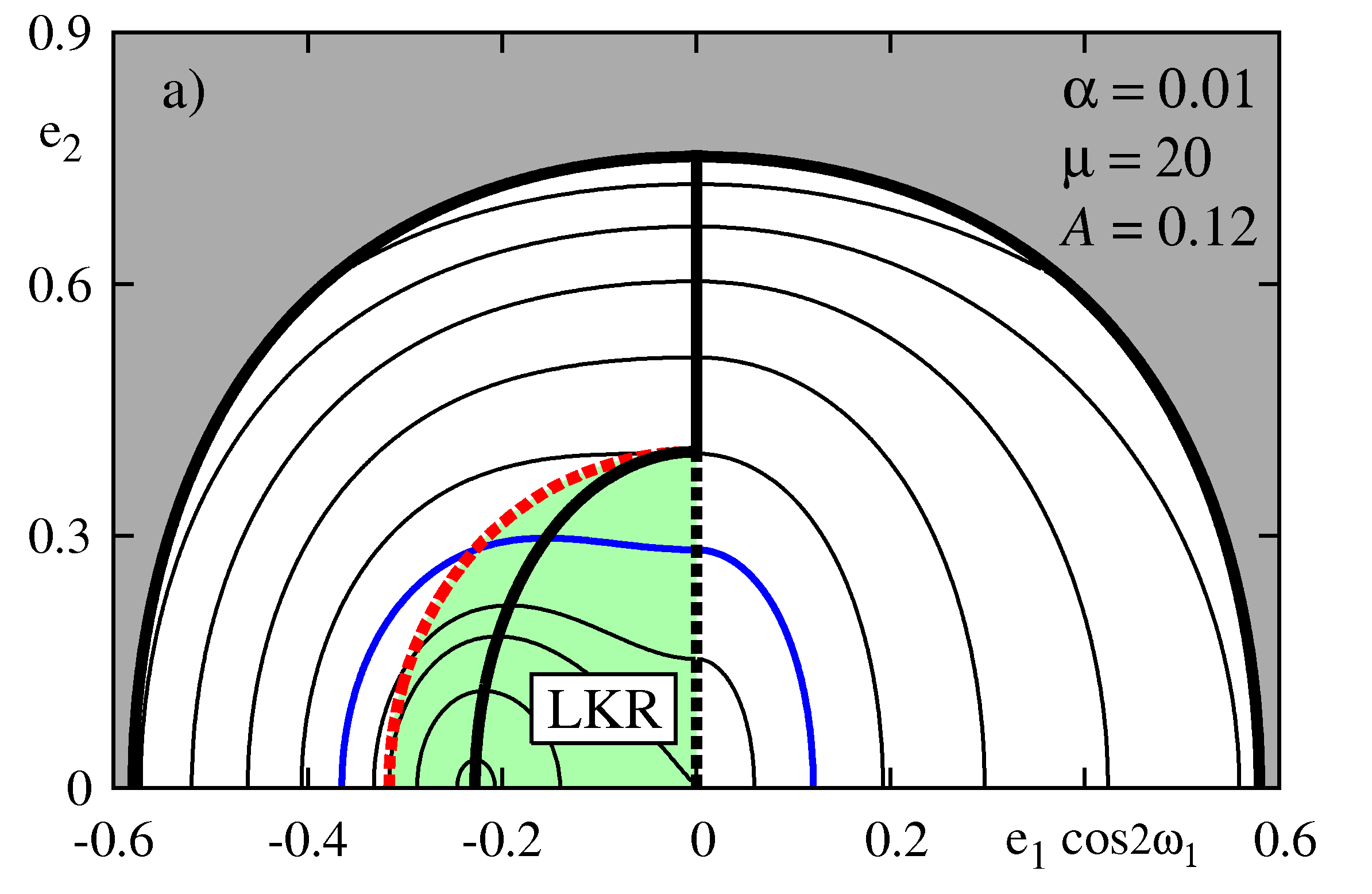}
          \includegraphics [height=56mm]{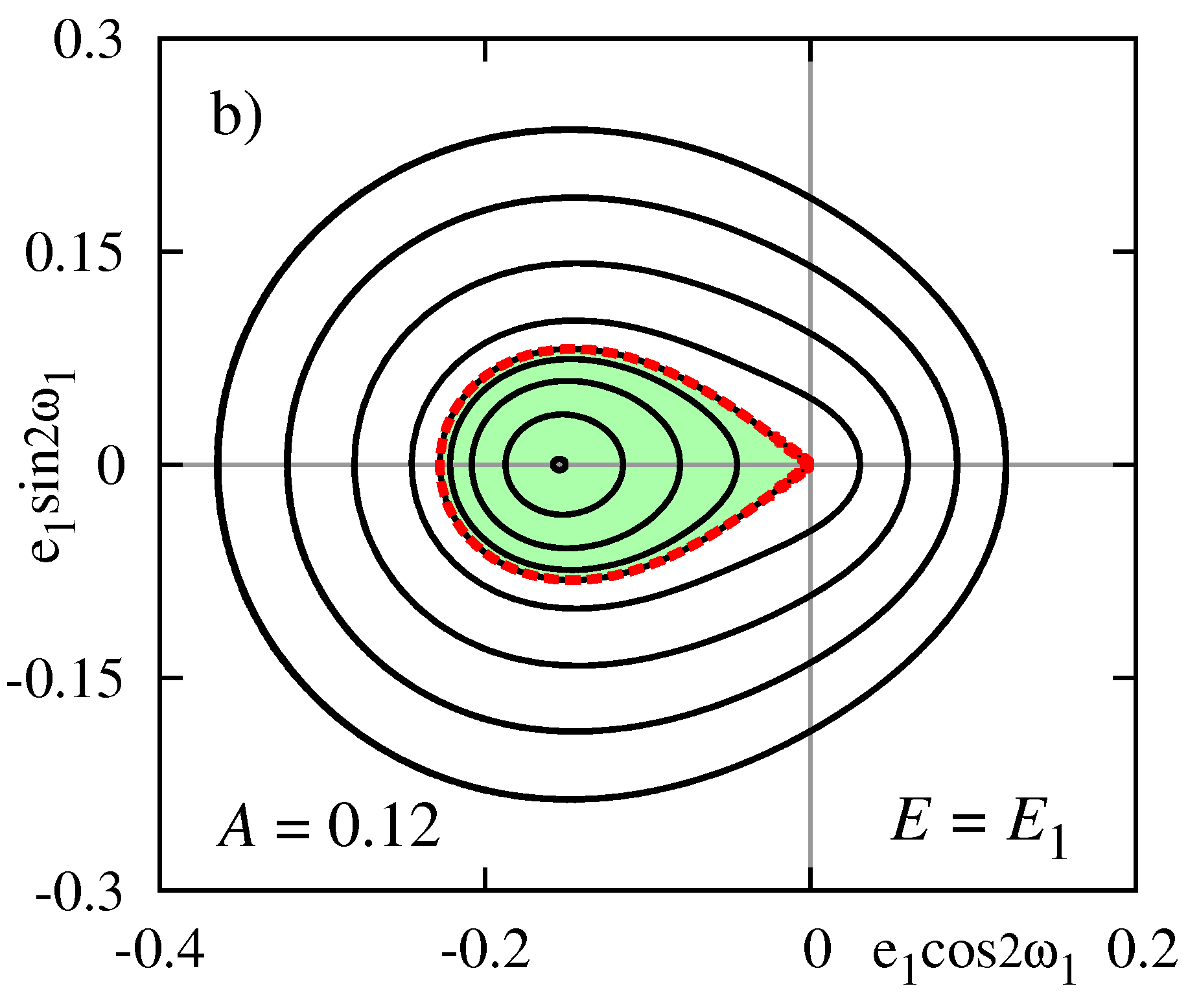}
	  }
 }
 }
\caption{
{\em The left panel} is for the levels of the 
secular energy  (thin curves) depicted in
the representative plane  of the quadrupole problem, \RP''{} $\equiv \{e_1 \cos
2\omega_1 \times e_2\}$-plane, $\alpha=0.01$, $\mu=20$, $\nAMD=0.12$. Recall that 
this problem is integrable and $e_2 \equiv \mbox{constant}$ is a parameter. 
The black, thick curve marks the LKR equilibrium for different
energies and $e_2$ integral.  The red, dotted curve marks {\em the separatrix}
between librations and rotations of $\omega_1$ for fixed values of
integrals.  The shaded (green) zone indicates \ed{librations of $\omega_1$ around
$\pi/2$}.
{\em The right hand panel} is for the phase diagram in the $\{e_1 \cos 
2\omega_1 \times e_1 \sin 2\omega_1 \}$-plane and a fixed energy level 
marked by blue curve in panel~a. The separatrix and the 
region of librations of $\omega_1$ are also marked with green colour.}
\label{fig7}
\end{figure*}

Here, we found a simple explanation of the  mechanism generating chaos, 
which is in fact the same as in the perturbed pendulum. To show this, let us 
recall that the expansion of $\Hsec$ to the second order in $\alpha$ is the 
celebrated integrable quadrupole-level model. The energy levels of this 
model are illustrated in Fig.~\ref{fig7}a. Because $e_2$ is the integral of 
motion, the representative plane may be constructed similarly to the 
integrable co-planar problem, $\RP'' \equiv \{ e_1 \cos 2\omega_1 \times 
e_2 $\}. In the region marked in green colour, any constant level of $e_2$ 
crosses the energy curve in two  turning points limiting the range of 
variation of $e_1$.   If the $e_2$ level is tangent to the given level of 
the energy, \corr{the dynamics must be then confined to fixed $e_1$,  hence 
we obtain an equilibrium point (stable or unstable)}. A set of these 
equilibria for increasing, fixed values of $e_2$ is marked by two thick 
curves that meet around of $(e_1=0, e_2 \sim 0.4)$. This is the bifurcation 
point, at which two families of equilibria emerge --- the stable branch of 
the LKR and the unstable origin $(e_1=0, e_2=0)$.

For a given value of constant $e_2$, we may then find the maximum range of 
$e_1$, which corresponds to librations of the canonical angle $\omega_1$, 
and this value determines the position of separatrix, see Fig.~\ref{fig7}b 
for the separatrix shown in the phase diagram in the $\{e_1 \cos 2\omega_1 
\times e_1 \sin 2\omega_1 \}$-plane, corresponding to the energy level 
marked by the blue curve on the levels diagram (Fig.~\ref{fig7}a). Collecting 
such points along increasing $e_2$, we  construct the red dashed curve in 
Fig.~\ref{fig7}a, showing the separatrix region globally in the whole $e_2$
-range (see the shaded area in Fig.~\ref{fig7}a). \corr{The dashed line
that marks the unstable origin, might be also understood as the other branch
of the global separatrix, which, for any value of $e_2$, corresponds to 
the equilibrium point.}

\begin{figure*}
 \centerline{
 \vbox{
    \hbox{\includegraphics [height=66mm]{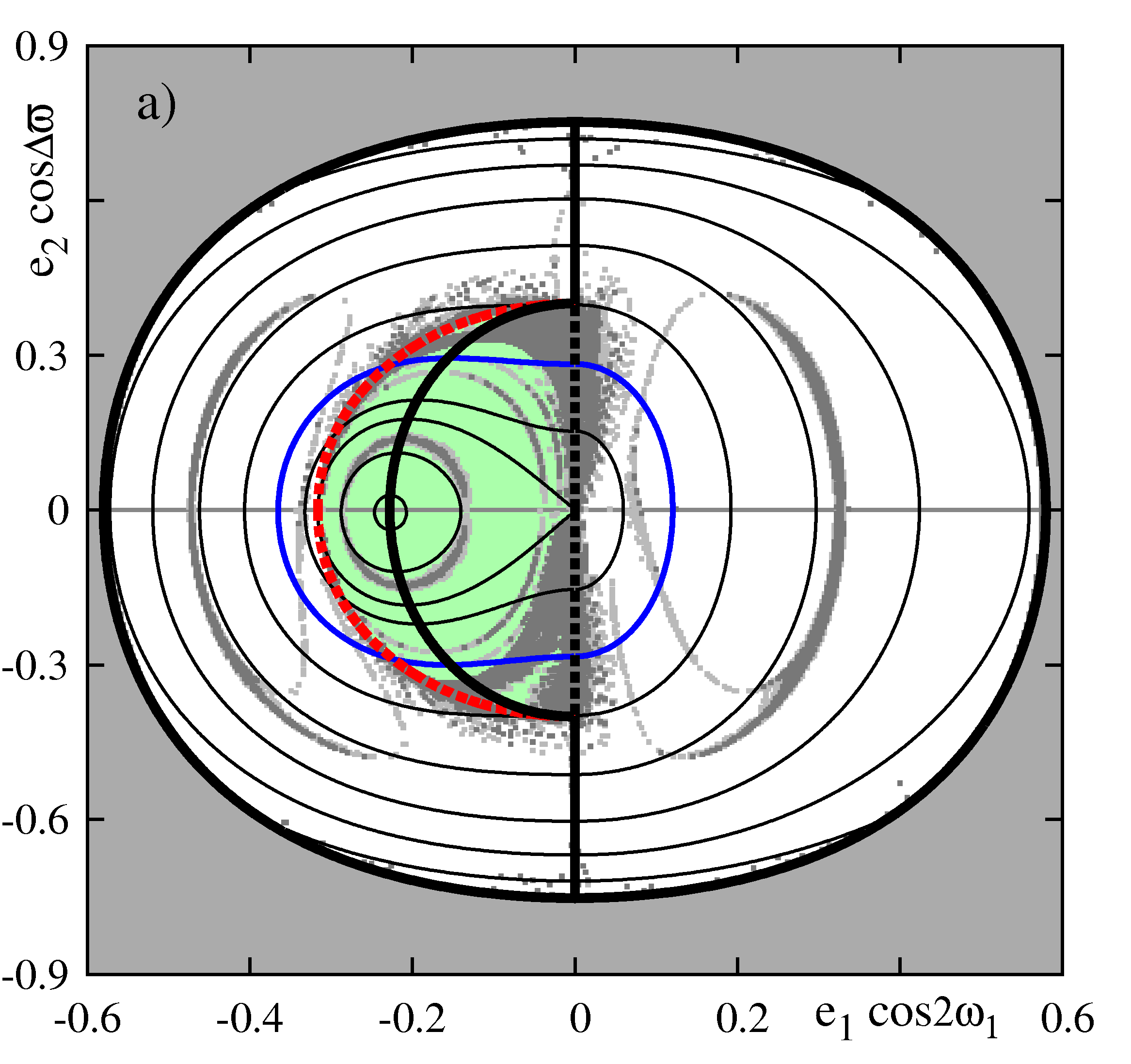}\hspace*{8mm}
          \includegraphics [height=66mm]{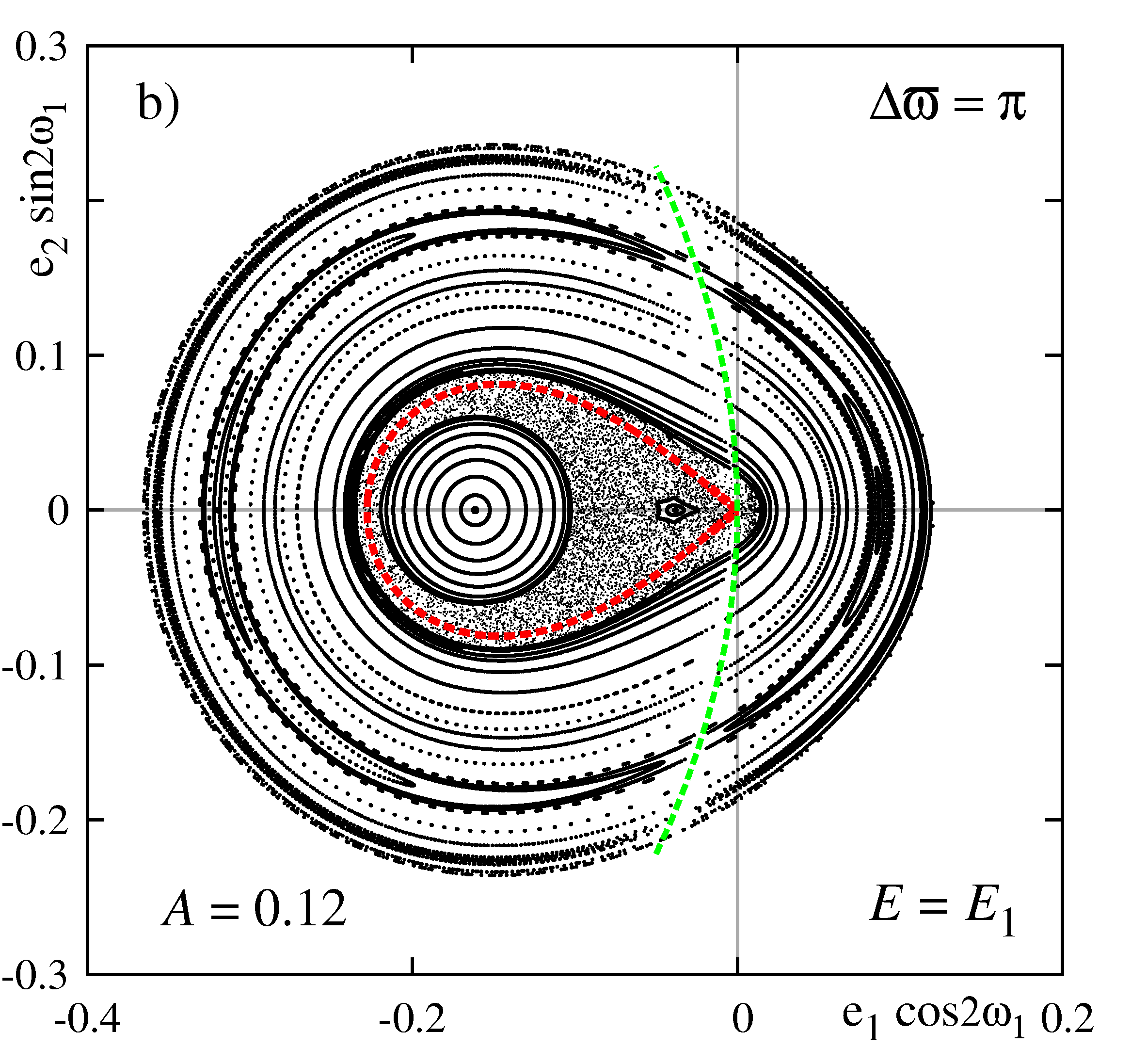}
	  }
 }
 }
\caption{
{\em The left panel} is for the secular energy levels (thin curves) shown in
the $\RP'{}$-plane of the octupole model, 
$\alpha=0.01$, $\mu=20$, $\nAMD=0.12$. Shaded areas are for initial conditions
leading to chaotic evolution of the secular model. 
The red curve is for the separatrix of the LKR resonance in the quadrupole model,
the black thick curve is for the equilibria in this model.
{\em The right hand panel} is for the Poincar\'e cross-section
$\Delta\varpi=\pi$ in the $\{ e_1 \cos 2\omega_1 \times e_1 \sin 2\omega_1
\}$-plane. It corresponds to  a fixed energy level marked with the blue curve in
panel~a. The separatrix of the quadrupole problem and a
region of librating $\omega_1$ are also marked.
}
\label{fig8}
\end{figure*}

Now, considering the perturbed quadrupole-model in terms of the 
octupole-level secular expansion, we may expect that chaotic motions will 
appear close to the separatrix, due to the perturbation. Indeed, this is 
illustrated in two panels of Fig.~\ref{fig8}. The left panel is for the 
frequency diffusion $\sigma$ computed for initial conditions in the $\RP''
\equiv \{e_1 
\cos 2\omega_1 \times e_2\}$-plane. This plane shows the energy levels of 
the octupole model, with stable (thick, solid curve) and unstable (thick, 
dashed line) equilibria in the quadrupole model over-plotted. The red curve 
is for the global separatrix of the LK resonance \ed{in the quadrupole 
model}, and the green region is for the librations of $\omega_1$ in the 
integrable case. Clearly, chaotic orbits are found along the branch of 
unstable equilibria, as well as close to the red separatrix curve. This may be 
better seen in Poincar\'e cross section computed for a fixed energy level 
marked with blue curve that is shown in the right-hand panel of Fig.~\ref
{fig8}b. This panel in fact comprises {\em two} sections $\Delta\varpi=\pi$ 
in the $\{ e_1 \cos 2\omega_1 \times e_1 \sin 2\omega_1 \}$-plane, for 
$\dot{\Delta\varpi}>0$ (the left half-plane), and for 
$\dot{\Delta\varpi}<0$ (the right half-plane). In place of the separatrix
curves, 
a wide chaotic regions appear. 

\begin{figure*}
 \centerline{
 \vbox{
    \hbox{\includegraphics[width=84mm]{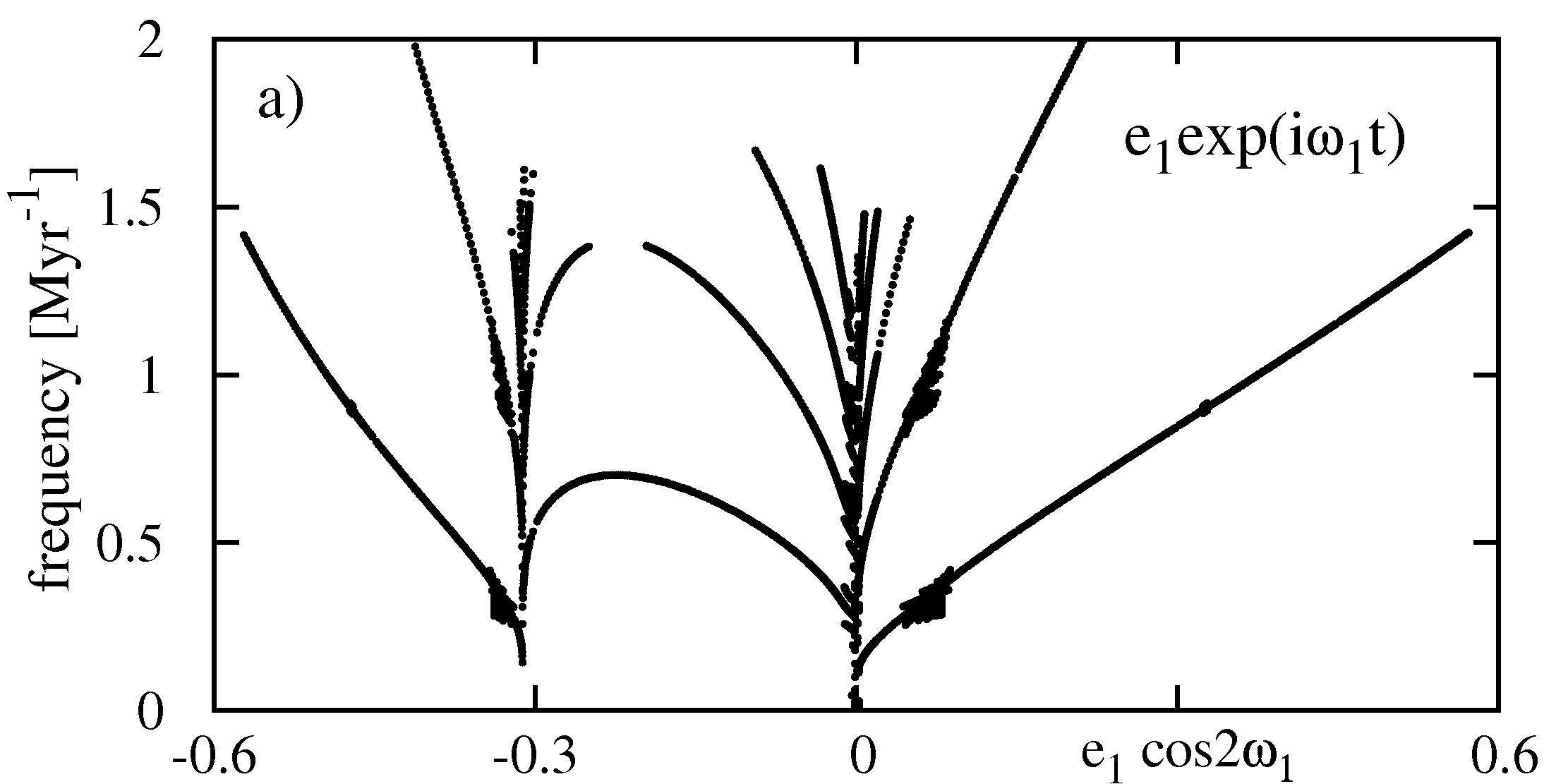}
          \includegraphics[width=84mm]{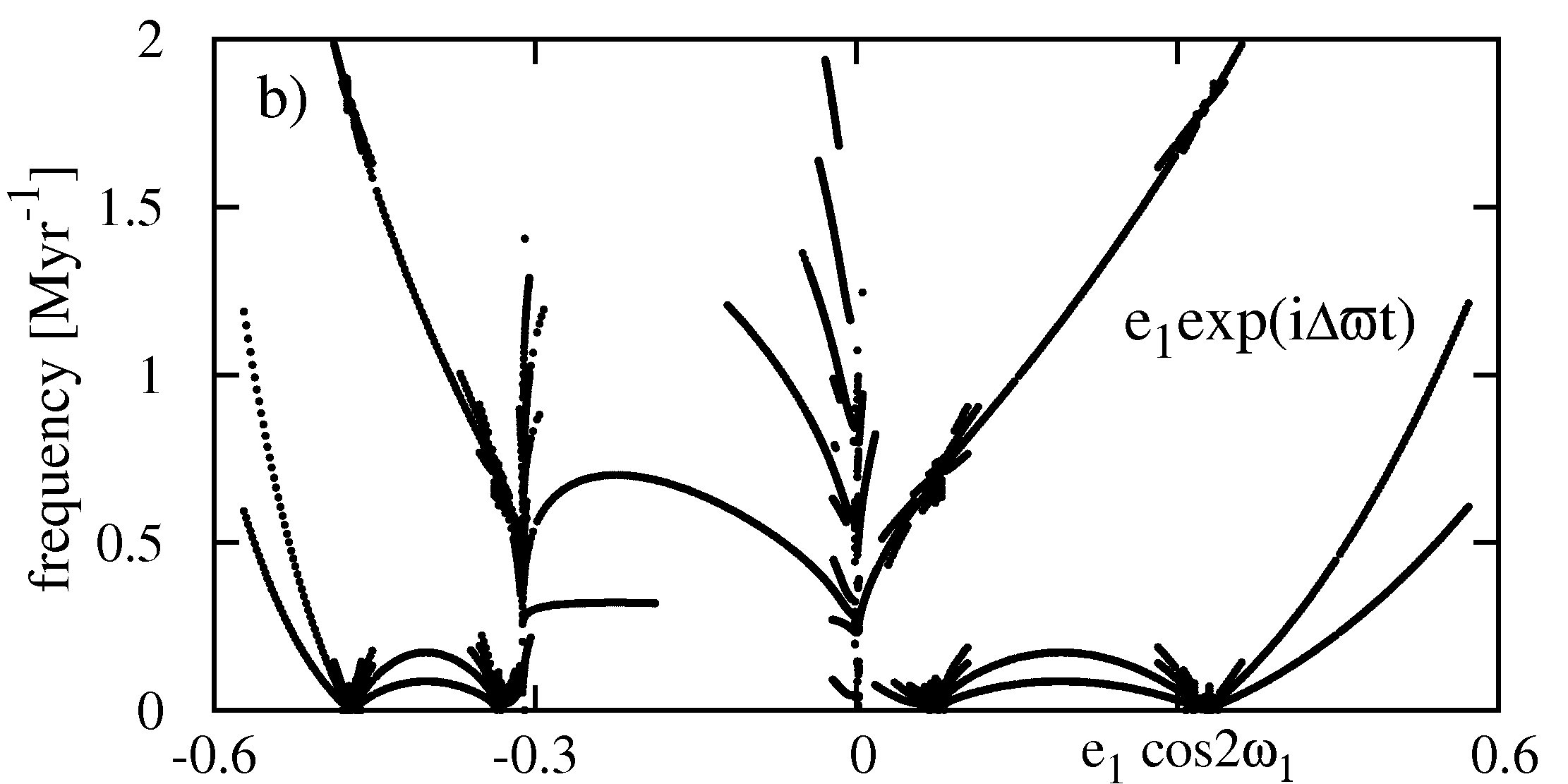}
	  }
 }
}
\caption{The spectrum of fundamental frequencies computed along
the $x$-axis of Fig.~\ref{fig8}, for the same set of parameters. 
Abscissas, in which the
frequencies approach zero, indicate the separatrices of secular
resonances of the respective angles, $\omega_1$ (panel a), and
$\Delta\varpi$ (panel b). \ed{Parameters are: $m_{0}= 1~m_{\sun}$,
$m_1= 40\mJ$, $m_2 = 2\mJ$, $a_1=0.1$~au, $a_2=10$~au.}
}
\label{fig8c}
\end{figure*}

\corr{A careful inspection of Fig.~\ref{fig8}a reveals additional narrow
banana-shaped chaotic structures.  They are related to the separatrix of the
librations of angle $\Delta\varpi$ around $\pi$ and simultaneous circulation
of angle $\omega_1$, which is classified as mode ${\bf 2}$ in \cite[][see
their Fig.~\ref{fig3}]{Michtchenko2006} .  Further, the plots of fundamental
frequencies computed along the $x$-axis of the \RP'-plane in
Fig.~\ref{fig8}a, which are shown in Fig.~\ref{fig8c}, reveal that indeed,
in this region the fundamental frequencies decrease to 0, indicating the
separatrix region.}

This example shows clearly that apparently very small 
perturbation (recall that $\alpha=0.01$) to the integrable model introduces 
extended chaotic behavior and qualitative change of the secular dynamics
of the system.

This interpretation is helpful to understand the source of chaotic motions 
shown in all dynamical maps (e.g., Figs.~\ref{fig4}--\ref{fig6}) -- \corr{usually, 
they appear close to the separatrices associated with unstable equilibria
or resonances of the secular angles (unstable periodic orbits)}.

\begin{figure}
\centerline{
\hbox{
          \includegraphics [width=80mm]{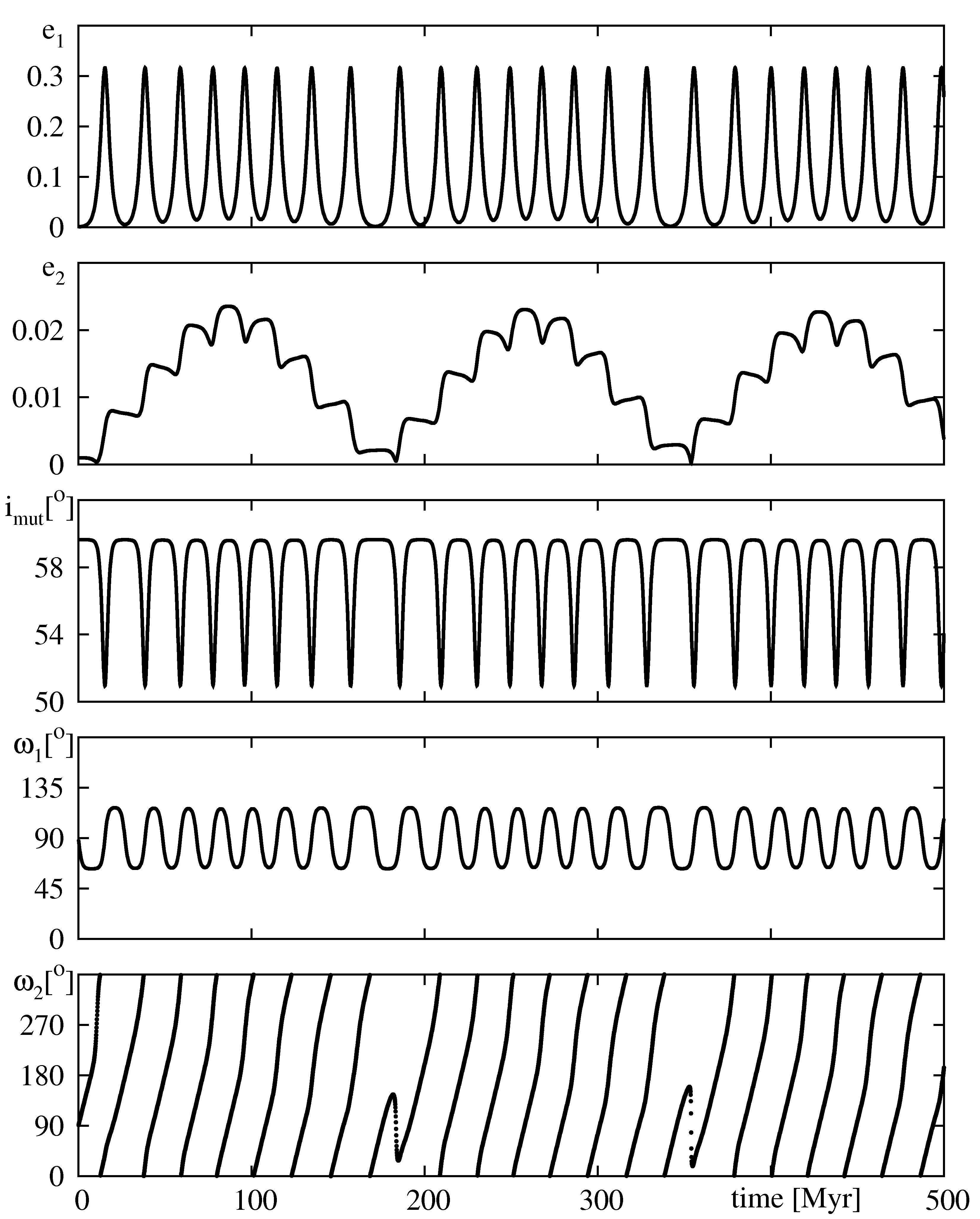}
 }
}
\caption{
The mean orbits with the following initial condition: $m_0 =
1~\msun$, $m_1 = 200~\mJ$, $m_2 = 10~\mJ$, $a_1 = 0.3~\au$, $a_2 = 30~\au$, $e_1
= 0.001$, $e_2 = 0.001$, $\omega_1 = \pi/2$, $\omega_2 = \pi/2$, $\imut =
59^{\circ}.64$, $\nAMD = 0.12$. Panels from the top to the bottom 
present the time-evolution of $e_1, e_2, \imut, \omega_1, \omega_2$, respectively.
}
\label{fig9}
\end{figure}

Figure~\ref{fig9} shows the temporal evolution of the mean orbits with initial
conditions  close to the origin in Fig.~\ref{fig4} (see its caption for
the details). The eccentricity of the inner orbit is
significantly perturbed while the outer orbit is almost unchanged. 
The mutual
inclination evolves in anti-phase with $e_1$, and $\omega_1$ librates around
$\pi/2$. These are typical features of the Lidov-Kozai resonance, still we
recall that in this instance, the smaller outer body forces 
the LK cycles on the orbits of much more (20 times) massive than 
the inner component
of the binary.
%
%
\section{Equilibria in the classic octupole model}
%
The dynamical maps and their analysis show that the equilibria constrain the 
secular dynamics of the perturbed model. In terms of the Newtonian, 
point-mass formulation, the stationary solutions depend on parameters 
$\alpha$ and $\mu$. Limiting our survey to $\imut < \pi/2$ (direct orbits), 
we perform a parametric survey of the equilibria in terms of the octupole 
expansion, in such a manner that a comparison with the results derived for 
the more general, relativistic model will be possible. Here, we consider a 
more extended  range of mass ratio $\mu$, covering a transition from the 
planetary regime (small $\mu$) to the circumbinary case (large $\mu$ ) than 
in \citep{Migaszewski2009a}. Yet the assumption of small $\alpha$ makes it 
possible to use the analytic formulation of the secular Hamiltonian, which 
is very precise in terms of the octupole-level approximation, instead of the 
numerical approach in \citep {Migaszewski2009a}. 

\begin{figure*}
\centerline{
\vbox{
    \hbox{\includegraphics [width=126mm]{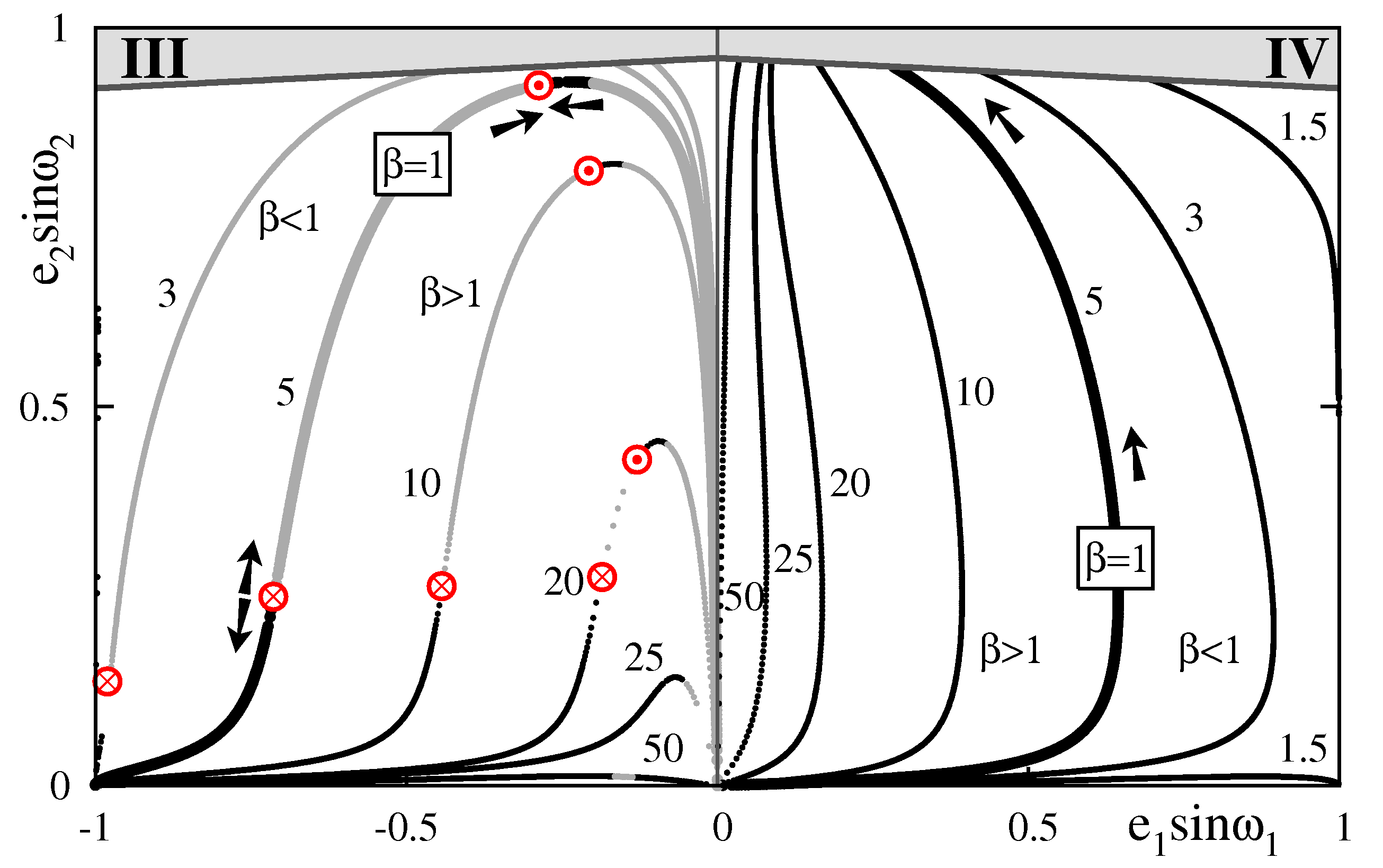}
          }}
      }
\caption{
Families of stationary solutions in the \RPs-plane. The semi-major axes ratio
$\alpha=0.04$, $\mu = \{1.5, 3, 5, 10, 20, 25, 50\}$
(each curve is labeled accordingly with the value of $\mu$). Dark dots are 
for stable equilibria, grey dots are for unstable equilibria. Crossed and  
dotted circles mark the positions on the equilibria curves where solutions 
bifurcate (see the text for details). Small arrows show the directions of
the  
evolution of particular equilibrium with increasing $\nAMD$. 
Parameter $\beta \equiv L_1/L_2$ (see the text). Stationary curves corresponding to 
$\beta=1$ are drown with thicker lines. Stationary solutions were
calculated  for the octupole classic model. See also Fig.~\ref{fig3}.
}
\label{fig12}
\end{figure*}

The parameter dependence of the equilibria is illustrated in Fig.~\ref 
{fig12} that  shows the \RPs-plane (note that due to the symmetry, only the 
upper half-plane is illustrated, see also Fig.~\ref{fig3}). We set 
$\alpha=0.04$, $\mu$ $\in [1.5, 3, 5, 10, 20, 25, 50]$, and then positions 
of the equilibria may be traced along increasing $\nAMD$, which might be 
understood as the natural curve parameter.

In fact, instead of $\mu$, we choose a new 
parameter $\beta$ \citep[see][]{Krasinsky1974,Migaszewski2010}
\[
\beta(\mu,\alpha) \equiv L_1/L_2 \sim \mu \sqrt{\alpha},
\]
that better reflects the dependence of the dynamics on {\em both} $\mu$ and 
$\alpha$ than on one of these parameters itself \ed{(we did a few numerical 
tests for different $\alpha$ which confirm this scaling very well).} \ed
{Actions $L_i$ reflect the angular momentum partition between both 
components}. If $\beta \sim 1$, $(\mu \sim 5)$, both  secondaries are 
dynamically equivalent, if $\beta > 1$  ($\mu > 5$) then the inner body 
``dominates'' dynamically in the system, and if $\beta < 1$ then the 
hierarchy is reversed. To illustrate the stability of the equilibria, the 
curves are marked with black dots are for Lyapunov stable solutions, and 
grey dots mean unstable solutions. The curves are labeled with both $\mu$ 
and $\beta$ parameters.

In the right-hand half-plane (quarter IV, $\omega_1=\omega_2 = \pi/2$), for $\imut 
< \pi/2$, only one solution appears that is the LK resonance. For small 
$\mu$, it  evolves with increasing $\nAMD$ along the axis of $e_2 \sim 0$  
between $e_1=0$ (when the first LK bifurcation appears, see Fig.~\ref
{fig2}b) and $e_1 \sim 1$. After the second bifurcation of the LKR 
(Fig.~ \ref{fig2}e), two solutions emerge. One of them is the LKR for 
$\imut>\pi/2$ \citep[see][this solution is not discussed 
here]{Migaszewski2009a}. The second new solution moves along the axis $e_1 \sim 1$ 
up to large $e_2$.

For $\mu \geq 2$, the LKR does not reach $e_1=1$ but it ``turns back'', 
with decreasing $e_1$ and increasing $e_2$. For larger mass ratio, the 
maximal $e_1$ is  smaller. The families of LKR  for $\mu = 1.5, 3$ 
correspond to systems with $\beta < 1$, hence the dynamical hierarchy is 
reversed, and the eccentricity of the inner, {\em more massive} body is  
forced by {\em the outer} companion.  The position of the LKR family moves 
to the region of smaller $e_1$, and for large enough $\mu$, this solution 
is confined to $e_1 \sim 0$-axis and tends to large $e_2$. The direction of 
parametric evolution of this equilibrium, corresponding to increasing $\nAMD$, 
is marked with an arrow. We see that for all $\mu$ this direction is the 
same.

The view of the left-hand half-plane is more complex. As we shown in \citep 
{Migaszewski2009a}, the first solution appearing in quarter III of the 
\RPs-plane is unstable equilibrium IIIa emerging \corr{due to the second 
bifurcation of the origin  $(e_1 = e_2 = 0)$ (marked with $I_1^{-}$ in our 
paper)}. Then, with increasing $\nAMD$ , this solution moves along $e_1 \sim 
0$ towards large $e_2$. For some value of $\nAMD$ (e.g., $\sim 0.195$ for 
$\alpha=0.01$, $\mu=20$; see Fig.~\ref{fig3}) solutions IIIb and IVb+ appear 
in the same point of the phase space (this is illustrated with crossed 
circle in the left-hand half-plane of Fig.~\ref{fig3}). Solution IVb+ then 
moves along $e_1 \rightarrow 1$ and $e_2 \rightarrow 0$. Simultaneously, 
solution IIIb evolves towards a point marked with dotted circle. Solution 
IIIa also moves to that point and for some critical $\nAMD$ both solutions 
merge and vanish. Figure~\ref {fig12} reveals that the 
parametric paths of equilibria depend on parameter $\mu$ ($\alpha=0.04$ was 
fixed in this test). The path of the LKR becomes closer to $e_1 \sim 0$-axis 
for larger mass ratio. Also families of stationary solutions that are 
present in quarter III, become closer to the origin. The range of 
eccentricities corresponding to these equilibria is very small for $\mu 
\geq 25$. We may also notice that the bifurcation points (crossed and 
dotted circles, respectively) tend to each other with increasing mass 
ratio, which is consistent with the transition between the planetary and
the circumbinary regime. For $\mu \geq 25$, these two points are merged. In 
this particular case, the structure of equilibria in quarter III of the 
\RPs-plane is more simple than in the general case because only one 
solution persists, a stable equilibrium corresponding to nearly 
circular outer orbit.

For different $\alpha < 0.1$, the general, global view of the families 
of equilibria is very similar to the results presented in Fig.~\ref{fig12}. 
In fact, as we noticed previously, the  parametric
evolution of the equilibria is reflected by parameter $\beta(\mu,\alpha)$,
and basically does not depend on the individual values of $\alpha$ and $\mu$. 
%
%
\section{The PN correction to the secular model}
%
The results presented in the previous section illustrate the
already well known feature of the classic model \citep[see, 
e.g.,][]{Michtchenko2004,Michtchenko2006}. In the approximation of small 
masses, the  secular dynamics of the Newtonian 2-planet, hierarchical model  
depend on the semi-major axes ratio and planetary mass ratio, and not on 
individual system parameters, $(a_1, a_2, m_1, m_2)$. In our works \citep
{Migaszewski2009b,Migaszewski2010}, we shown that this feature is not 
preserved in the more general model, including relativistic, rotational and
tidal corrections to 
the Newtonian point-mass interactions. In these settings, the secular 
dynamics depend on the individual semi-major axes, as well as individual 
planetary masses. Because the overall structure of the phase space is 
characterized by the equilibria, we now attempt to show that deviations between these 
equilibria in the classic and relativistic models become more important when 
the system dimension scales down ($a_1, a_2$ decrease when $\alpha$ is
$\mbox{const}$), and masses $m_1, m_2$ are smaller, when their ratio $\mu$ is 
kept constant.

\begin{figure*}
\centerline{
\vbox{
    \hbox{\includegraphics [width=126mm]{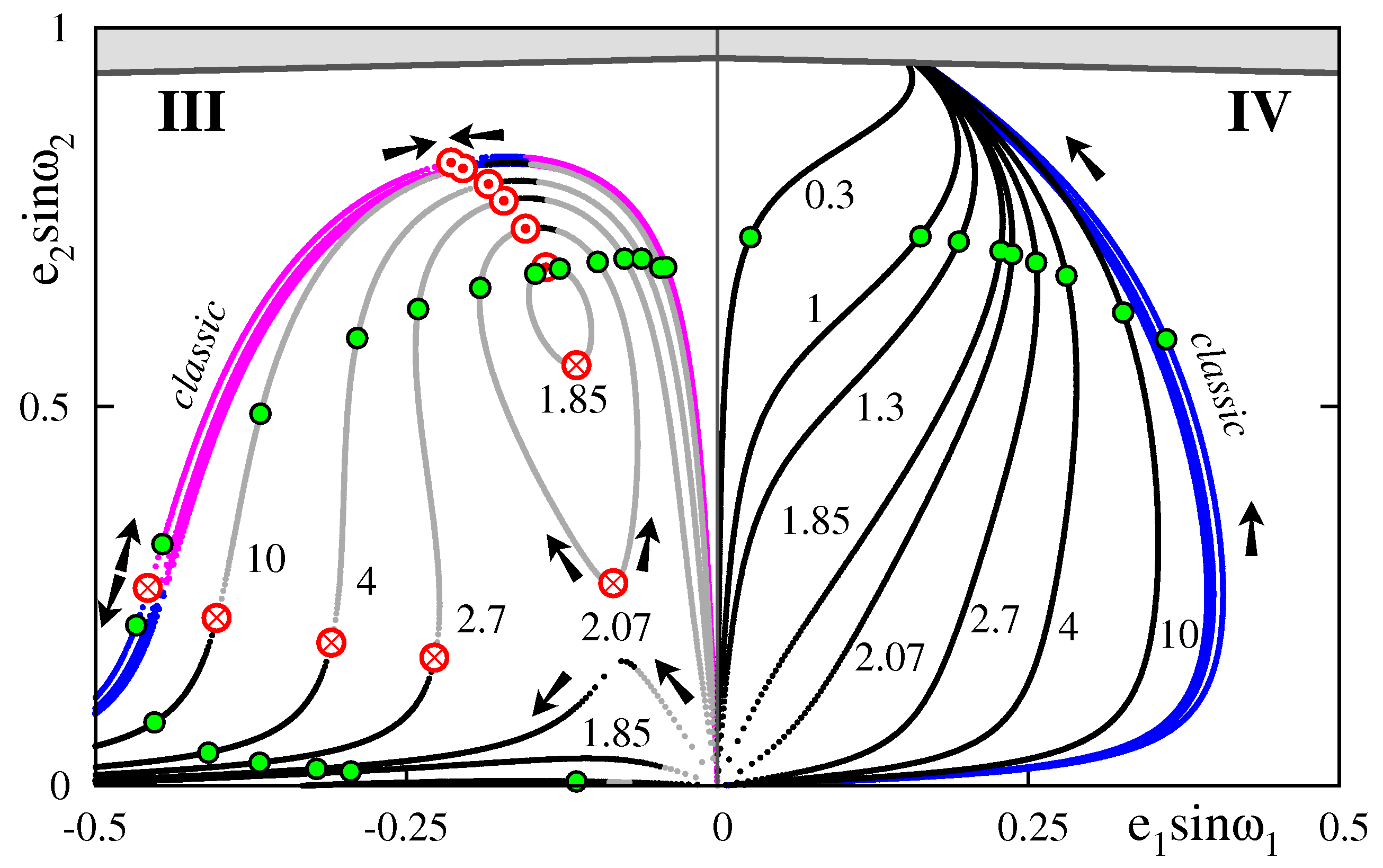}
          }}
      }
\caption{
Families of stationary solutions presented in the \RPs-plane, calculated for 
$\alpha=0.04, \mu=10, a_1=0.2~\mbox{au}, a_2=5.0~\mbox{au}, 
m_0=1~\mbox{m}_{\odot}$ and varied $m_2=10, 4, 2.7, 2.07, 1.85, 1.3, 1.0,
0.3~\mJ$. Equilibria in the classic model 
are compared with the equilibria in the relativistic model. The 
mass of the outer body $m_2$ labels each particular curve. Black dots are 
for stable equilibria, grey dots are for unstable equilibria of the relativistic 
model. Equilibria of the classic model are plotted with blue and violet dots for 
stable and unstable solutions, respectively. Positions of equilibria were 
calculated with the help of the octupole theory.
}
\label{fig13}
\end{figure*}

The differences between the two coplanar models manifest itself through the 
shapes and localization of stationary solutions and depend on the {\em 
individual} masses and semi-major axes \citep {Migaszewski2009b}. Now we can 
observe the same feature in the spatial planetary system, corrected for the 
relativistic perturbation.  Figure~\ref{fig13} presents families of 
equilibria in the same manner as Fig.~\ref{fig12} (due to the symmetry, only 
the $y$-positive half-plane of \RPs-plane is presented). Families of 
stationary solutions in the classic model are drawn with blue and violet 
curves for stable and unstable solutions, respectively. Solutions in the 
relativistic model are plotted with black (stable equilibria) and grey 
(unstable equilibria) curves.  In this experiment, we varied the individual 
masses of the secondary bodies, still keeping their ratio $\mu=10$ as a 
constant.  The masses are changed between $(m_1 = 100~\mJ, m_2 = 10~\mJ)$ to 
$(m_1 = 3~\mJ, m_2 = 0.3~\mJ)$, and the primary mass is $m_0 = 1~\msun$.

The results illustrated in Fig.~\ref{fig13} confirm that even for 
large secondary masses,  the parametric curves of stationary solutions in 
the realm of the classic model depends weakly on the individual masses. Yet 
similarly to the co-planar case, curves of equilibria calculated with the 
relativistic corrections differ significantly from the results obtained for 
the classic model. The deviations between both models are more significant 
for smaller masses of the secondary bodies. For instance, the family IVa 
moves to the range of much smaller $e_1$. Solutions of this type exist even 
for very small $m_1$ and $m_2$, which is just not possible in the Newtonian model \citep
{Migaszewski2009a}. When the mutual perturbations between secondaries are 
small enough, the critical inclination in the relativistic model, that leads 
to the LK bifurcation, becomes larger than $\pi/2$ . Thus $\icrit$ increases 
with decreasing masses.

In quadrant III of the \RPs-plane ($\omega_2 = -\omega_1 = \pm \pi/2$), the 
structure of equilibria is even more complex. For some critical values of 
masses ($m_2 \sim 2.07~\mJ$) the parametric paths divide into two parts. The 
top part, characterized by larger $e_2$, comprises of two unstable equilibria 
emerging from one bifurcation point, which then meet in another bifurcation 
point. The bottom part of the equilibria curve represents a saddle point, 
that changes its stability, from unstable IIIa-type solution to the stable 
solution IVb+. This branch  is very similar to equilibria in the classic 
system with large $\mu$ (or large $\beta > 5$) observed \ed{already in Fig.~
\ref{fig12}}, however this  takes place for quite a different 
value of $\beta = 2$.

\begin{figure*}
\centerline{
\vbox{
    \hbox{\includegraphics [width=52mm]{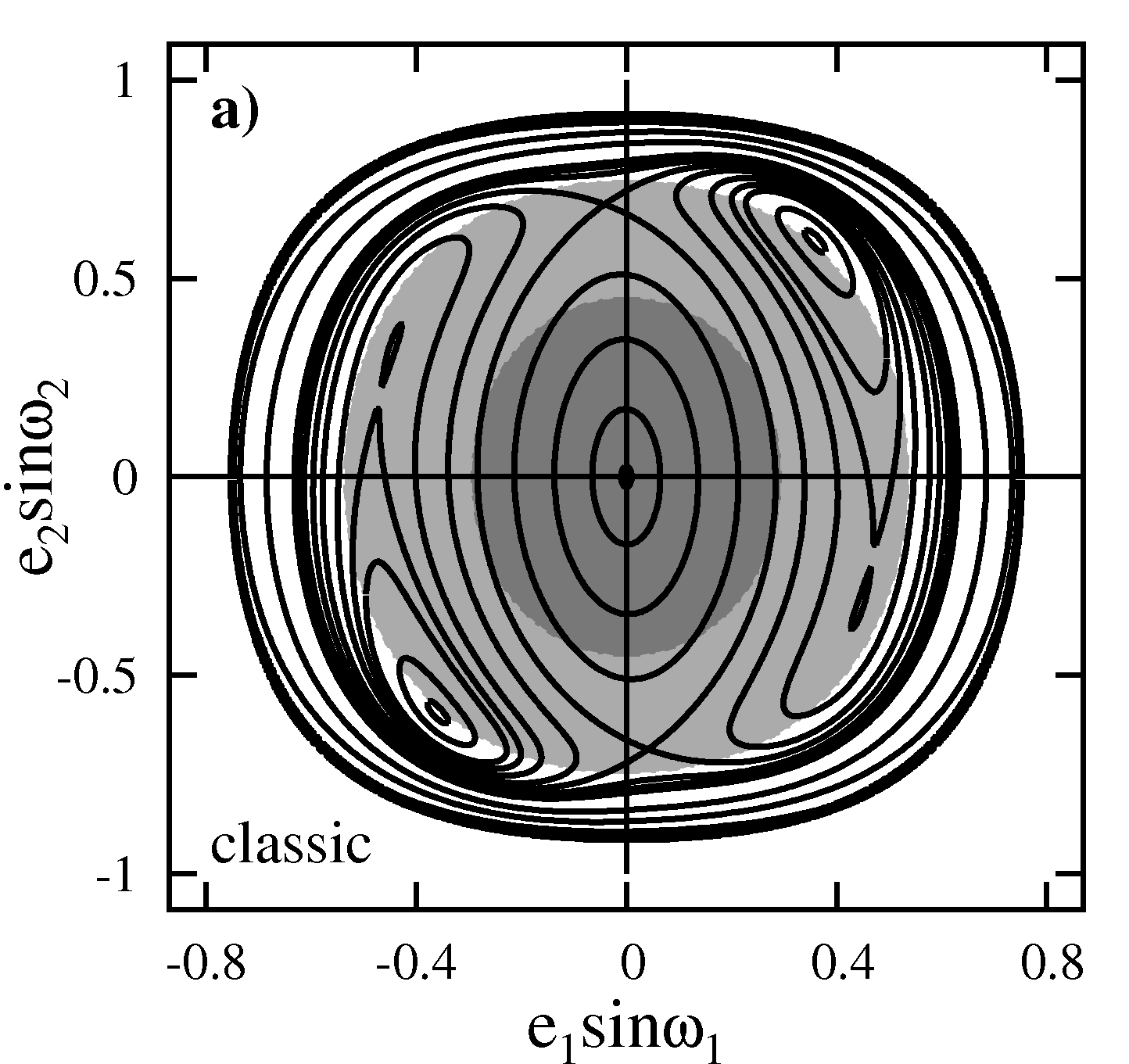} \hspace*{5mm}
          \includegraphics [width=52mm]{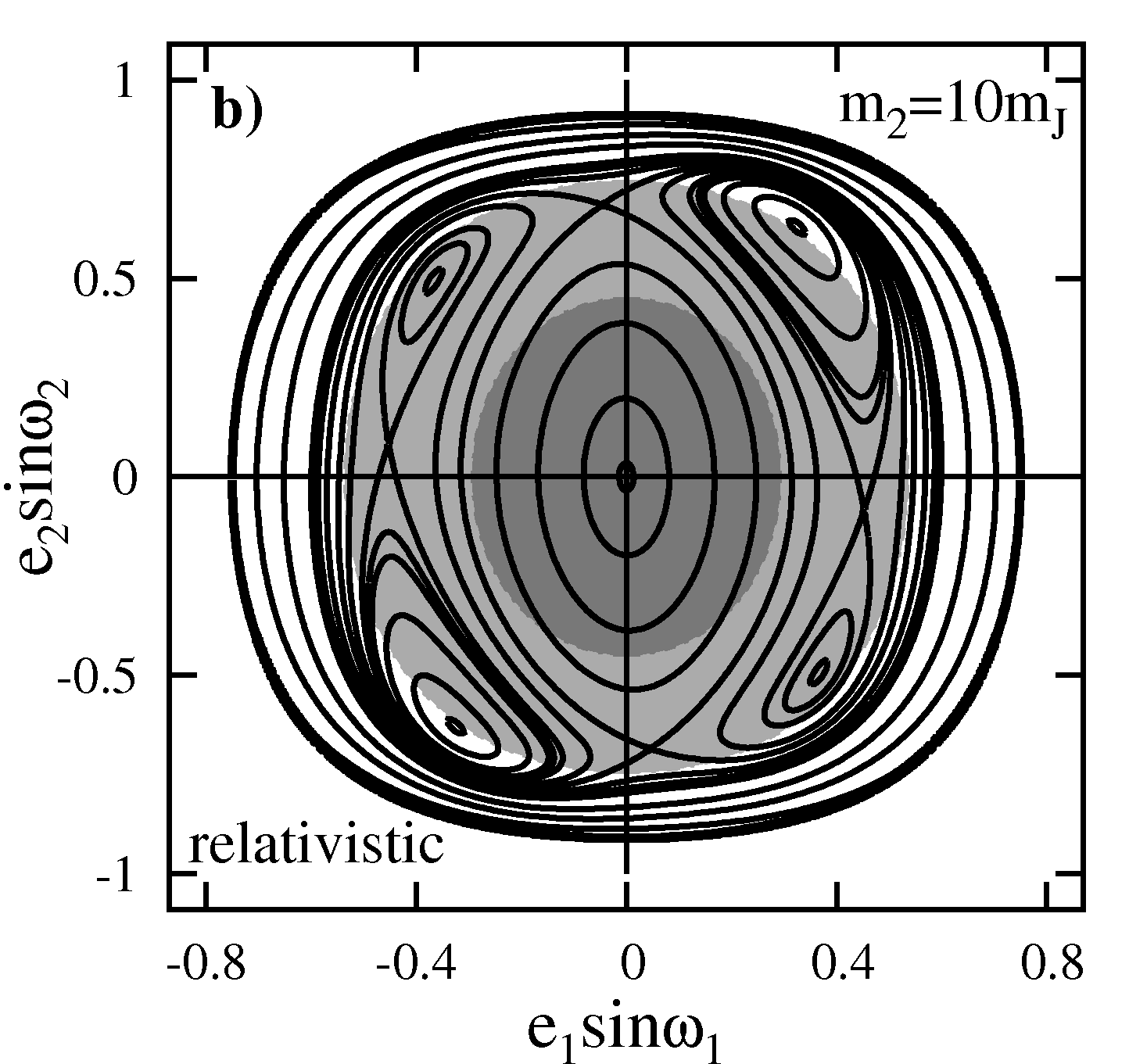}
         }
    \hbox{
          \includegraphics [width=52mm]{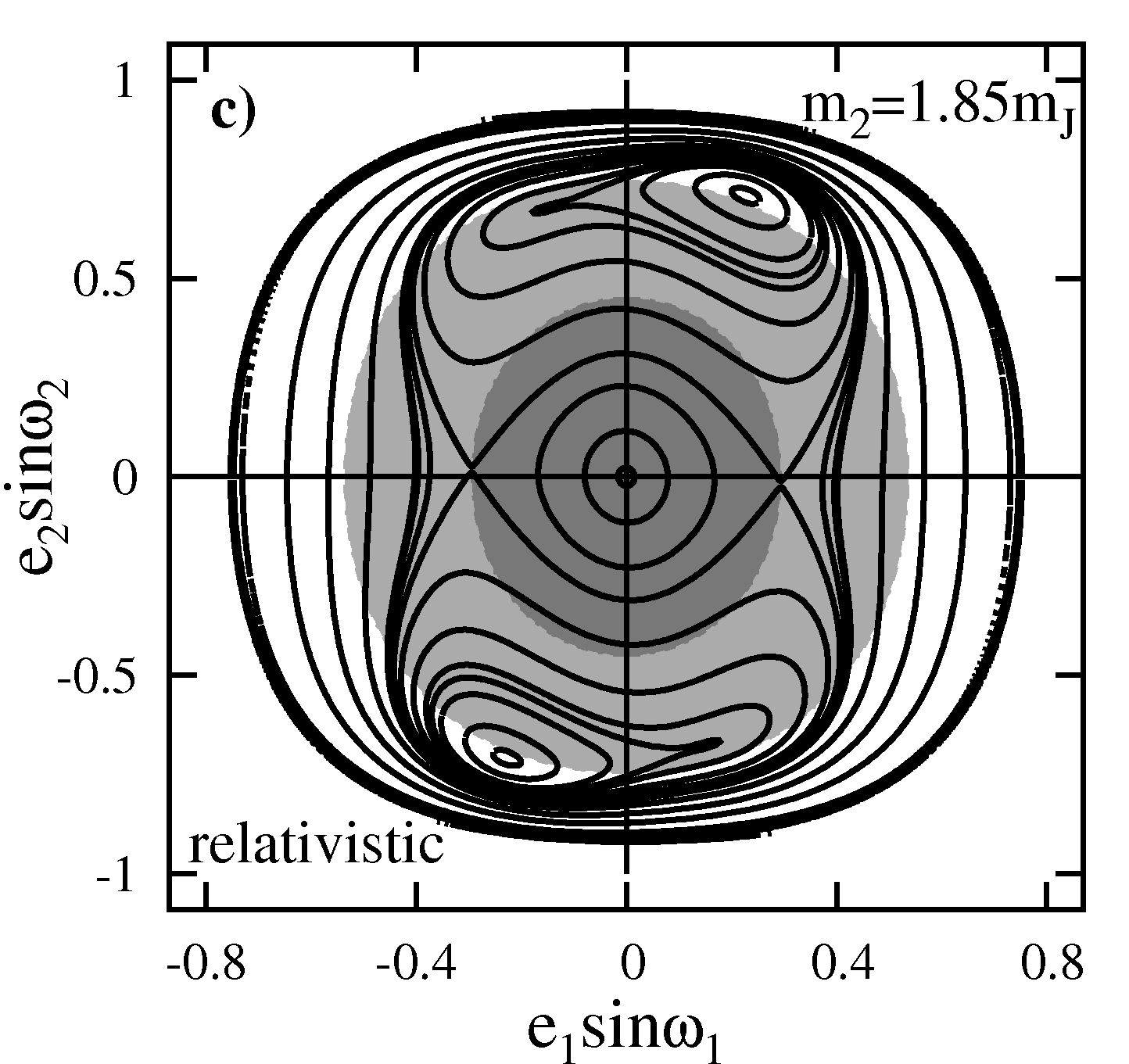}\hspace*{5mm}
          \includegraphics [width=52mm]{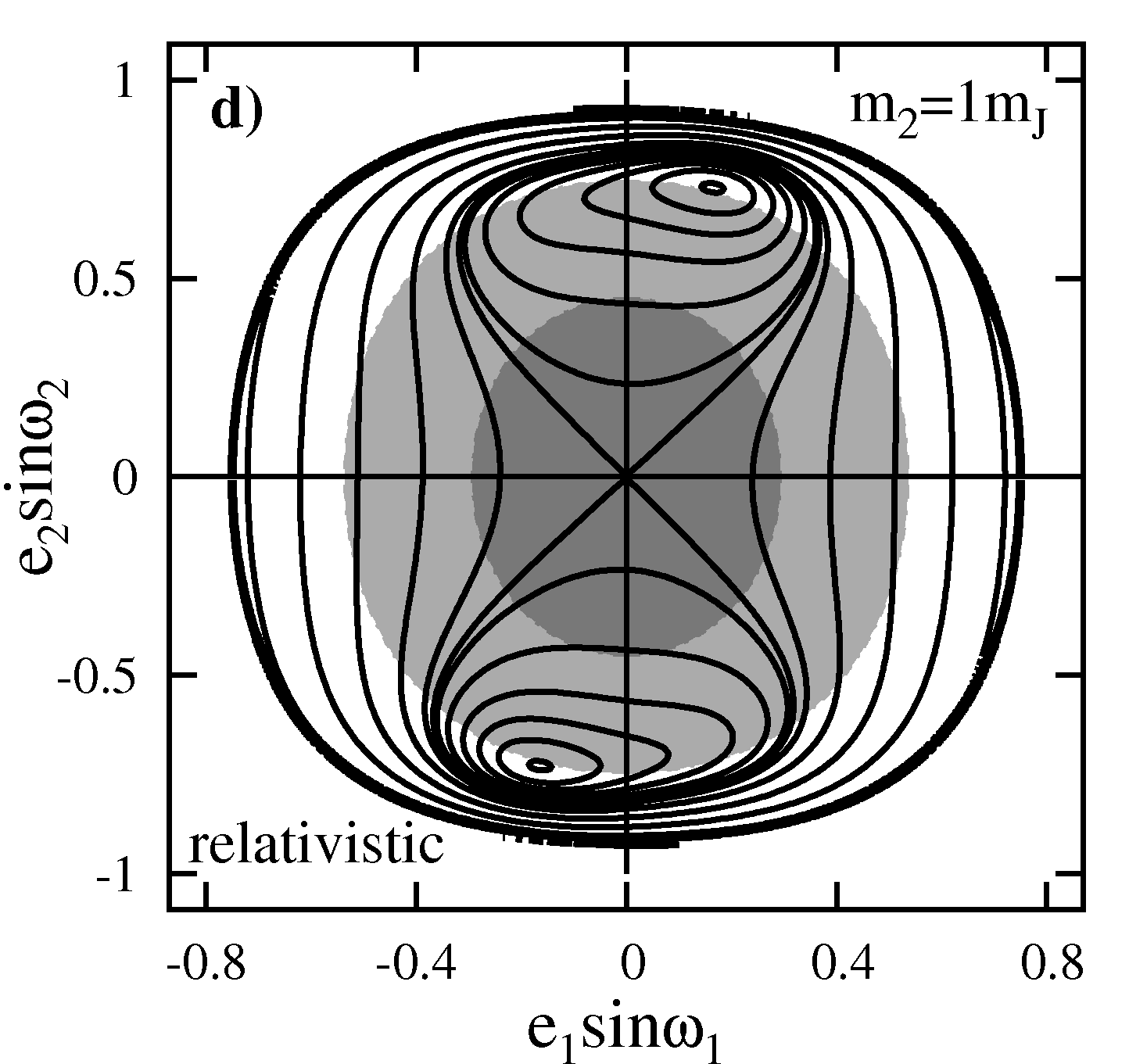}}
         }
 }
\caption{
Energy levels presented in the \RPs-plane, and calculated for the following
parameters $\alpha=0.04, \mu=10, \nAMD=0.215, i_0=84^{\circ}$, 
$a_1=0.2~\mbox{au}, a_2=5.0~\mbox{au}, m_0=1~m_{\odot}$. A sequence of
panels demonstrates the view of the phase space when
masses of the secondary bodies decrease (their ratio is  constant). From
panel b) to panel d) $(m_1,m_2)[\mJ]$ are $(100,10), (18.5,1.85), (10,1)$. 
Shaded areas mark ranges of $\imut$ larger than $60^{\circ}, 75^{\circ}$, 
(the light-- and dark-- grey, respectively). 
}
\label{fig14}
\end{figure*}

The stationary solutions are determined by the shape of the secular 
Hamiltonian. Its levels are plotted in the \RPs-plane, Fig.~\ref{fig14}. Each 
panel in this figure is calculated for the same parameters $\mu$, $\alpha$, 
$\nAMD$ (but in this case, particular values of masses and semi-major 
axes are varied). Let us recall, that Fig.~\ref{fig14}a shows the phase 
space calculated in classic model, while subsequent panels~\ref{fig14}b,c,d 
are for relativistic model, and different masses and semi-major axes, 
labeled in subsequent panels.

If the masses are large (Fig.~\ref{fig14}b), we can see four elliptic points 
separated by four saddles (let us recall that the \RPs-plane is redundant, and the 
energy levels are reflected with respect to the origin, thus in fact we 
have only two unique elliptic points and only two saddles). The elliptic 
points may be identified with solutions IVa ($\omega_1 = \omega_2 = \pm 
\pi/2$) and IIIb ($\omega_1 = -\omega_2 = \pm \pi/2$). The saddles 
correspond to solutions IIIa ($e_1 \sim 0$) and IVb+ ($e_2 \sim 0$). When 
the masses decrease (still, $\mu$ is kept constant), the structure 
surrounding solution IIIb becomes smaller and moves towards solution IIIa. 
Simultaneously, the saddle point IVb+ tends to the origin. For the masses 
small enough, (Fig.~\ref{fig13}), solutions IIIa and IIIb merge and vanish, 
while IVb+ ``falls'' into the origin.

\begin{figure*}
\centerline{
\vbox{
    \hbox{\includegraphics [width=84mm]{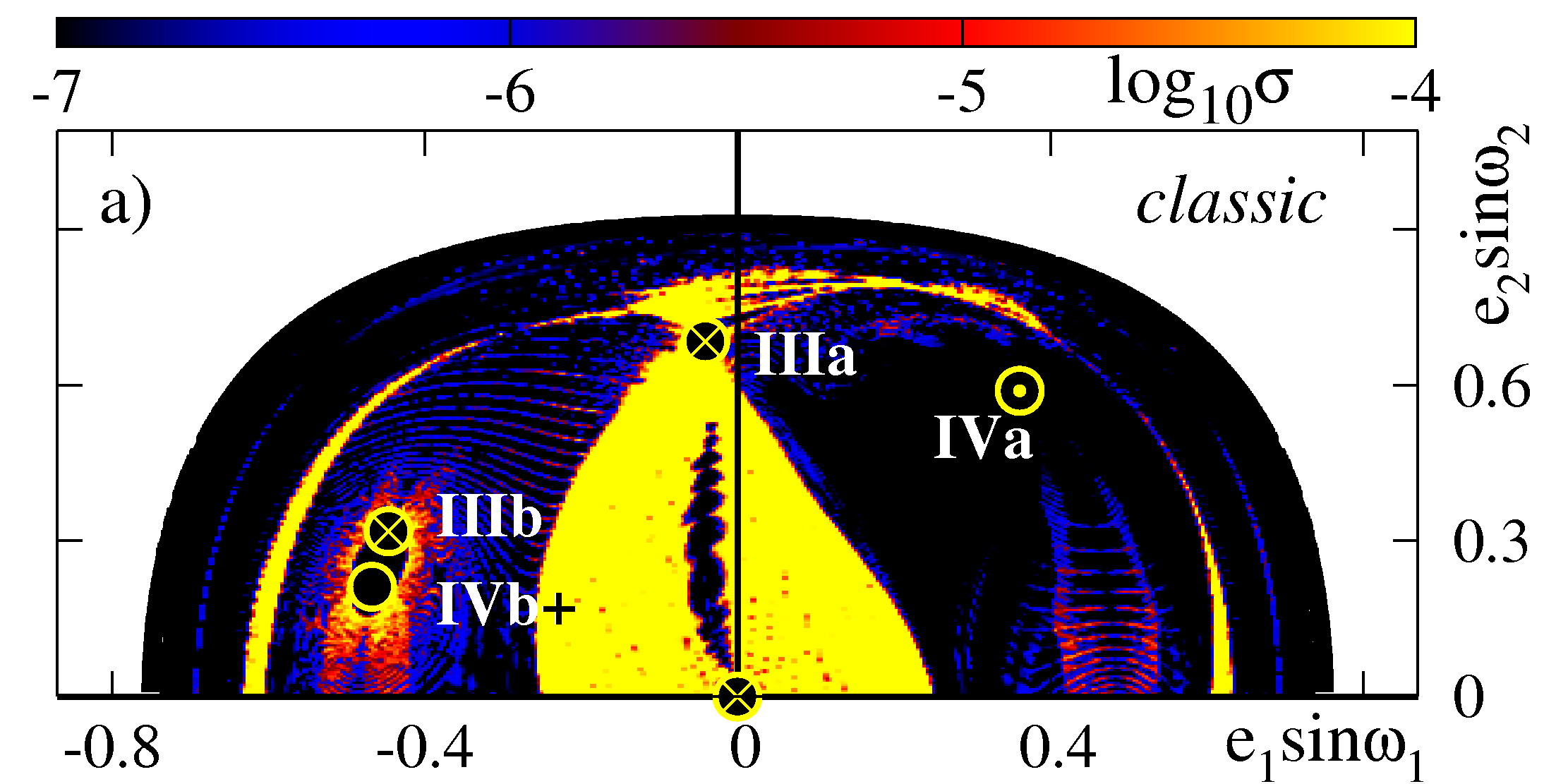}\hspace*{2mm}
          \includegraphics [width=84mm]{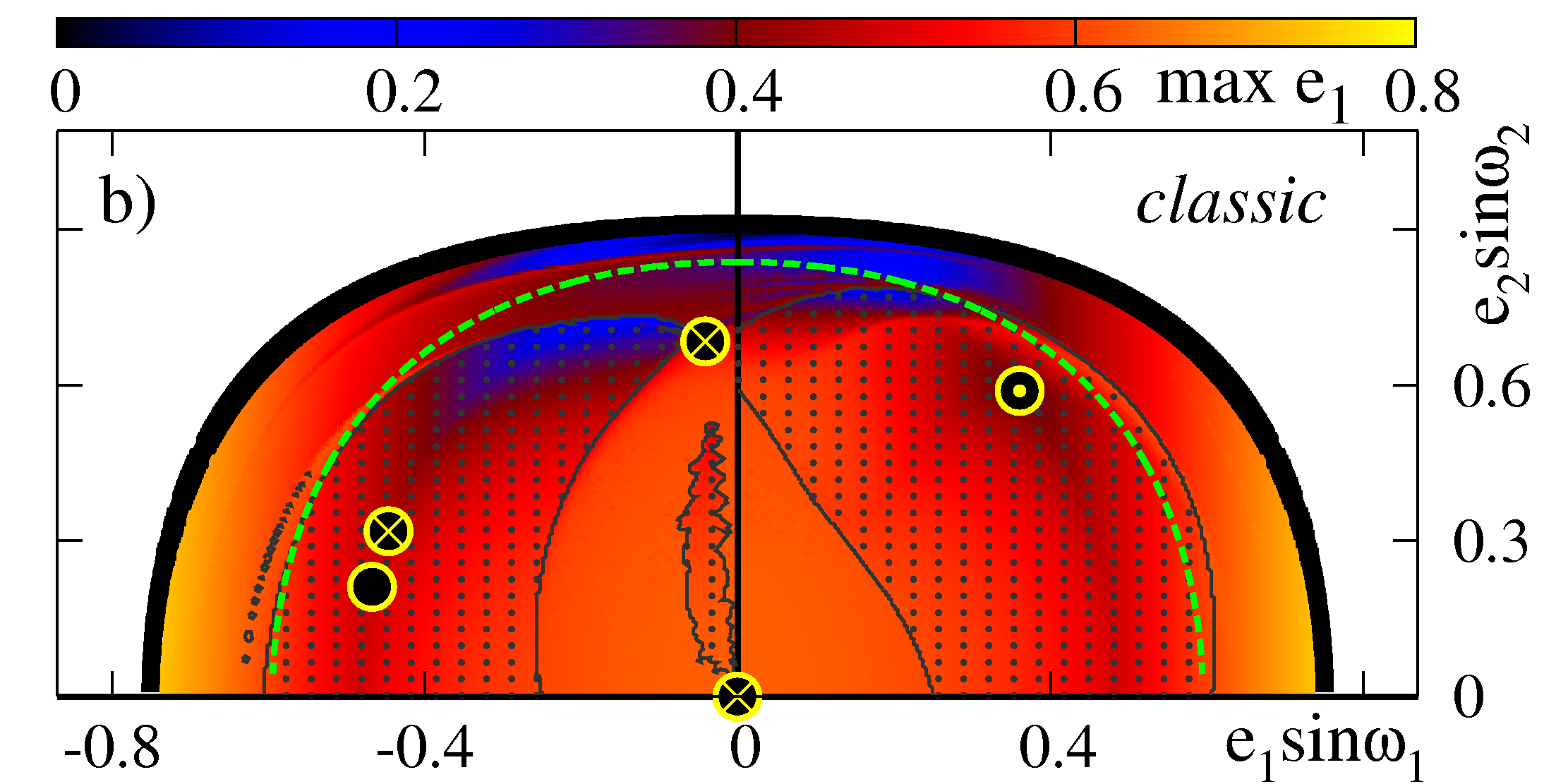}
	  }
    \hbox{\includegraphics [width=84mm]{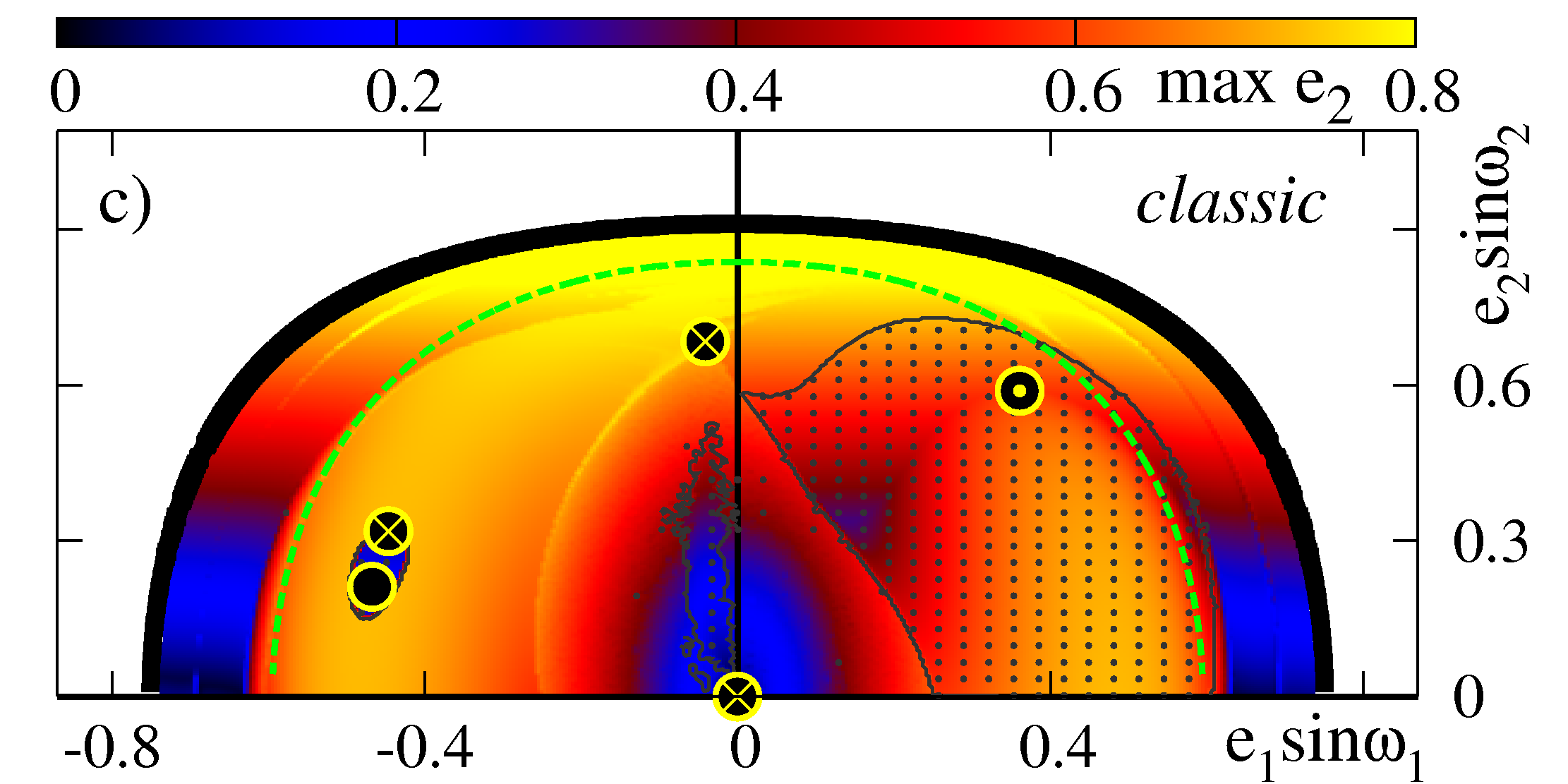}\hspace*{2mm}
          \includegraphics [width=84mm]{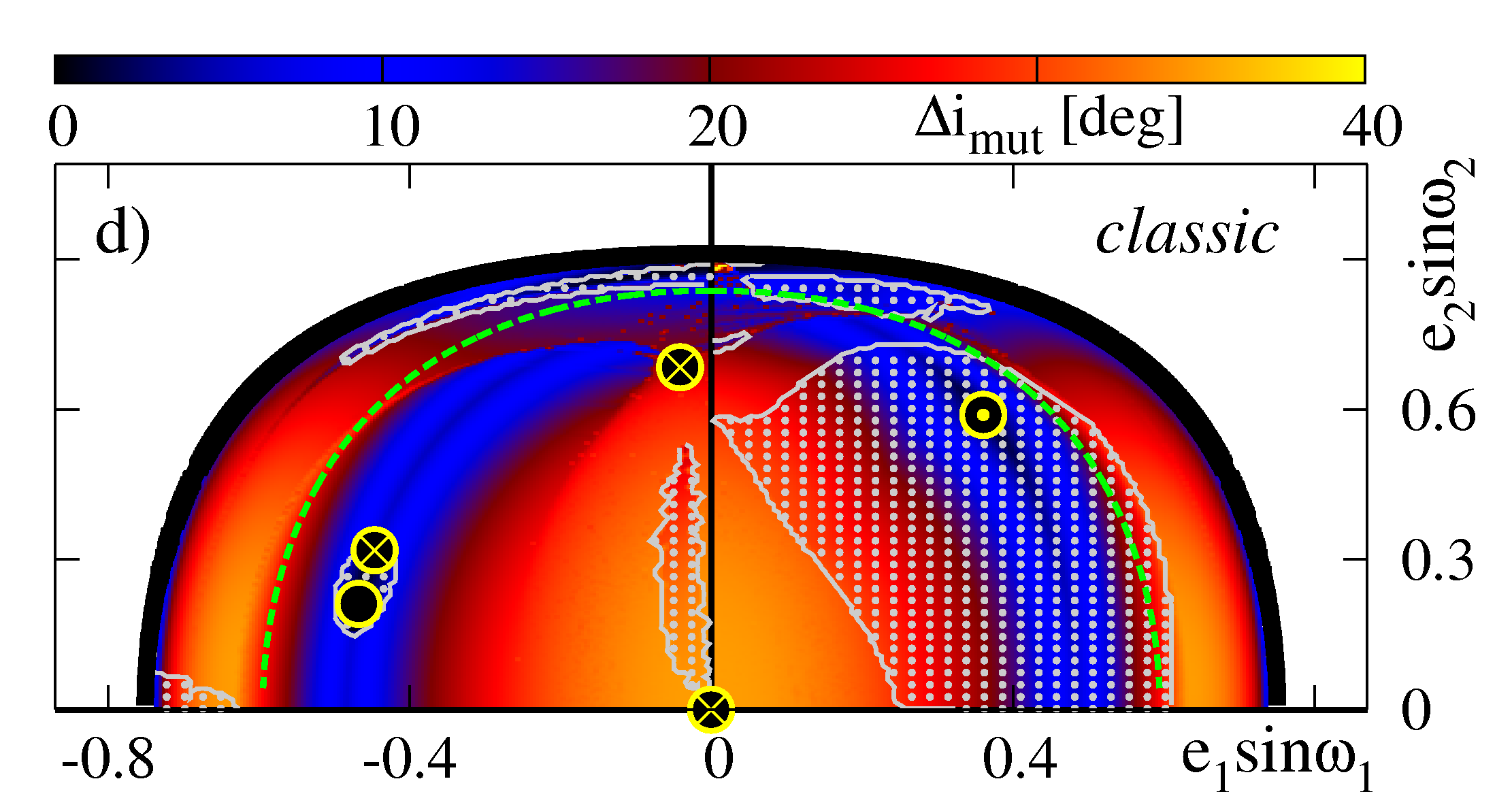}
	  }
 }
 }
\caption{\corr{Dynamical maps for the classic, Newtonian model in 
the \RPs-plane. 
The model parameters are  $\alpha=0.04, \mu=10, \nAMD=0.215,
i_0=84^{\circ}$, $a_1=0.2~\mbox{au}, a_2=5.0~\mbox{au}, m_0=1~m_{\odot}$,
\ed{$m_1 = 100~\mJ$, $m_2 = 10~\mJ$}.
See the text and caption to Fig.~\ref{fig4} for the details.
}
}
\label{fig15}
\end{figure*}

\begin{figure*}
\centerline{
\vbox{
    \hbox{\includegraphics [width=83.5mm]{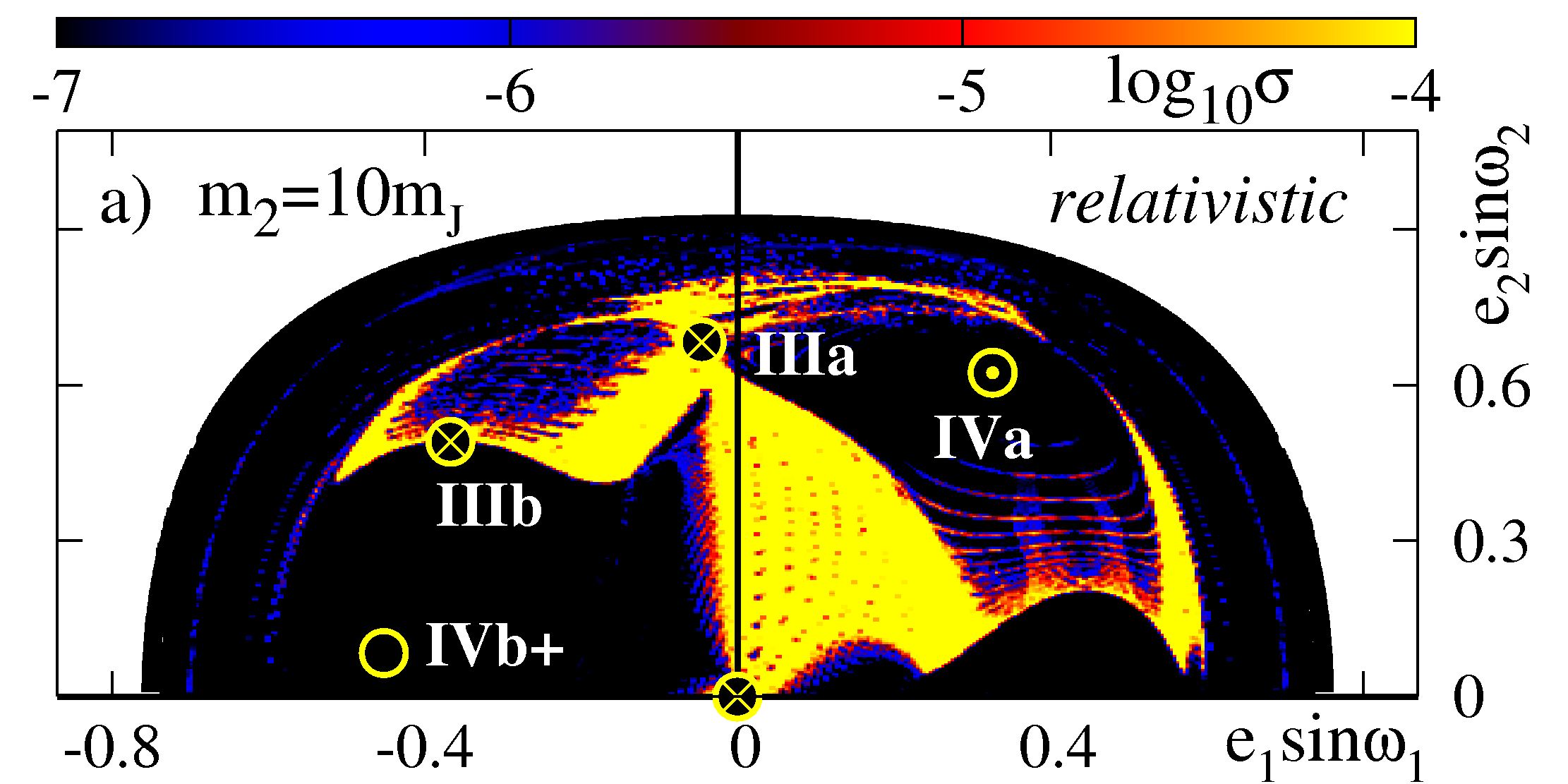}\hspace*{2mm}
          \includegraphics [width=83.5mm]{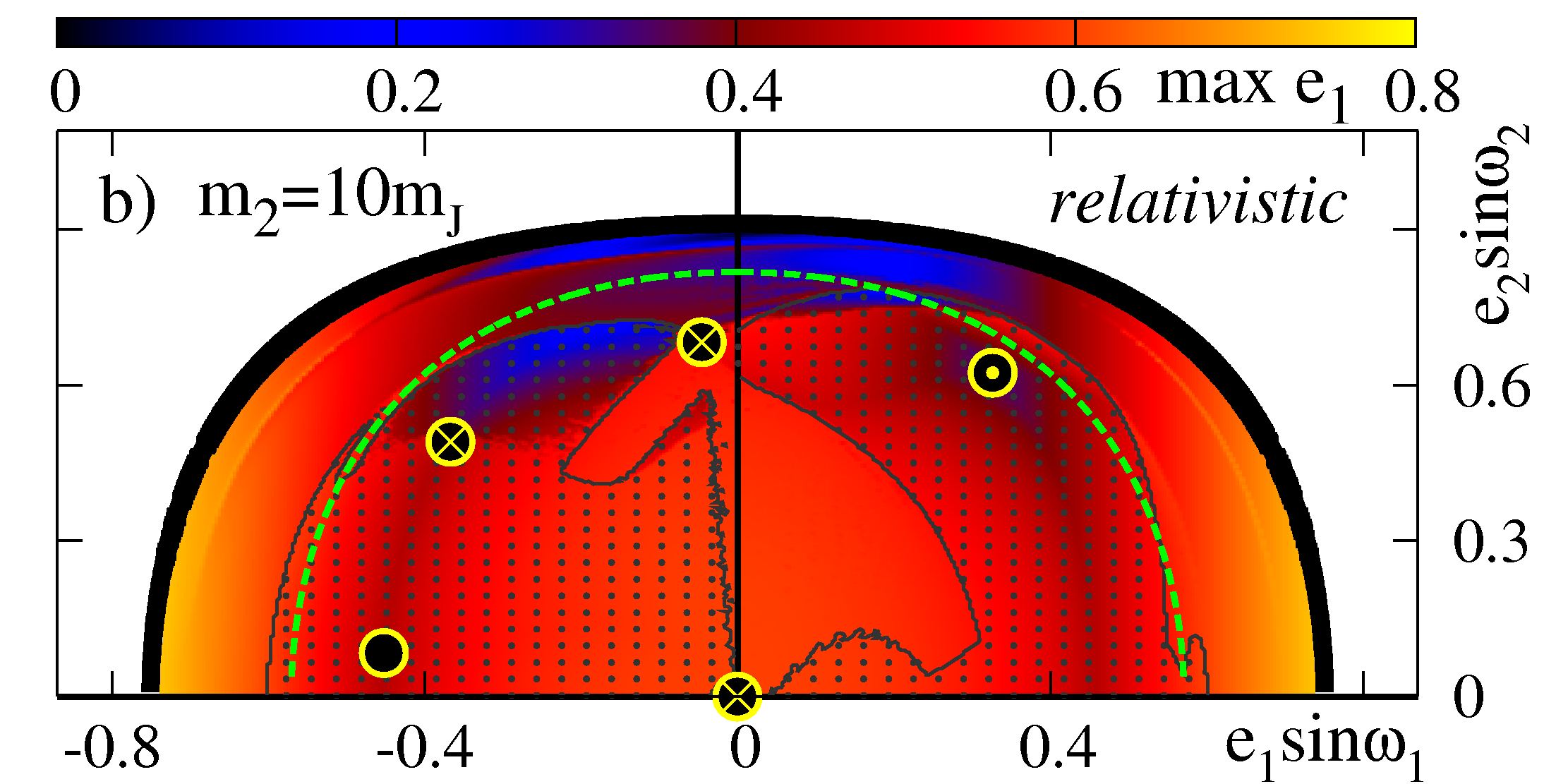}
	  }
    \hbox{\includegraphics [width=83.5mm]{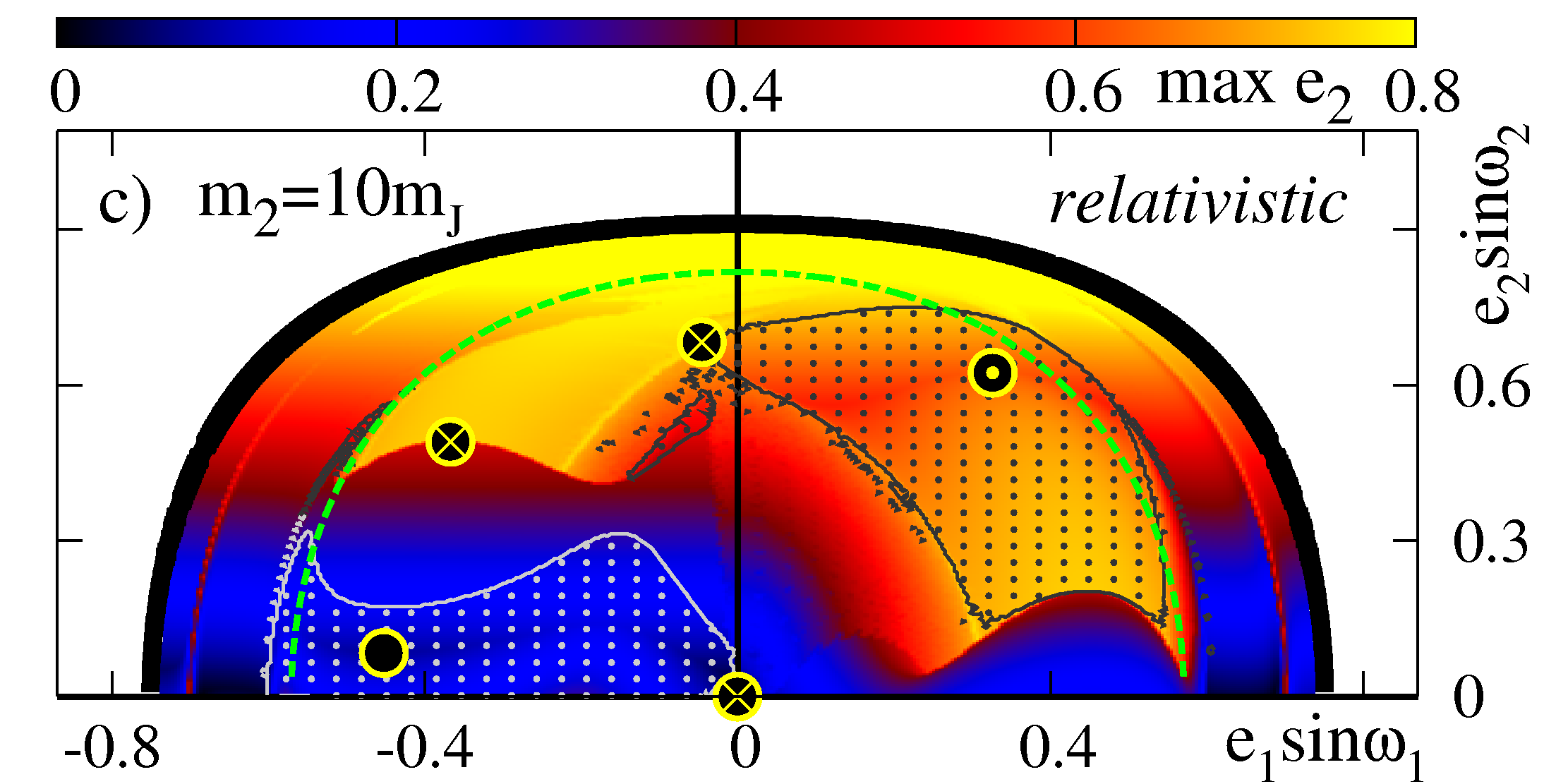}\hspace*{2mm}
          \includegraphics [width=83.5mm]{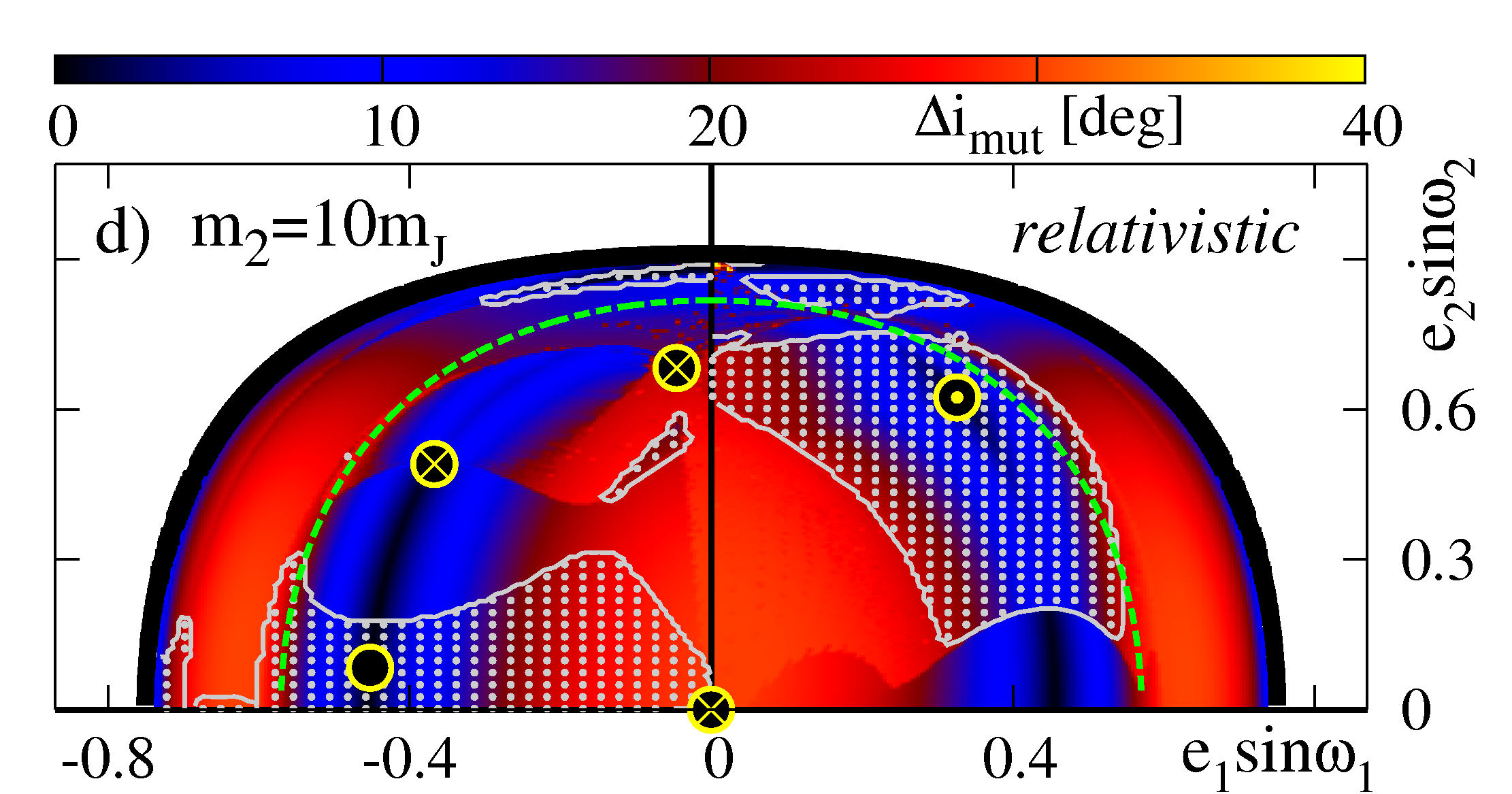}
	  }
}
}
\caption{\corr{
The dynamical maps for the octupole model with relativistic corrections.
Masses of the secondary bodies are $m_1 = 100\,\mJ$,
$m_2 = 10\,\mJ$. See the text and caption to Fig.~\ref{fig4} for
more details.
}
}
\label{fig16}
\end{figure*}

\begin{figure*}
\centerline{
\vbox{
    \hbox{\includegraphics [width=83.5mm]{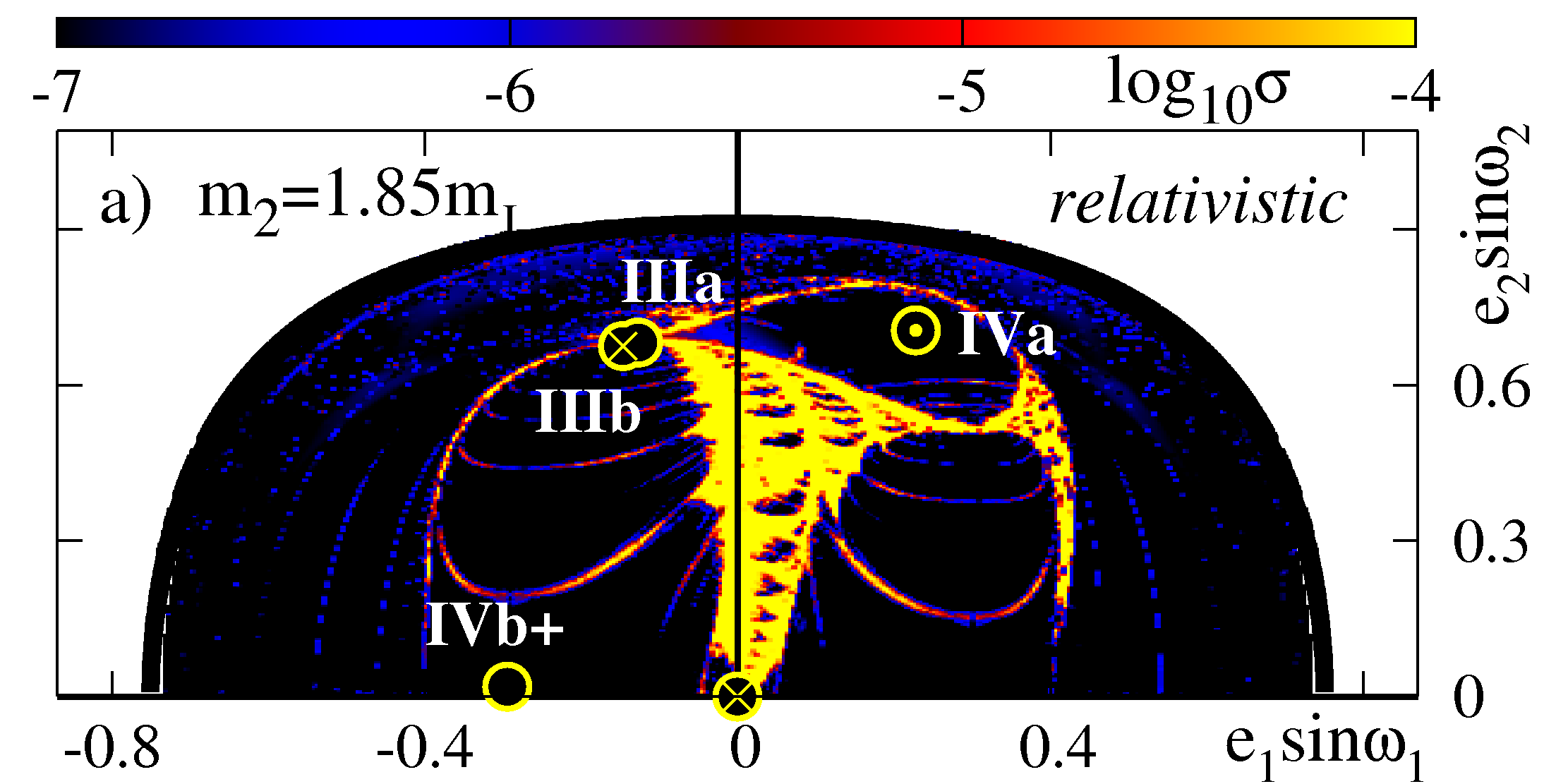}\hspace*{2mm}
          \includegraphics [width=83.5mm]{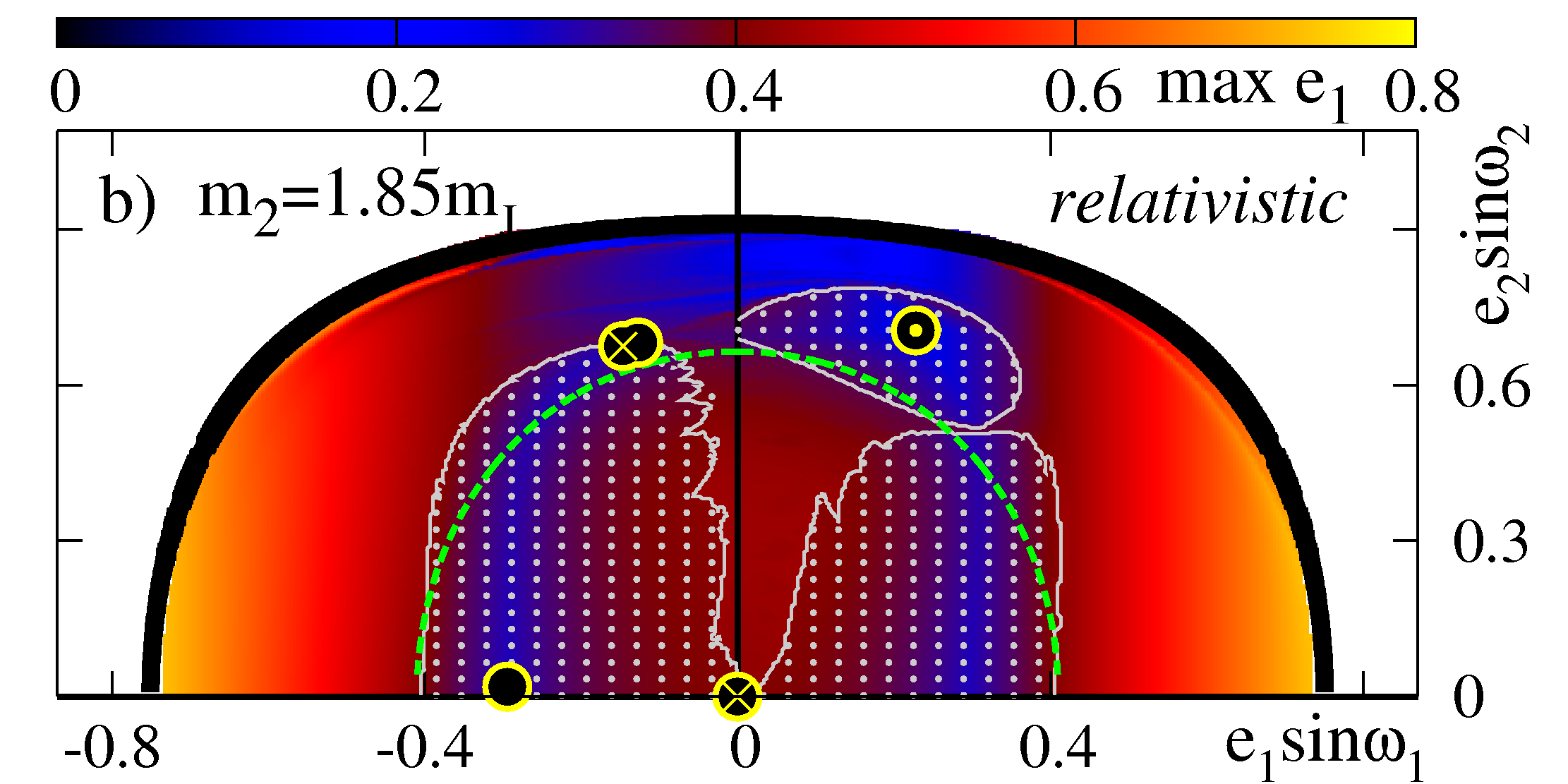}
	  }
    \hbox{\includegraphics [width=83.5mm]{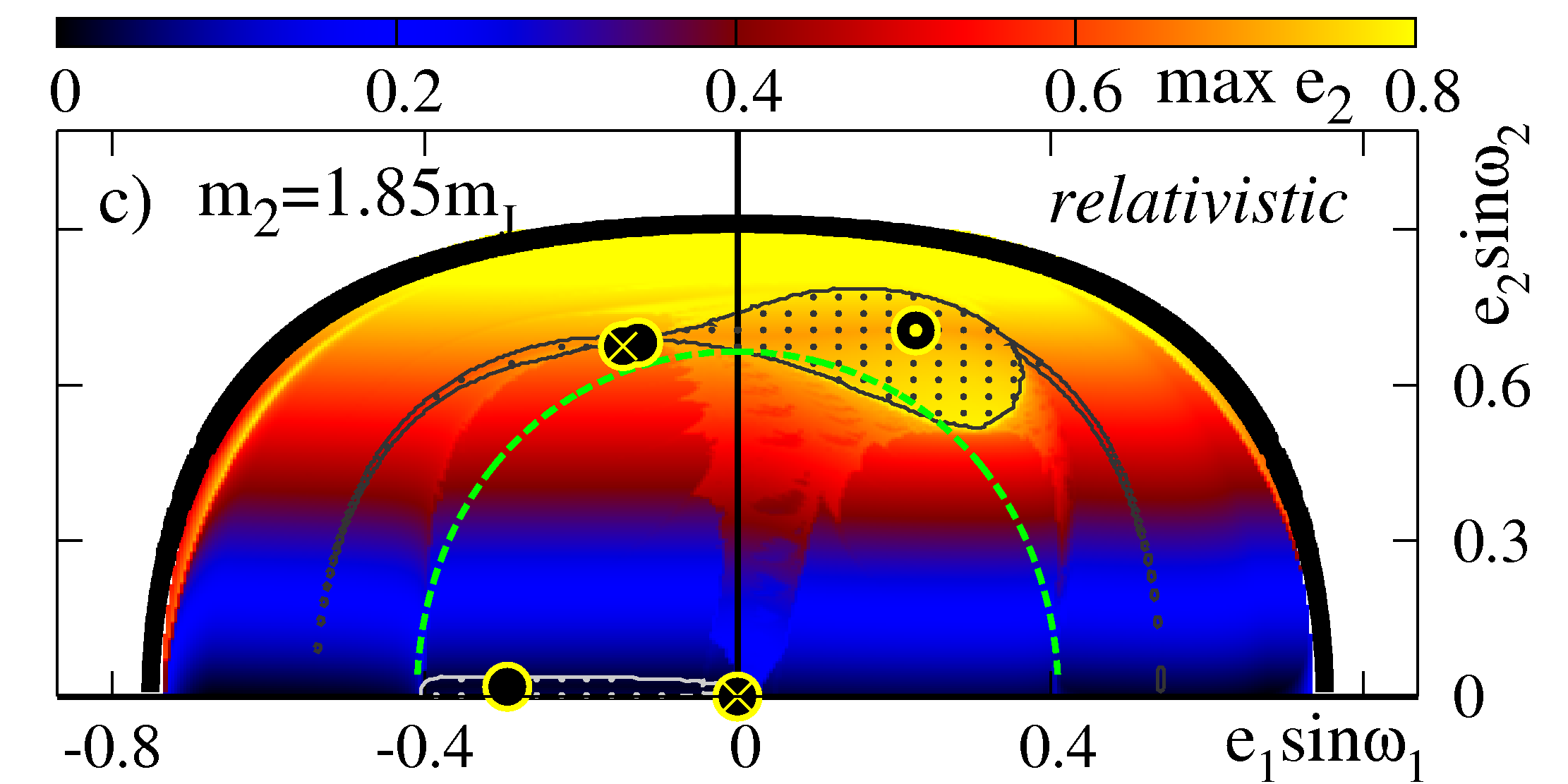}\hspace*{2mm}
          \includegraphics [width=83.5mm]{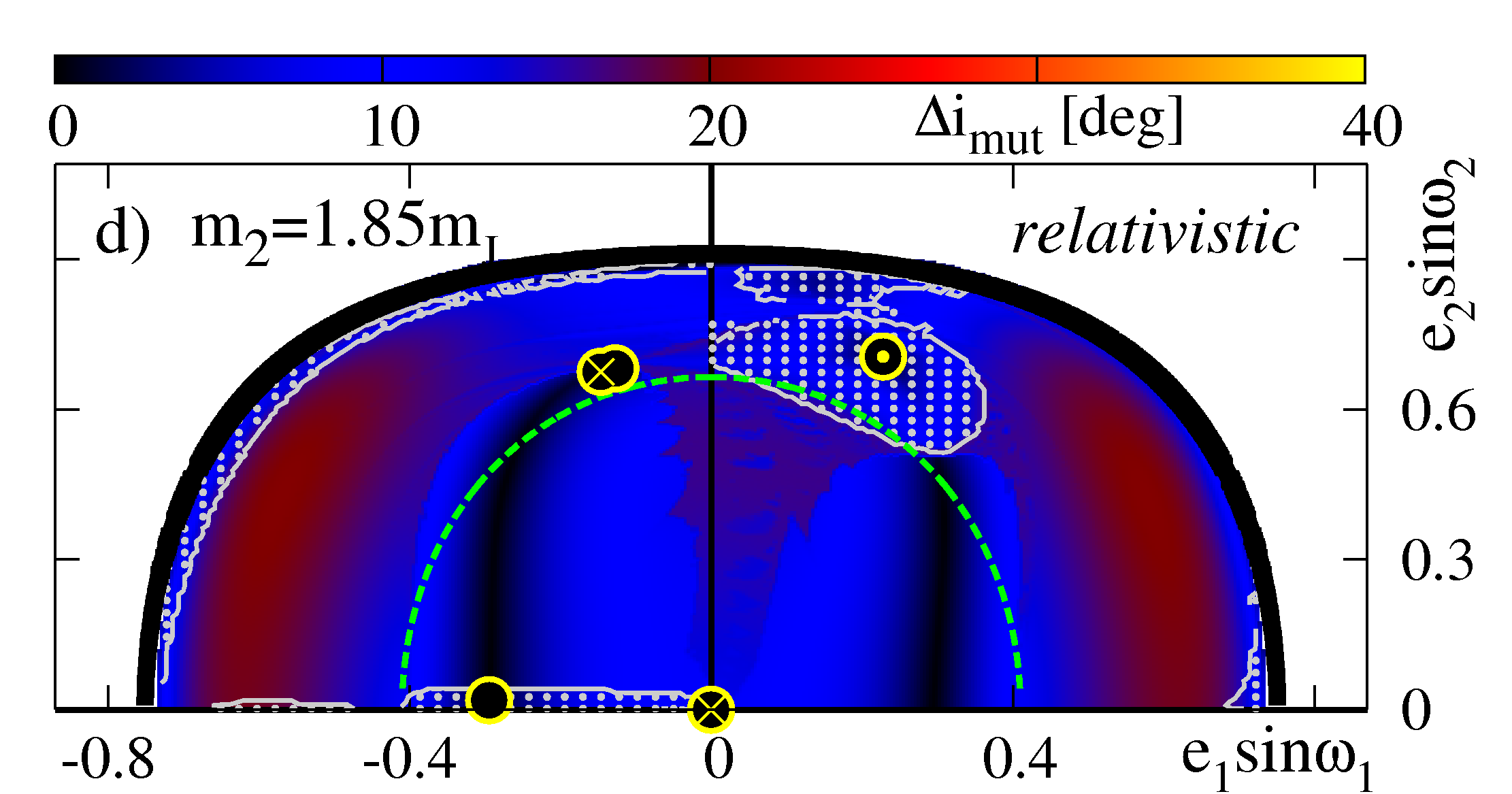}
	  }
 }
 }
\caption{\corr{
The dynamical maps for the octupole model with relativistic corrections.
Masses of the secondary bodies are  $m_1 = 18.5\,\mJ$, $m_2 = 1.85\,\mJ$. 
See the text and caption to Fig.~\ref{fig4} for more details.
}
}
\label{fig17}
\end{figure*}

Figures~\ref{fig15}--\ref{fig17} shows the dynamical maps for the 
relativistic model, and are constructed in the same  manner as maps in Figs.~
\ref{fig4}--\ref{fig6}. Subsequent figures  correspond to masses and $\nAMD$
used to plot the energy levels in Figs.~\ref{fig14}a,c,d respectively: Fig.~
\ref{fig15} is for classic model, Figs.~\ref{fig16} and~ \ref{fig17} are for 
the relativistic model with $m_2 = 10~\mJ$ and $m_2 = 1.85~\mJ$, 
respectively. The mass ratio is $\mu=10$ in both instances. The order of 
panels and symbol-coding of equilibria, as well as coding libration zones of the 
angles $\omega_1$, $\omega_2$, and $\Delta\varpi$ are the same as in Figs.~\ref{fig4}--\ref
{fig6}; in particular, dotted, crossed and empty circles mark the Lyapunov 
stable, unstable and linearly stable equilibria, respectively. Clearly, the 
overall view of the phase space is different in all cases. The regions of 
chaotic motions (yellow areas in the $\sigma$-maps) obtained for the classic 
and relativistic model are significantly different. Also the dependence on 
the masses of secondaries in the relativistic models is evident. The 
structure of chaotic/regular secular evolution is reflected in
$\max e_{1,2}$-maps and through librational regions of $\omega_1$
and $\omega_2$. We note, that in 
the case of the regular solutions, \corr{$\omega_1$ librates around $\pm 
\pi/2$ when $\imut > \icrit$. In  some part of this region also $\omega_2$ 
librates around $\pm \pi/2$}.

\begin{figure}
\centerline{
\hbox{
          \includegraphics [width=84mm]{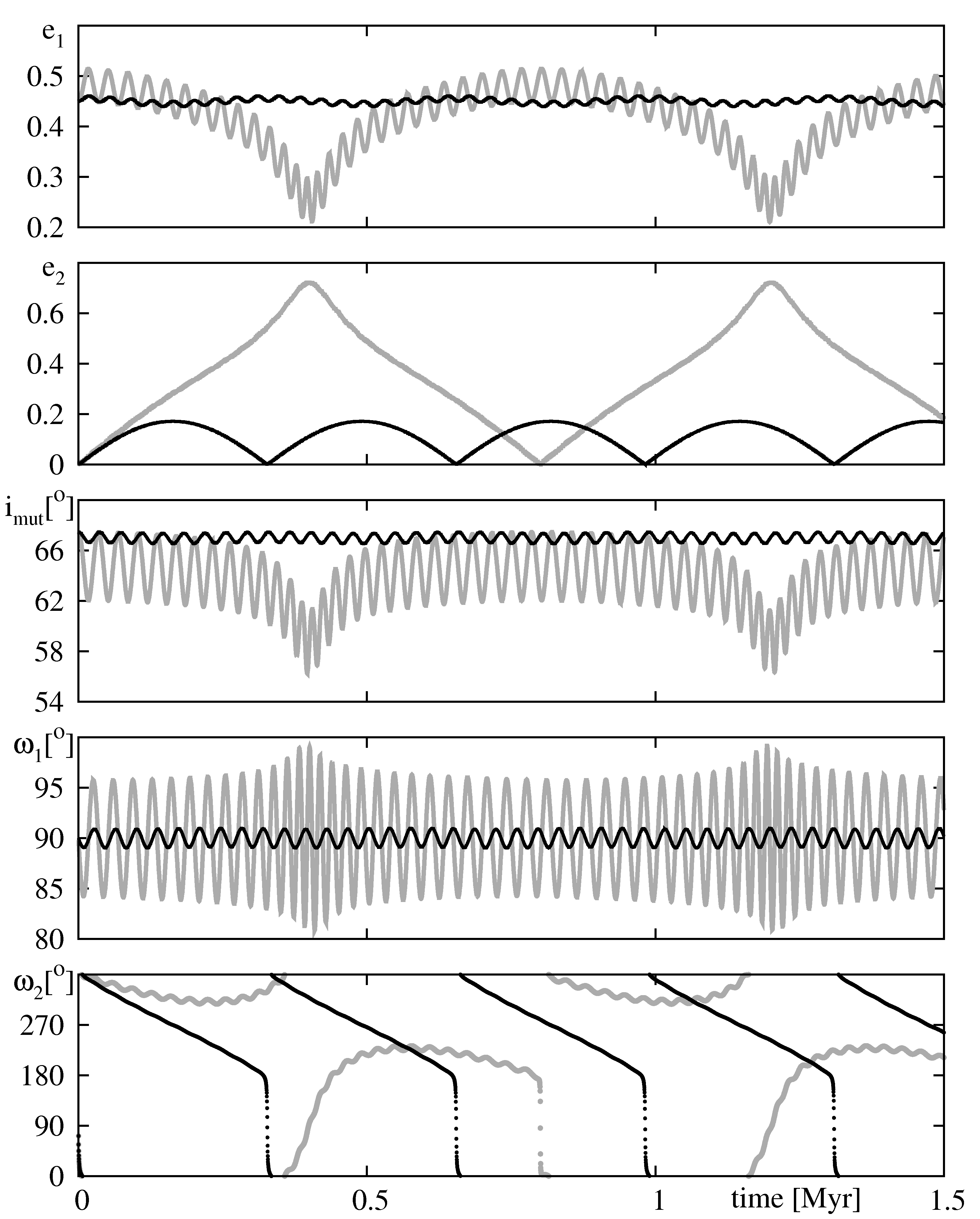}
}
}
\caption{
Evolution of the mean orbits in the following system: $m_0 =
1~\msun$, $m_1 = 100~\mJ$, $m_2 = 10~\mJ$, $a_1 = 0.2~\au$, $a_2 = 5~\au$, $e_1
= 0.45$, $e_2 = 0.001$, $\omega_1 = \pi/2$, $\omega_2 = \pi/2$, $\imut =
67^{\circ}.46$, $\nAMD = 0.215$. Panels from the top to 
the bottom illustrate the evolution
of $e_1, e_2, \imut, \omega_1, \omega_2$, respectively. The grey curves are for
the classic secular model, 
the black curves are for the model including the relativistic
corrections to the potential of the inner binary.
}
\label{fig18}
\end{figure}

Figure~\ref{fig18} illustrates the temporal evolution for an initial 
condition written in the caption. The parameters of this model are the same 
as in  Fig.~\ref{fig15}. We chose the same initial eccentricities $e_1 = 0.45$ 
and $e_2 = 0.001$, and integrate the secular equations of motion of classic 
(grey curves) and relativistic (black curves) models. We note qualitative 
differences between both configurations. The classic model leads to much larger 
variations of the elements. Particularly, the outer eccentricity $e_2$ is 
strongly amplified, compared to the variations in the relativistic 
model. Still, this is not a rule. Inspecting the bottom row of Fig.~\ref 
{fig17}, we can find regions in the \RPs-plane, in which,  for the same 
initial condition, $\max e_2$ becomes larger in the relativistic model than 
in the Newtonian model. Also the secularly chaotic
configurations  appear in quite  
different zones of the phase space in both models. Moreover, the 
relativistic corrections may transform the regular evolution in the classic 
model into the chaotic evolution in the relativistic
systems, and {\em vice versa}. Remarkably, the configuration
illustrated in Fig.~\ref{fig15} has very large masses $m_1 = 100~\mJ$ and 
$m_2 = 10~\mJ$.

%
\section{Conclusions}
%
In this work, we attempt to show that the global features of the secular 
dynamics of the 3-D, non-resonant planetary system depend qualitatively on 
the apparently subtle relativistic corrections to the Newtonian gravity. A 
lesson, which we learned studying the co-planar case \citep
{Migaszewski2009b} is that the {\em non-Newtonian point-mass interactions} 
might be very important for the global dynamics, because, in 
contrary to intuition one may have, the corrections to the Newtonian 
interactions might be {\em not small}, as compared to the mutual 
point-to-point gravity. The numerical analysis of this multi-parameter 
problem is complex, hence the underlining idea of this paper lies in a 
construction of possibly precise analytical model. Essentially simple 
averaging of the perturbing Hamiltonian in \citep {Migaszewski2008}, makes 
it possible to derive the analytic secular theory up to the order of $10$ 
in the semi-major axes ratio $\alpha$. The accuracy of this model may be 
compared with the results of the numerical averaging of the perturbation 
\citep {Michtchenko2004}.  Moreover, for a class of hierarchical systems considered 
in our paper, already the third-order model is precise enough to find and 
investigate the qualitative features of the system. In this work, we focus 
on three-body configurations, e.g., a star and two massive planets or a binary 
stars and one planet. ``Fortunately'', the second-order model is 
integrable, hence the octupole-level approximation might be considered as 
the first order perturbation to this analytically soluble case. This makes it 
possible to understand the sources of instabilities appearing in the full 
(non-averaged) model.

The averaging over mean anomalies reduces the dynamics to a system having 
two DOF, which then may be investigated with the help of rich geometrical 
tools, like the Poincar\'e cross section and the representative planes of 
initial conditions introduced in \cite{Michtchenko2006}. Unfortunately, all 
additional corrections that increase the DOF number must be neglected here, 
as for instance the rotational quadrupole moment of the star and/or of the planets. 
This is the price that must be paid for the possibly global model of the 
dynamics. \edi{In the assumed range of $a_1$ and masses}, the relativistic 
``corrections'' in fact compare with the Newtonian point-mass mutual 
interactions, and are much larger than other perturbations, like the tidal 
and rotational distortions of the bodies. \edi{The analysis of the 
secular frequencies introduced by various corrections justify that 
the model only takes into account the PN perturbation to the potential of 
star and the inner secondary}. 

Having the two DOF model, we investigate the simplest class of 
solutions that are the equilibria. We focus on the Lidov-Kozai resonance 
(LKR) in systems characterized by large range of the mass ratio $\mu$. This 
part extends the results derived for small $\mu$ in \citep 
{Migaszewski2009a}. We found that even much smaller outer body (planetary 
mass $m_2$ with respect to sub-stellar values of $m_1$) moving 
in a wide, highly inclined ($\imut > \sim 60^{\circ}$) orbit may significantly 
perturb the inner orbit. In turn, the restricted model of the circumbinary 
planet, when we assume that the planet does not influence the binary, is not 
generally valid, even if the inner mass $m_1$ is $100$ times larger than the 
outer body $m_2$.

We also studied the parametric structure of families of particular equilibria 
classified as IVa, IIIa, IIIb, IVb+ in \citep{Migaszewski2009a}
for small $\alpha$. Thanks to this assumption, the analytic model 
makes it possible to investigate the transition 
between the planetary regime (small $\mu$) 
and the circumbinary regime ($\mu \sim 50, 100$). This study shows that 
solutions in the planetary problem may disappear for 
large mass ratio.

A particularly interesting feature of the ocupole model is the appearance of 
the {\em secular chaos}. We found that if  $\imut$ exceeds the  critical 
inclination $\icrit(\mu,\alpha)$, the long term evolution of the system may 
be strongly chaotic, leading to large amplification of the eccentricities. 
In the regular regions of the phase space,  the mean angle $\omega_1$ 
librates around $\pm \pi/2$. In some parts of these regions also the second 
secular angle $\omega_2$ librates around $\pm \pi/2$. The initial 
conditions satisfying $\imut > \icrit$ lead to strong amplification of the 
inner eccentricity $e_1$. \corr{Simultaneously, for the same values of the 
angular momentum of the system, we may observe strong amplification of $e_2$
in some region of the phase space, with almost {\em constant} relative 
inclination of the orbits. This behaviour may be attributed to unstable 
equilibrium IIIb emerging in the secular system.} The amplification of $e_1$
happens not only for librations of $\omega_1$ around $\pm \pi/2$ (in the 
LK regime), but also and particularly when this angle behaves chaotically, 
varying in the whole range of $[0,2\pi]$. The dynamical maps reveal that 
the primarily source of the chaotic motions are the unstable equilibria and 
unstable periodic orbits in the full system, following the appearance of 
separatrices in the integrable, quadrupole-level model.

Thanks to the simple analytic model, the influence of relativistic 
correction $\Hrel$ on the global secular dynamics of the problem might be 
clearly demonstrated. A simple proof of this influence is 
provided by the analysis of the equilibria in the perturbed model. 
The differences between the Newtonian and relativistic 
models are larger when the mutual interactions between the 
secondaries are weaker, e.g., when companion masses are smaller. \corr{Yet
the dynamics are basically very simple up to the limit of the critical
inclination, when the first bifurcation of the origin ($e_1=e_2=0$) occurs,
and this feature of the dynamics is preserved in both models.}

\edi{
We stress that although the analysis is done for specific, discrete mass 
ratios, the results are valid as far the assumptions of the averaging 
theorem are fulfilled and the corrections besides general relativity are 
negligible. We demonstrated that similarly to the co-planar problem, the global 
3-D dynamics  of the classic, Newtonian model essentially depend only on the 
ratios of semi-major axes and masses of the secondaries. Hence, although we 
consider mostly the circumbinary configurations, the results may be are 
easily extrapolated to the ``typical'' planetary regime, investigated 
already \citep {Michtchenko2006}. Moreover, when the PN corrections are 
added to the model, the dynamics are much more complex. Still  the global 
picture of the phase space is determined by the ratio of these corrections 
and the Newtonian mutual interactions. Hence, if the system scales down, 
while this ratio is roughly preserved, the structure of the phase space, 
determined by stationary solutions of the secular system should not 
essentially change. 
}

Our approach may be generalized for other perturbations, like the rotational 
and conservative tidal distortion of the bodies in the system. 
Unfortunately, in the most general case, the dimension of the hierarchical 
system cannot be generally reduced to two DOF. Moreover, these perturbations 
lead to even more interesting and intriguing dynamics, which we investigated 
in the co-planar and spatial case \citep 
[e.g.,][]{Fabrycky2007,Mardling2007,Ragozzine2009,Migaszewski2009b}. We work 
on a global approach, suitable for the 3D systems, \edi{aiming to publish 
these results in future papers.} 
%
\section*{Acknowledgements}
%
We thank the anonymous referee for comments that improved the manuscript.
This work is supported by the Polish Ministry of Science and 
Higher Education, through grants
N/N203/402739  and 92/N-ASTROSIM/2008/0. CM is a recipient of the stipend of the Foundation for Polish Science (programme START, edition 2010).
%
\bibliographystyle{mn2e}
%
\label{lastpage}
\bibliography{ms}

\begin{thebibliography}{}

\bibitem[\protect\citeauthoryear{{Adams} \& {Laughlin}}{{Adams} \&
  {Laughlin}}{2006}]{Adams2006}
{Adams} F.~C.,  {Laughlin} G.,  2006, International Journal of Modern Physics
  D, 15, 2133

\bibitem[\protect\citeauthoryear{{Arnold}, {Kozlov} \& {Neishtadt}}{{Arnold}
  et~al.}{1993}]{Arnold1993}
{Arnold} V.~I.,  {Kozlov} V.~V.,    {Neishtadt} A.~I.,  1993, {Dynamical
  systems III. Mathematical aspects of classical and celestial mechanics}.
Encyclopaedia of mathematical sciences, Springer Verlag

\bibitem[\protect\citeauthoryear{{Batygin}, {Bodenheimer} \&
  {Laughlin}}{{Batygin} et~al.}{2009}]{Batygin2009}
{Batygin} K.,  {Bodenheimer} P.,    {Laughlin} G.,  2009, ApJL, 704, L49

\bibitem[\protect\citeauthoryear{{Brouwer} \& {Clemence}}{{Brouwer} \&
  {Clemence}}{1961}]{Brouwer1961}
{Brouwer} D.,  {Clemence} G.~M.,  1961, {Methods of celestial mechanics}.
New York: Academic Press, 1961

\bibitem[\protect\citeauthoryear{{Brumberg}}{{Brumberg}}{2007}]{Brumberg2007}
{Brumberg} V.,  2007, Celestial Mechanics and Dynamical Astronomy, 99, 245

\bibitem[\protect\citeauthoryear{{Eggenberger}}{{Eggenberger}}{2010}]{Eggenber%
ger2010}
{Eggenberger} A.,  2010, in {K.~Go{\'z}dziewski, A.~Niedzielski, \&
  J.~Schneider} ed., EAS Publications Series Vol.~42.
pp 19--37

\bibitem[\protect\citeauthoryear{{Fabrycky} \& {Tremaine}}{{Fabrycky} \&
  {Tremaine}}{2007}]{Fabrycky2007}
{Fabrycky} D.,  {Tremaine} S.,  2007, ApJ, 669, 1298

\bibitem[\protect\citeauthoryear{{Farago} \& {Laskar}}{{Farago} \&
  {Laskar}}{2010}]{Farago2010}
{Farago} F.,  {Laskar} J.,  2010, MNRAS, 401, 1189

\bibitem[\protect\citeauthoryear{{Ferraz-Mello}}{{Ferraz-Mello}}{2007}]{Ferraz%
Mello2007}
{Ferraz-Mello} S.,  ed. 2007, {Canonical Perturbation Theories - Degenerate
  Systems and Resonance} Vol.~345 of Astrophysics and Space Science Library

\bibitem[\protect\citeauthoryear{{Ferrer} \& {Os{\'a}car}}{{Ferrer} \&
  {Os{\'a}car}}{1994}]{Ferrer1994}
{Ferrer} S.,  {Os{\'a}car} C.,  1994, Celestial Mechanics and Dynamical
  Astronomy, 60, 187

\bibitem[\protect\citeauthoryear{{Ford}, {Kozinsky} \& {Rasio}}{{Ford}
  et~al.}{2000}]{Ford2000}
{Ford} E.~B.,  {Kozinsky} B.,    {Rasio} F.~A.,  2000, ApJ, 535, 385

\bibitem[\protect\citeauthoryear{{Gronchi} \& {Milani}}{{Gronchi} \&
  {Milani}}{1998}]{Gronchi1998}
{Gronchi} G.~F.,  {Milani} A.,  1998, Celestial Mechanics and Dynamical
  Astronomy, 71, 109

\bibitem[\protect\citeauthoryear{{Harrington}}{{Harrington}}{1968}]{Harrington%
1968}
{Harrington} R.~S.,  1968, AJ, 73, 190

\bibitem[\protect\citeauthoryear{{Kozai}}{{Kozai}}{1962}]{Kozai1962}
{Kozai} Y.,  1962, AJ, 67, 579

\bibitem[\protect\citeauthoryear{{Krasinskii}}{{Krasinskii}}{1972}]{Krasinsky1%
972}
{Krasinskii} G.~A.,  1972, Celestial Mechanics, 6, 60

\bibitem[\protect\citeauthoryear{{Krasinskii}}{{Krasinskii}}{1974}]{Krasinsky1%
974}
{Krasinskii} G.~A.,  1974 Vol.~62 of IAU Symposium.
pp 95--116

\bibitem[\protect\citeauthoryear{{Laskar}}{{Laskar}}{1990}]{Laskar1990}
{Laskar} J.,  1990, Icarus, 88, 266

\bibitem[\protect\citeauthoryear{{Laskar}}{{Laskar}}{2000}]{Laskar2000}
{Laskar} J.,  2000, Physical Review Letters, 84, 3240

\bibitem[\protect\citeauthoryear{{Lee} \& {Peale}}{{Lee} \&
  {Peale}}{2003}]{Lee2003}
{Lee} M.~H.,  {Peale} S.~J.,  2003, ApJ, 592, 1201

\bibitem[\protect\citeauthoryear{{Libert} \& {Henrard}}{{Libert} \&
  {Henrard}}{2007}]{Libert2007b}
{Libert} A.-S.,  {Henrard} J.,  2007, Icarus, 191, 469

\bibitem[\protect\citeauthoryear{{Lidov}}{{Lidov}}{1962}]{Lidov1962}
{Lidov} M.~L.,  1962, Planetary and Space Science, 9, 719

\bibitem[\protect\citeauthoryear{{Lidov} \& {Ziglin}}{{Lidov} \&
  {Ziglin}}{1976}]{Lidov1976}
{Lidov} M.~L.,  {Ziglin} S.~L.,  1976, Celestial Mechanics, 13, 471

\bibitem[\protect\citeauthoryear{{Mardling}}{{Mardling}}{2007}]{Mardling2007}
{Mardling} R.~A.,  2007, MNRAS, 382, 1768

\bibitem[\protect\citeauthoryear{{Mardling}}{{Mardling}}{2010}]{Mardling2010}
{Mardling} R.~A.,  2010, MNRAS, 407, 1048

\bibitem[\protect\citeauthoryear{{Michtchenko}, {Ferraz-Mello} \&
  {Beaug{\'e}}}{{Michtchenko} et~al.}{2006}]{Michtchenko2006}
{Michtchenko} T.~A.,  {Ferraz-Mello} S.,    {Beaug{\'e}} C.,  2006, Icarus,
  181, 555

\bibitem[\protect\citeauthoryear{{Michtchenko} \& {Malhotra}}{{Michtchenko} \&
  {Malhotra}}{2004}]{Michtchenko2004}
{Michtchenko} T.~A.,  {Malhotra} R.,  2004, Icarus, 168, 237

\bibitem[\protect\citeauthoryear{{Migaszewski} \&
  {Go{\'z}dziewski}}{{Migaszewski} \&
  {Go{\'z}dziewski}}{2008}]{Migaszewski2008}
{Migaszewski} C.,  {Go{\'z}dziewski} K.,  2008, MNRAS, 388, 789

\bibitem[\protect\citeauthoryear{{Migaszewski} \&
  {Go{\'z}dziewski}}{{Migaszewski} \&
  {Go{\'z}dziewski}}{2009a}]{Migaszewski2009a}
{Migaszewski} C.,  {Go{\'z}dziewski} K.,  2009a, MNRAS, 395, 1777

\bibitem[\protect\citeauthoryear{{Migaszewski} \&
  {Go{\'z}dziewski}}{{Migaszewski} \&
  {Go{\'z}dziewski}}{2009b}]{Migaszewski2009b}
{Migaszewski} C.,  {Go{\'z}dziewski} K.,  2009b, MNRAS, 392, 2

\bibitem[\protect\citeauthoryear{{Migaszewski} \&
  {Go{\'z}dziewski}}{{Migaszewski} \&
  {Go{\'z}dziewski}}{2010}]{Migaszewski2010}
{Migaszewski} C.,  {Go{\'z}dziewski} K.,  2010, in {K.~Go{\.z}dziewski,
  A.~Niedzielski, \& J.~Schneider} ed., EAS Publications Series Vol.~42, {}.
pp 385--391

\bibitem[\protect\citeauthoryear{{Murray} \& {Dermott}}{{Murray} \&
  {Dermott}}{2000}]{Murray2000}
{Murray} C.~D.,  {Dermott} S.~F.,  2000, {Solar System Dynamics}.
Cambridge Univ. Press

\bibitem[\protect\citeauthoryear{{Rabl} \& {Dvorak}}{{Rabl} \&
  {Dvorak}}{1988}]{Rabl1988}
{Rabl} G.,  {Dvorak} R.,  1988, A\&A, 191, 385

\bibitem[\protect\citeauthoryear{{Ragozzine} \& {Wolf}}{{Ragozzine} \&
  {Wolf}}{2009}]{Ragozzine2009}
{Ragozzine} D.,  {Wolf} A.~S.,  2009, ApJ, 698, 1778

\bibitem[\protect\citeauthoryear{{Richardson} \& {Kelly}}{{Richardson} \&
  {Kelly}}{1988}]{Richardson1988}
{Richardson} D.~L.,  {Kelly} T.~J.,  1988, Celestial Mechanics, 43, 193

\bibitem[\protect\citeauthoryear{{Takeda}, {Kita} \& {Rasio}}{{Takeda}
  et~al.}{2009}]{Tekada2009}
{Takeda} G.,  {Kita} R.,    {Rasio} F.~A.,  2009, in IAU Symposium Vol.~253 of
  IAU Symposium, {}.
pp 181--187

\bibitem[\protect\citeauthoryear{{{Tamuz}, O., {et al.,}}}{{{Tamuz}, O., {et
  al.,}}}{2008}]{Tamuz2008}
{{Tamuz}, O., {et al.,}} 2008, A\&A, 480, L33

\bibitem[\protect\citeauthoryear{{{\v S}idlichovsk{\'y}} \&
  {Nesvorn{\'y}}}{{{\v S}idlichovsk{\'y}} \&
  {Nesvorn{\'y}}}{1996}]{Sidlichovsky1996}
{{\v S}idlichovsk{\'y}} M.,  {Nesvorn{\'y}} D.,  1996, Celestial Mechanics and
  Dynamical Astronomy, 65, 137

\end{thebibliography}

\end{document}